\documentclass[aps,prd,nofootinbib,twocolumn,reprint,groupedaddress,showpacs,showkeys,floatfix]{revtex4-1}
\usepackage {amsmath}
\usepackage {amssymb}
\usepackage {amsfonts}
\usepackage {amsthm}
\usepackage {mathrsfs}
\usepackage {natbib}
\usepackage {latexsym}
\usepackage {graphicx}
\usepackage {txfonts}
\usepackage {rotating}
\usepackage {wasysym}
\usepackage {multirow}
\usepackage {hhline}
\usepackage {hyperref}
\usepackage[usenames,dvipsnames]{xcolor}
\usepackage {bm}
\usepackage{appendix}
\usepackage{acronym}
\usepackage{dcolumn}   
\usepackage {url}
\usepackage {caption}
\usepackage {subcaption}
\usepackage {mathtools}
\usepackage {enumerate}
\usepackage {acronym}
\pdfoutput=1
\newcommand{\dcc}{LIGO-P1500247-v1}

\begin{document}

\title{Systematic errors in estimation of gravitational-wave candidate significance}
\author{C.~Capano}
\affiliation{Max Planck Institute for Gravitational Physics,
  Callinstra\ss e 38, D-30167, Hannover, Germany}
\author{T.~Dent}
\affiliation{Max Planck Institute for Gravitational Physics,
  Callinstra\ss e 38, D-30167, Hannover, Germany}
\author{C.~Hanna}
\affiliation{The Pennsylvania State University, University Park, PA 16802, USA}
\author{Y.-M.~Hu}
\email{yiming.hu@aei.mpg.de}
\affiliation{Max Planck Institute for Gravitational Physics,
  Callinstra\ss e 38, D-30167, Hannover, Germany}
\affiliation{School of Physics and Astronomy, Kelvin Building,
  University of Glasgow, Glasgow, G12 8QQ, UK}
\affiliation{Tsinghua University, Beijing, 100084, China}
\author{M.~Hendry}
\affiliation{School of Physics and Astronomy, Kelvin Building,
  University of Glasgow, Glasgow, G12 8QQ, UK}
\author{C.~Messenger}
\affiliation{School of Physics and Astronomy, Kelvin Building,
  University of Glasgow, Glasgow, G12 8QQ, UK}
\author{J.~Veitch} 
\affiliation{School of Physics and Astronomy,
  University of Birmingham, Edgbaston, Birmingham, B15 2TT, UK}
\date{\today\\\mbox{\dcc}}

\begin{abstract}
  We investigate a critical issue in determining the statistical 
  significance of candidate transient gravitational-wave events in 
  a ground-based interferometer network.  Given the presence of 
  non-Gaussian noise artefacts in real data, the noise background 
  must be estimated empirically from the data itself. However, the
  data also potentially contains signals, thus the background 
  estimate may be overstated due to contributions from signals.  
  It has been proposed to mitigate possible bias by removing 
  single-detector data samples that pass a multi-detector 
  consistency test (and thus form coincident events) from the 
  background estimates. 
  We conduct a high-statistics Mock Data Challenge to evaluate the 
  effects of removing such samples, modelling a range of scenarios 
  with plausible detector noise distributions and with a range of 
  plausible foreground astrophysical signal rates; thus, we are able 
  to calculate the exact false alarm probabilities of candidate events 
  in the chosen noise distributions. 
  We consider two different modes of selecting the samples used for 
  background estimation: one where coincident 
  samples are removed, and one where all samples are retained and used.
  Three slightly different algorithms for calculating the false alarm 
  probability of candidate events are each deployed in these two 
  modes. The three algorithms show good consistency with each other; however, 
  discrepancies arise between the results obtained under the 
  `coincidence removal' and `all samples' modes, for false alarm 
  probabilities below a certain value. 
  In most scenarios the median of the \ac{FAP} 
  estimator under the `all samples' mode is consistent with the 
  exact \ac{FAP}.  On the other hand the 
  `coincidence removal' mode is found to be unbiased for the mean of 
  the estimated \ac{FAP} over realisations. 
  While the numerical values at which discrepancies become apparent
  are specific to the details of our numerical experiment, 
  we believe that the qualitative differences in the behaviour of
  the median and mean of the \ac{FAP} estimator have 
  more general validity.  On the basis of our study we suggest that
  the \ac{FAP} of candidates for the first detection 
  of gravitational waves should be estimated \emph{without} removing 
  single-detector samples that form coincidences. 
\end{abstract}

\maketitle 

\acrodef{GW}[GW]{gravitational wave}
\acrodef{MDC}[MDC]{mock data challenge}
\acrodef{SNR}[SNR]{signal-to-noise ratio}
\acrodef{CBC}[CBC]{compact binary coalescence}
\acrodef{NS}[NS]{neutron-star}
\acrodef{LIGO}[LIGO]{Laser Interferometer Gravitational-wave
  Observatory}
\acrodef{LSC}[LSC]{LIGO Scientific Collaboration}
\acrodef{LVC}[LVC]{LSC (LIGO Scientific Collaboration) and Virgo collaborations}
\acrodef{CDF}[CDF]{cumulative distribution function}
\acrodef{PDF}[PDF]{probability distribution function}
\acrodef{FAR}[FAR]{false alarm rate}
\acrodef{FAP}[FAP]{false alarm probability}
\acrodef{aLIGO}[aLIGO]{Advanced LIGO}
\acrodef{adV}[adV]{Advanced Virgo}
\acrodef{ROC}[ROC]{Receiver Operating Characteristic}
\acrodef{FPR}[FPR]{false positive rate}
\acrodef{TPR}[TPR]{true positive rate}
\acrodef{IQR}[IQR]{inter-quartile range}
\acrodef{APC}[APC]{all possible coincidences}
\acrodef{CMB}[CMB]{cosmic microwave background}

%
%
\section{Introduction\label{sec:introduction}}
%
%
The global network of advanced \ac{GW} detectors is poised to make its first
direct detection \cite{Harry:2010zz,TheLIGOScientific:2014jea,TheVirgo:2014hva,Aasi:2013wya}.  The
coalescence of binary systems containing neutron stars and/or black holes is
the most likely source of transient gravitational waves \cite{ratesdoc} and the
detection of such a \ac{CBC} event would open the new window of \ac{GW}
astronomy \cite{SathyaSchutzLRR}.  The observation of \ac{CBC} events would not
only allow us to test General Relativity, but also help to give hint on the
equation-of-state of \acp{NS}.  The study of populations of \ac{CBC} events
would help to deepen our understanding of stellar evolution for binary massive
stars, especially the very late stages \cite{Dominik2015,Morscher2015,Antonini2015}.

%
%
To support possible detections, it will be necessary to determining the 
confidence that the candidate signals are associated with 
astrophysical sources of \acp{GW}~\cite{GW100916} rather than spurious 
noise events.  Thus, a \ac{FAP} estimate is produced in 
order to classify candidate events. 
Claims for the detection of previously undetected or unknown physical 
phenomena have been held up to high levels of scrutiny, e.g.\ the Higgs 
boson~\cite{Aad2012} and B-modes in the polarization of the \ac{CMB}~\cite{Ade2014}.
The same will be true for the direct detection of \acp{GW}: a high level of
statistical confidence will be required as well as a thorough understanding of 
the accuracy and potential systematic bias of procedures to determine the 
\ac{FAP} of a candidate event.

%
%
Existing \ac{CBC} search pipelines (\emph{e.g.}\ \cite{ihope,GstLAL,Cannon2013FAR}) 
assess the \ac{FAP} 
of a detection candidate by estimating its probability under the
null hypothesis.  The null hypothesis is that candidate events
are caused by non-gravitational-wave processes acting on and within the
interferometers.  In order to determine this probability, we assign a test
statistic value to every possible result of a given observation or 
experiment performed on the data: larger values indicate a higher deviation 
from expectations under the null hypothesis.  We compute the \ac{FAP} as the 
probability of obtaining a value of the test statistic
equal to, or larger than, the one actually obtained in the experiment.  The
smaller this probability, the more significant is the candidate.

%
%
In general the detector output data streams are the sum of terms due to
non-astrophysical processes, known as background noise, and of astrophysical
\ac{GW} signals, labelled foreground.  If we were able to account for all
contributions to noise using predictive and reliable physical models, we
would then obtain an accurate calculation of the \ac{FAP}
of any observation. However, for real \ac{GW} detectors, in addition to terms
from known and well-modelled noise processes, the data contains large numbers
of non-Gaussian noise transients (``glitches'', see for example
\cite{Aasi2014b,Chassande-Mottin2013}) whose sources are either unknown or not
accurately modelled, and which have potentially large effects on searches for 
transient \acp{GW}.  Even given all the information available from
environmental and auxiliary monitor channels at the detectors, many such noise
transients cannot be predicted with sufficient accuracy to account for their
effects on search algorithms. Thus, for transient searches in real detector
noise it is necessary to determine
the background noise distributions empirically, i.e.\ directly from the strain data. 

%
%
Using the data to empirically estimate the background has notable potential
drawbacks.  It is not possible to operate \ac{GW} detectors so as to `turn
off' the astrophysical foreground and simply measure the background; if the
detector is operational then it is always subject to both background and
foreground sources.  In addition, our knowledge of the background noise
distribution is limited by the finite amount of data available.  This
limitation applies especially in the interesting region of low event
probability under the noise hypothesis, corresponding to especially rare
noise events.

%
%
\ac{CBC} signals are expected to be well modelled by the predictions of
Einstein's general relativity~\cite{Cutler:1994}.  The detector data are
cross-correlated (or matched filtered)~\cite{Finn1992} against a bank of
template \ac{CBC} waveforms, resulting in an \ac{SNR} time series for each
template~\cite{Allen:2005fk}. If this time series crosses a predetermined
threshold, the peak value and the time of the peak are recorded as a 
{\it trigger}. Since \acp{GW} propagate at the speed of light (see e.g.\ 
\cite{SathyaSchutzLRR}), the arrival times of signals will differ between 
detectors by ${\simeq}40$\,ms or less, i.e. Earth's light crossing time. 
Differences in arrival times are governed by the direction of the source on the
sky in relation to the geographical detector locations.  We are thus able to
immediately eliminate the great majority of background noise by rejecting any
triggers which are not coincident in two or more detectors within a predefined
coincidence time window given by the maximum light travel time plus 
trigger timing errors due to noise.  Only triggers coincident in multiple
detectors with consistent physical parameters such as binary component 
masses are considered as candidate detections.  In order to make a
confident detection claim one needs to estimate the rarity, i.e., the {\em
\ac{FAP}\/} of an event, and claim detection only when the
probability of an equally loud event being caused by noise is below 
a chosen threshold.

%
%
The standard approach used in \ac{GW} data analysis for estimating the
statistical properties of the background is via analysis of time-shifted data,
known as ``time slides''~\cite{Babak:2012zx,Abbott:2009qj}.  This method exploits 
the coincidence requirement of foreground events by time-shifting the data 
from one detector relative to another.  Such a shift, if larger than the 
coincidence window, would prevent a zero-lag signal remaining untouched
in a time-shifted analysis.  Therefore, from a single time-shifted
analysis (``time slide'') the output coincident events should represent one
realisation of the background distribution of coincident events, given
the sets of single-detector triggers, assuming that the background 
distributions are not correlated between detectors. 
By performing many time slide
analyses with different relative time shifts we may accumulate instances of
background coincidences and thus estimate their rate and distribution. 

%
%
The time-slides approach has been an invaluable tool in the analysis and
classification of candidate \ac{GW} events in the initial detector era. 
Initial LIGO made no detections, however note that
in 2010, the \ac{LVC} performed a blind injection challenge~\cite{GW100916}
to check the robustness and confidence of the pipeline.  
A signal was injected in ``hardware'' (by actuating the test masses) in 
the global network of interferometers and analysed by the collaboration
knowing only that there was the possibility of such an artificial event. 
The blind injection was recovered by the templated binary inspiral search
with a high significance~\cite{Colaboration:2011np}; however, the blind 
injection exercise highlighted potential issues with the use of 
time-shifted analysis to estimate backgrounds in the presence of 
astrophysical (or simulated) signals. 

%
%
Simply time-shifting detector outputs with respect to each other does not
eliminate the possibility of coincident events resulting from foreground 
(signal) triggers from one detector passing the coincidence test with
random noise triggers in another detector. Thus, the ensemble of samples
generated by time-shifted analysis may be ``contaminated'' by the presence of
foreground events in single-detector data, compared to the ensemble that would
result from the noise background alone. The distribution of events caused by
astrophysical signals is generally quite different from that of noise: it is 
expected to have a longer tail towards high values of \ac{SNR} (or other
event ranking test-statistic used in search algorithms). Thus, depending on the
rate of signals and on the realisation of the signal process obtained in any
given experiment, such contamination could considerably bias the estimated
background.  If the estimated background rate is raised by the presence of 
signals, the \ac{FAP} of coincident search events (in 
non-time-shifted or ``zero-lag'' data) may be overestimated, implying
a conservative bias in the estimated \ac{FAP}.  The expected 
number of false detection claims will not increase due to the presence of 
signals in time-slide analyses, however some signals may fail to be detected 
due to an elevated background estimate. \footnote{The
reader may ask what a ``false detection claim'' means if signals are present.
This refers to cases where the search result contains both foreground events, 
and background events with comparable or higher ranking statistic values. 
In particular, if the loudest event in the search is due to background a 
detection claim would be misleading.}

%
%
Besides the standard time-shift approach, other background estimation
techniques have been developed~\cite{Cannon2013FAR,Farr:2015,Capano:2015}. 
All are variants on one key 
concept: events that are not coincident within the physically allowed time 
window cannot be from the same foreground
source. All therefore use non-coincident triggers as proxies for the 
distribution of background events due to detector noise.  Differences between
methods occur in implementation, and fall into two categories. In the 
standard scheme, many time slides are applied to a set of triggers from a
multi-detector observation and all resultant coincidences are retained. We
label this the `all samples' mode of operation.
Concern about the potential conservative bias of including foreground \ac{GW}
events of astrophysical original in the time-shifted
distribution motivates a modification to this procedure: 
one can instead choose to first
identify all coincidences between different detectors in zero-lag and exclude
them from the subsequent time-slide analysis.
We label this the `coincidence removal' mode 
of operation. 

%
%
In this paper we describe the results of an \ac{MDC} in which participants
applied three different background estimation algorithms, each applied to 
the \ac{MDC} data in the two modes of operation described above: the 
`coincidence removal' and `all samples' modes. Since some 
aspects of the \ac{MDC} data generation were not fully realistic, the 
algorithms were simplified compared to their use on real data; two were based
on methods currently in use in the \ac{LVC}, while the third introduces a new
approach.
The \ac{MDC} consisted of simulated realisations of single-detector triggers --
maxima of matched filter \ac{SNR} $\rho$ above a fixed threshold value chosen
as $\rho_0=5.5$.  The trigger \ac{SNR} values were generated according to
analytically modelled background and foreground distributions unknown to the
\ac{MDC} participants.  The background distributions spanned a range of 
complexity including a realistic Initial detector distribution \cite{Briggs:2012ce}.  
The foreground triggers were chosen to model one of the most likely first 
sources, binary neutron stars, with an expected detection rate from zero 
to the maximum expected for the first Advanced detector science runs~\cite{ratesdoc}.
Participants were asked to apply their algorithms to these 
datasets and report, in both modes, the estimated \ac{FAP}
of the loudest event found in each realisation.  The \ac{MDC} 
used analytic formulae to generate the background distributions; although 
we will not, in reality, have access to such formulae, they allow us to 
compute an ``exact'' \ac{FAP} semi-analytically.  We use 
these exact values, alongside other figures of merit, as a benchmark for 
comparing the results obtained in the `coincidence removal' and `all 
samples' modes, both within each background estimation algorithm and 
between the different algorithms.

%
%
In the following section of this paper we provide details of the \ac{MDC},
describing the background and foreground distributions, the data 
generating procedure, and the ``exact'' \ac{FAP} 
calculation.  In Sec.~\ref{sec:estimation} we
then describe the different methods of background estimation used in this
\ac{MDC}.  We report the results of the challenge in Sec.~\ref{sec:results},
comparing and contrasting the results obtained from each of the algorithms in
each of their 2 modes of operation.  Finally in Sec.~\ref{sec:discussion} we
provide a summary and present our conclusions.

%
%
\section{The mock data challenge\label{sec:MDC}}
%
%
Our \ac{MDC} is designed to resolve the question of whether to remove
(`coincidence removal') or not to remove (`all samples') coincident zero-lag 
events from the data used to estimate \ac{FAP}.  Simulated 
single detector triggers are generated from
known background and 
foreground distributions: challenge participants are then given lists of 
trigger arrival times and single detector \acp{SNR}.  The ranking statistic 
used for each coincident trigger is the combined multi-detector \ac{SNR}. The 
calculation of \ac{FAP} could be carried out with any 
appropriate ranking statistic.  When applied to real data, different 
pipelines may define their ranking statistics in slightly different ways; in 
order to make a fair comparison in the challenge we use the multi-detector 
\acp{SNR} as a
simplified ranking statistic. The \acp{CDF} of the \ac{SNR} in each detector
are described analytically and hidden from the challenge participants.  With
this analytic form, the challenge designers are able to compute the exact 
\ac{FAP} at the ranking statistic value of the loudest coincident
event in each realisation. 

%
%
The challenge consisted of 14 independent experiments, each with a different
foreground and background distribution for which $10^5$ observational 
realisations are generated (see section \ref{sec:triggerParams}). 
This number is large enough to reach the interesting statistics region while computationally feasible.
Each realisation contained, on average,
${\sim}10^4$ single-detector triggers in each of two detectors.  A
realisation should be considered analogous to a single \ac{GW} observing run.
Participants were asked to estimate the \ac{FAP} of the loudest
coincident event in each of the $10^5$ realisations for each experiment.  The 
14 experiments cover a variety of background distributions; the foregrounds 
differ in that the astrophysical rate ranges between zero, i.e., no
events, and a relatively high value corresponding to ${\sim}3$ detections per
realisation.  Analysis of the differences between the estimated and exact
\ac{FAP} values enable us to quantify the effect of
removing coincident zero-lag triggers in the background estimate, as opposed
to retaining all samples.  It also allows us to directly
compare implementations between different algorithms in each mode of 
operation.  Finally, it allows us to quantify the accuracy and limiting 
precision of our \ac{FAP} estimates and thus their uncertainties.

%
%
We divide our simulations into three groups according to astrophysical rate, 
and independently into three groups according to background distribution 
complexity (``simple'', ``realistic'' and ``extreme'').  To this nine 
combinations we have appended an additional four simulations, three of which 
have exactly the same background distributions as three of the original nine 
but contain no signals.  The final simulation contains a background
distribution with a deliberately extended tail such that the generation of
particularly loud background triggers is possible.  The primary properties of
each experiment are given in Table~\ref{tab:matrix} and details are listed 
in Tables~\ref{tab:paramsBkgd},~\ref{tab:paramsTail} and~\ref{tab:paramsRate}.
\begin{table}
    \begin{tabular}{c p{0.2cm} c p{0.1cm} c p{0.1cm} c p{0.1cm} c}
    & &\multicolumn{7}{c}{Background property} \\
    Foreground rate & & simple & & realistic & & extreme & & ext.~tail\\
    \cline{1-1}\cline{3-9}
        zero	& & 1,3   & & 12    & & 14   & &  - \\
        low 	& &  -    & & 10    & & 2    & &  7 \\
        medium  & & 9,13  & & 8     & & 6    & &  - \\
        high    & & 5     & & 11    & & 4    & &  - \\
   \end{tabular}
   \caption{The classification of each experiment in the \ac{MDC} in 
     terms of background complexity and astrophysical foreground rate. 
     See main text for definitions.
\label{tab:matrix}}
\end{table}.
A \emph{low} foreground rate corresponds to ${<}0.01$ expected coincident
signal triggers per realisation, a \emph{medium} rate corresponds to
$0.01$--$1$ coincidences per realisation, and \emph{high} rate corresponds to
${>}1$ per realisation.  We do not consider foreground rates above ${\sim}3$
coincidences per realisation since we are motivated by \ac{FAP}
estimation for the first advanced era \ac{GW} detections.

%
%
\subsection{Modelling the detector noise backgrounds\label{sec:background}}
%
%
The \ac{CDF} of the background single-detector \ac{SNR} triggers is modelled as
the exponential of a piecewise polynomial function in the \ac{SNR} $\rho$ via
\begin{equation}\label{eq:CDF}
  \mathcal{C}(\rho) =
  \begin{dcases*}      
    1-\exp{\left(\sum\limits_{i=0}^{6}
        a_{i}\left(\rho-\rho_{\text{th}}\right)^{i}\right)}, & for
    $\rho\leq\rho_{\text{sp}}$ \\
    1-C_{\text{sp}}\exp{\left(b\left(\rho-\rho_{\text{sp}}\right)\right)},
    & for $\rho>\rho_{\text{sp}}$,
  \end{dcases*}
\end{equation}
where the trigger generation threshold is set as $\rho_{\text{th}}=5.5$.  The
polynomial coefficients $a_i$ must satisfy the constraint that the \ac{CDF}
remains monotonic in $\rho$; additionally, $a_0$ is determined by the 
constraint that the \ac{CDF} should range between $0$ and $1$.  We define the
\ac{CDF} differently in the regions below and above a switching-point
$\rho_{\text{sp}}$ value in order to satisfy the constraints on the \ac{CDF}
model, such that the \ac{CDF} and its derivative with respect to $\rho$ are 
continuous at the switching point.  Hence, a choice of $C_{\text{sp}}$ 
determines the values of $\rho_{\text{sp}}$ and $b$.
Details of the background distribution parameters chosen for each simulation
can be found in Appendix~\ref{sec:triggerParams}; here we describe the broader
properties of the chosen distributions.  

\begin{figure*}
  \centering
    \begin{subfigure}[b]{0.33\textwidth}
      \centering
      \includegraphics[width=\textwidth]{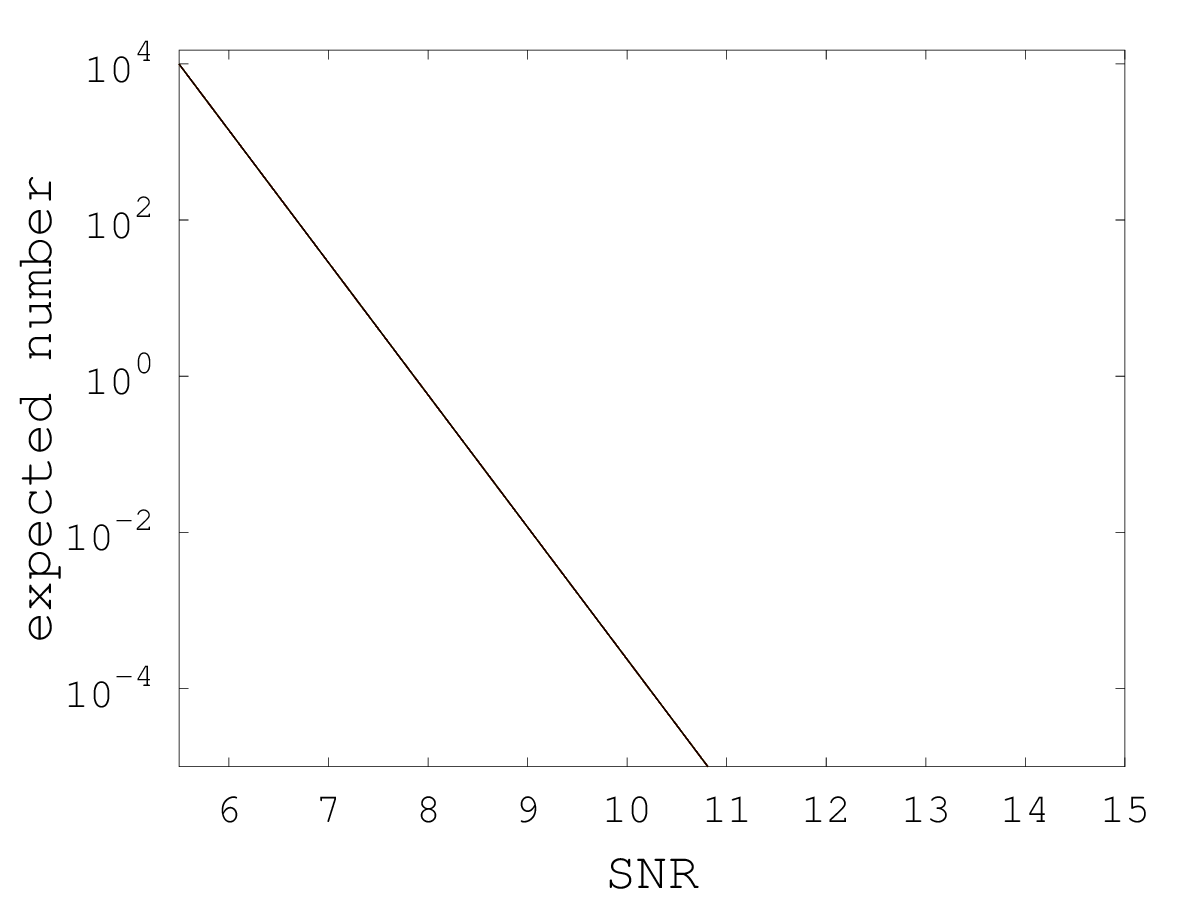}
      \caption{A ``simple'' example background (experiment 3).\label{fig:examplesimple}}
    \end{subfigure}
    \begin{subfigure}[b]{0.33\textwidth}
      \centering
      \includegraphics[width=\textwidth]{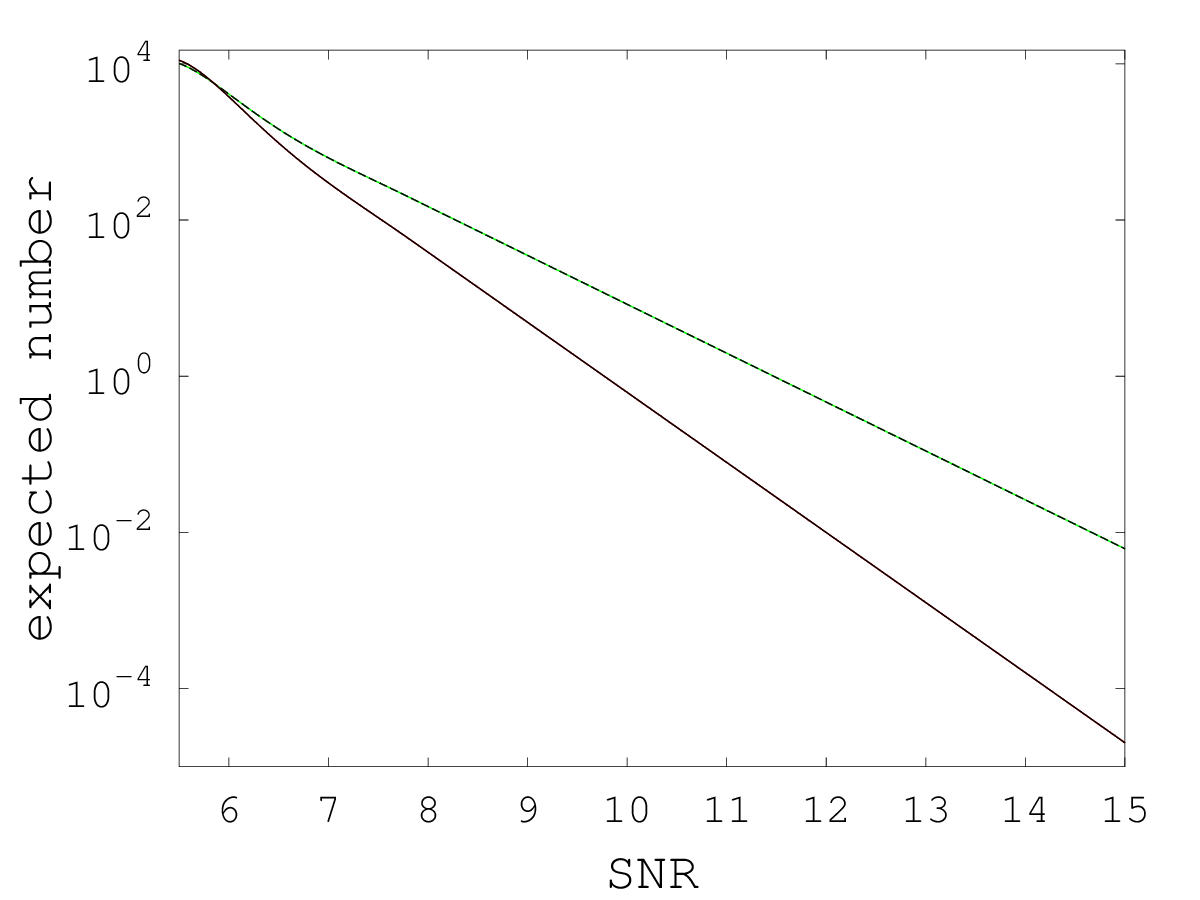}
      \caption{A ``realistic'' example background (experiment 12).\label{fig:examplerealistic}}
    \end{subfigure}
    \begin{subfigure}[b]{0.33\textwidth}
      \centering
      \includegraphics[width=\textwidth]{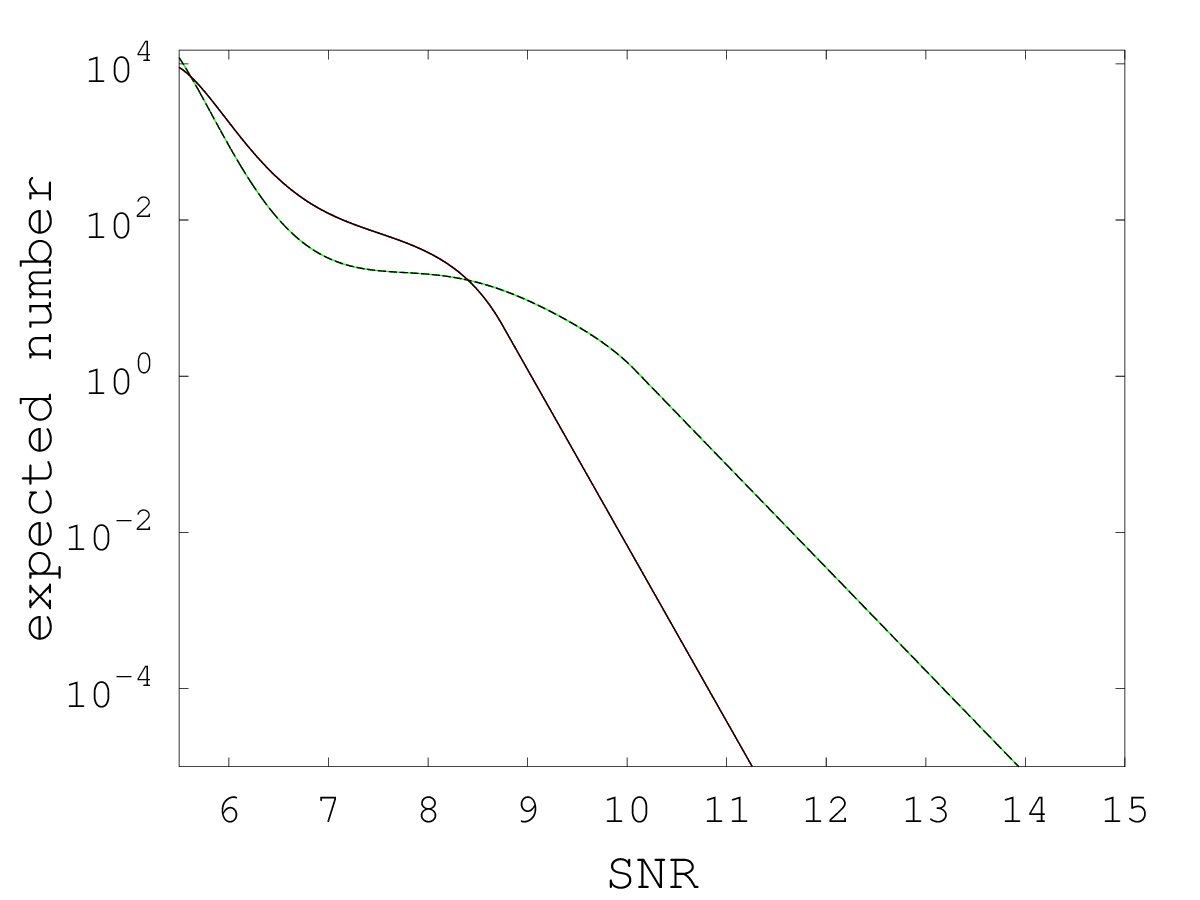}
      \caption{An ``extreme'' example background (experiment 14).\label{fig:exampleextreme}}
    \end{subfigure}    
    \caption{Examples of different background distributions used in
the \ac{MDC}. In each of the three examples we show the complementary \ac{CDF}
$(1-\mathcal{C})$ versus the single-detector \ac{SNR} for each detector. There
were no foreground distributions present in these experiments.}
  \label{fig:exampleBkgd}
\end{figure*}

%
%
In cases with a ``simple'' background, the coefficients of our model 
(Eq.~\ref{eq:CDF}) are all zero with the exception of $a_0$ and 
$\rho_{\text{sp}}=\infty$. The \ac{CDF} then follows the simple
form $\mathcal{C} = 1-\exp(-a_{0}(\rho-\rho_{\text{th}}))$ for the
single-detector \ac{SNR}.  A ``realistic'' background is modelled by basing 
our analytic \ac{CDF} model on distributions of existing \ac{GW} trigger
data~\cite{Colaboration:2011np}. The ``extreme'' backgrounds attempt to
model distributions containing multiple independent populations of
detector noise artefacts resulting in \acp{CDF} that exhibit large 
variations in their gradients as a function of \ac{SNR}. We give examples
of each type of background distribution in Fig.~\ref{fig:exampleBkgd}.
The single experiment described as containing an ``extended tail'' is similar to
the extreme cases in the sense that its gradient varies substantially as a
function of \ac{SNR}. However, this variation occurs at much smaller values of
$1-\mathcal{C}$, thus it is rare that realisations have events generated from the
``tail''.  This rarity and shallowness of the tail are designed to mimic the
behaviour of an astrophysical foreground (with the exception of being
coincident between detectors).

%
%
The trigger time of a background event is a random variable generated from a
uniform distribution spanning the the length of an observation realisation. The
number of such triggers within a realisation is drawn from a Poisson
distribution with parameter $\lambda_j$, the expected number of triggers in the
$j$'th detector. The two detectors are treated independently and for events to
be labelled coincident, their trigger times must satisfy
\begin{equation}\label{eq:window} 
  |t_{1}-t_{2}|\leq \delta t, 
\end{equation}
where $t_{1}$ and $t_{2}$ are the times associated with a trigger from the
first and second detectors respectively and $\delta t$ is our allowed
coincidence window. We can therefore estimate the expected number of coincident
events $n$ within a single realisation as
\begin{equation}\label{eq:bkgdRate} 
  n = \frac {2\lambda_1\lambda_2\delta t}{T},
\end{equation}
where $T$ is the total time of the observation and we have assumed that
$\lambda_{j}\delta t/T\ll 1$ in both detectors. In order to generate a large
enough number of background triggers to adequately model a \ac{GW}
observation, we use $\lambda_{1}{\sim}\lambda_{2}{\sim}10^{4}$.  This choice is
also motivated by the expected dependence of the uncertainty in estimation of 
\ac{FAP} on the numbers of triggers, and the computational cost
to challenge participants. We set the coincidence window $\delta t=50$\,ms to
broadly model realistic values and in order to obtain a desired ${\sim}10$
coincidences per realisation the total observation time is set to
$T=10^{6}$\,s. 

Note, however, that the MDC does not model some aspects of a real search
for \ac{CBC} signals, most notably the need for many thousands of different
template waveforms to optimise the sensitivity to different possible source
parameters.  The multiplicity of templates 
has 
various implications for estimating \ac{FAP}.  The numbers of
single-detector background triggers will increase, but the probability that 
any given pair of triggers will form a coincident noise event will drop since
random noise triggers are unlikely to have consistent physical source 
parameters across different detectors.  The complexity and computational load
of modelling sets of single-detector triggers would be considerably increased,
since the times and \acp{SNR} of triggers will be nontrivially correlated 
between templates with similar waveforms.

%
%
\subsection{Modeling an astrophysical foreground\label{sec:foreground}}
%
%
In the majority of experiments (10 of the 14) an astrophysical foreground 
was simulated.  We model the astrophysical signal distribution as
originating from the inspiral of equal mass $1.4-1.4 M_\odot$ binary neutron
stars having a uniform distribution in volume and in time.  For each source the
binary orientation is selected from an isotropic distribution (uniform in the
cosine of the inclination angle $\iota$), the polarisation angle $\psi$ is
uniform on the range $[0,2\pi)$ and the sky position is distributed
isotropically on the sphere parametrised by right ascension $\alpha$ (with 
range $[0,2\pi)$) and declination $\delta$ (range $[-\pi/2,\pi/2)$). Given a 
set of these parameters we can compute the optimal single-detector \ac{SNR} 
$\rho_{\text{opt}}$ as
\begin{equation}\label{eq:optsnr}
  \rho^{2}_{\text{opt}} = 4\int\limits_{f_{\text{min}}}^{f_{\text{ISCO}}}\,
  \frac{|\tilde{h}(f)|^{2}}{S_{\text{n}}(f)}df
\end{equation}
where the lower and upper integration limits are selected as $10$\,Hz and as the
innermost stable circular orbit frequency $=1570$\,Hz for our choice of total
system mass.  The detector noise spectral density $S_{\text{n}}(f)$
corresponds to the advanced LIGO design~\cite{TheLIGOScientific:2014jea}, and the frequency 
domain signal in the stationary phase approximation is given by
\begin{equation}\label{eq:stationaryphase}
  \tilde{h}(f) = \frac{Q(\{\theta\})\mathcal{M}^{5/6}}{d}\sqrt{\frac{5}{24}}\pi^{-2/3}
  f^{-7/6}e^{i\Psi(f)}.
\end{equation}
Here the function $Q(\{\theta\})$, where $\{\theta\} = (\alpha,\delta,\psi,
\cos\iota)$, describes the antenna response of the detector; $d$ is the 
distance to the source, and $\mathcal{M}$ is the ``chirp mass'' of the system
given by $\mathcal{M}=(m_{1}m_{2})^{3/5}/(m_{1}+m_{2})^{1/5}$. Since we 
consider that such signals, if present in the data, are recovered with exactly
matching templates, the phase term $\Psi(f)$ does not influence the optimal 
\ac{SNR} of Eq.~\ref{eq:optsnr}. Hence the square of the observed (or matched
filter) \ac{SNR} $\rho$ is drawn from a non-central $\chi^2$ distribution with
2 degrees of freedom and non-centrality parameter equal to $\rho_{\rm opt}^2$.

%
%
We generate foreground events within a sphere of radius 1350\,Mpc such that
an optimally oriented event at the boundary has ${<}0.3\%$ probability of
producing a trigger with \ac{SNR}\,$>\rho_{\text{th}}=5.5$.  Each event is given
a random location (uniform in volume) and orientation from which we calculate
the corresponding optimal \ac{SNR} and relative detector arrival times. 
The matched filter \ac{SNR} is modelled as a draw from the non-central
chi-squared distribution. For each detector, if the matched filter \ac{SNR} is
larger than $\rho_{\text{th}}$, independently of the other detector, it is 
recorded as a single detector trigger.  The arrival time in the first detector
(chosen as the LIGO Hanford interferometer) is randomly selected uniformly
within the observation time and the corresponding time in the second detector
(the LIGO Livingston interferometer) is set by the arrival time difference
defined by the source sky position. We do not model statistical uncertainty in
the arrival time measurements, hence when a foreground event produces a trigger
in both detectors the trigger times will necessarily lie within the time window
and will generate a coincident event.  

\subsection{The definition of false alarm probability (FAP) for the MDC\label{sec:sigdef}}

%
%
In order to define the \ac{FAP} for any given realisation of an
experiment we require a ranking statistic which is a function of the 
coincident triggers within a realisation. In this \ac{MDC} the chosen ranking
statistic was the combined \ac{SNR} of coincident events, defined as
\begin{equation}\label{eq:coincSNR} 
   \rho^2 = \rho_{1}^{2}+\rho_{2}^{2},
\end{equation}
where $\rho_{1,2}$ are the \acp{SNR} of the single-detector triggers
forming the coincident event.  Challenge participants were required to estimate
the \ac{FAP} of the ``loudest'' coincident event within each
realisation, i.e.\ the event having the highest $\rho$ value, independent of 
its unknown origin (background or foreground). The \ac{FAP} of 
an outcome defined by a loudest event $\rho^{\ast}$ is the probability of
obtaining at least one background event having $\rho \geq \rho^\ast$
within a single realisation. 
Given that single-detector 
background distributions fall off with increasing $\rho_{1,2}$, the louder a
coincident event is, the less likely it is for a comparable or larger $\rho$ 
value to be generated by noise, and the smaller the \ac{FAP}.

%
%
With access to the analytical description of the backgrounds from both
detectors we may compute the single trial \ac{FAP} $\mathcal{F}_{1}$ as
\begin{align}
    1-\mathcal{F}_{1}(\rho)&=\int\limits_{\rho_{\text{th}}}^{\sqrt{\rho^{2}-\rho_{\text{th}}^{2}}}\hspace{-0.2cm}d\rho_{1}\hspace{-0.2cm}\int\limits_{\rho_{\text{th}}}^{\sqrt{\rho^{2}-\rho_{1}^{2}}}\hspace{-0.2cm}d\rho_{2}\,p_{1}(\rho_{1})p_{2}(\rho_{2}),\nonumber\\
    &=\int\limits_{\rho_{\text{th}}}^{\sqrt{\rho^{2}-\rho_{\text{th}}^{2}}}\hspace{-0.2cm}d\rho_{1}p_{1}(\rho_{1})\,\mathcal{C}_{2}\left(\sqrt{\rho^{2}-\rho_{1}^{2}}\right),
\end{align} 
where $p_{1}(\rho_{1})$ and $p_{2}(\rho_{2})$ are the \acp{PDF} of the
background distributions (obtained by differentiating the corresponding
\acp{CDF} with respect to $\rho_{j}$), and $\mathcal{C}_2(\rho_{2})$ is the
\ac{CDF} for the second detector.

To account for the fact that we are interested in the ``loudest'' coincident
event within each realisation we must perform an additional marginalisation
over the unknown number of such coincident events.  To do this we model the
actual number of coincidences as drawn from a Poisson distribution with known
mean $n$ (values for the different \ac{MDC} experiments are given in
Table~\ref{tab:paramsRate}). The \ac{FAP} of the ``loudest'' event is modelled as
the probability of obtaining one or more coincident events with a combined
\ac{SNR} $\geq\rho$ and is given by
\begin{equation}
\mathcal{F}(\rho) = \sum_{j=0}^{\infty}\left(1 -
\left(1-\mathcal{F}_{1}(\rho)\right)^{j}\right)\frac{n^{j}e^{-n}}{j!}\label{eq:exact}.
\end{equation}
Challenge participants only had access to the trigger $\rho_{1,2}$ values and
trigger times in each realisation and were not given the distributions from
which they were drawn.  Estimates of the loudest coincident event \ac{FAP}
$\mathcal{F}$ from all participants will be compared to the ``exact'' values
computed according to Eq.~\ref{eq:exact}.

\subsection{The expected error on estimated false alarm probability (FAP)\label{sec:sigerror}}
%
%

Inferring the \ac{FAP}, as defined above, from a finite 
sample of data will have associated uncertainty, i.e., the computed values 
will be approximate. 
Methods to estimate the \ac{FAP} at a given combined
SNR value $\mathcal{F}(\rho)$ involve counting the number of noise events
$N(\rho)$ above that value:
\begin{align}
    N(\rho) &= \iint_{\rho\geq\rho_{\rm{th}}} n_1(\rho_1) \, n_2(\rho_2) 
               \, d\rho_1 d\rho_2 \nonumber \\
    &= \Lambda_{1} \Lambda_{2} - 
               \iint_{\rho<\rho_{\rm{th}}} n_1(\rho_1) \, n_2(\rho_2) 
               \, d\rho_1 d\rho_2 \label{eq:allcoincs},
\end{align}
where $n_i(\rho_i)$ is the number density of background triggers from detector
$i$ and $\Lambda_{i}$ is the total number of background triggers from detector $i$.
The region of integration is bounded by a threshold on
the coincident \ac{SNR} statistic of Eq.~\ref{eq:coincSNR}, though in 
general one may choose other functional forms for $\rho(\rho_1,\rho_{2},\dots)$. 

It is possible to compute (either analytically or numerically) the error on 
$N(\rho)$ given any functional form for $\rho$.  However, we seek a 
simple ``rule of thumb'' as a general approximation. We replace the region 
$\rho{<\rho_{\rm{th}}}$ with an equivalent hyper-cuboid with lengths $\rho_{i}^{*}$, such that for
an event to be counted towards the \ac{FAP} it must have a SNR greater than $\rho_{i}^{*}$ 
in either detectors.  In this case, the number of louder
triggers as a function of $\rho$ can be approximated by 
\begin{align}
    N(\rho) &\approx \Lambda_1 \Lambda_2 - \int_{0}^{\rho_{1}^{*}}d\rho_{1}
    \int_{0}^{\rho_{2}^{*}} d\rho_{2}\,n_1(\rho_1) n_2(\rho_2)
    \label{eq:re-express-allcoincs} \nonumber \\ 
    &\approx \Lambda_{1} \Lambda_{2}-N'_{1}(\rho_{1}^{*}) N'_{2}(\rho_{2}^{*}), 
\end{align}
where
\begin{align} 
    N'_{i}(\rho_{i}^{*}) &\equiv \int_{0}^{\rho_{i}^{*}} n_{i}(\rho_{i}) \, d\rho_{i},
\end{align}
is the cumulative number of triggers from detector $i$. We then define the
inferred \ac{FAP} as 
%
\begin{align} 
    \mathcal{F}(\rho) &\approx \frac{N(\rho)}{\Lambda_1 \Lambda_2} \nonumber\\ 
    &\approx1-\frac{N'_1(\rho_1^*) N'_2(\rho_2^*)}{\Lambda_1 \Lambda_2}\label{eq:FAP1} 
\end{align}

We wish to characterise the error in $\mathcal{F}(\rho)$ given the error in the
number of triggers counted above $\rho_i$ in each detector. We expect that this
error will increase when fewer triggers are available to estimate the single
detector counts.  Transforming Eq.~\ref{eq:FAP1} to use the counts above a
threshold $\rho_i^*$ in each detector via $N_{i}(\rho_i)\equiv\Lambda_i-
N'_i(\rho_i)$, we have
\begin{align}
\mathcal{F}(\rho) &\approx 1 - \frac{\big( \Lambda_1 - N_1(\rho_1^*) \big) \big( \Lambda_2 - N_2(\rho_2^*) \big)} {\Lambda_1 \Lambda_2}.
\label{eq:FAP2}
\end{align}
Assuming a negligible error on the total count of triggers in each detector 
$\Lambda_i$, we 
can then write
\begin{align}
\sigma^{2}_{\mathcal{F}(\rho)} &\approx \sum_{i} \left( \frac{\partial \mathcal{F}(\rho)}{\partial N_i(\rho_i^*)}\right)^{2} \sigma^{2}_{N_i(\rho_{i}^{*})}.
\end{align}
Taking the distribution of counts $N_i(\rho)$ to be Poisson, we have standard
errors $\sigma^2_{N_i(\rho_i^*)} = N_i(\rho_i^*)$; for the two-detector 
case we then find
\begin{align}
\sigma^{2}_{\mathcal{F}(\rho)} &\approx\frac{\big(\Lambda_2 - N_2(\rho_2^*)
\big)^2 N_{1}(\rho_1^*) + \big(\Lambda_1 - N_1(\rho_1^*) \big)^2 N_{2}(\rho_2^*)}{\Lambda_1^2\Lambda_2^2},
\end{align}
hence the fractional error is
\begin{align}
\frac{\sigma_{\mathcal{F}(\rho)}}{\mathcal{F}(\rho)} &\approx
	\frac{\sqrt{\frac{(\Lambda_2 - N_2(\rho_2^*))^2} {N_1(\rho_1^*) {N_2}^2(\rho_2^*)} +
	\frac{(\Lambda_1 - N_1(\rho_1^*))^2} {N_2(\rho_2^*)
{N_1}^2(\rho_1^*)}}}{\frac{\Lambda_2}{N_2(\rho_2^*)} + \frac{\Lambda_1}{N_1(\rho_1^*)} - 1}.
\end{align} 
In the limit of low \acp{FAP}, $N'_1(\rho_1^*) \ll
\Lambda_1$ and $N'_2(\rho_1^*) \ll \Lambda_2$, our expression simplifies
to
\begin{align} 
    \frac{\sigma_{\mathcal{F}(\rho)}}{\mathcal{F}(\rho)} &\approx \frac{\sqrt
    {\Bigg(\frac{\Lambda_2}{{N_2}(\rho_2^*)}\Bigg)^2\frac{1}{N_1(\rho_1^*)} +
    \Bigg(\frac{\Lambda_1}{{N_1}(\rho_1^*)}\Bigg)^2 \frac{1}{N_2(\rho_2^*)} }
    }{\frac{\Lambda_2}{N_2(\rho_2^*)} + \frac{\Lambda_1}{N_1(\rho_1^*)}}.
\end{align} 
Now we consider two limiting cases. First, when the distribution of counts is
similar in each detector such that $\Lambda_1 \approx \Lambda_2$ and
$N_1(\rho_1^*) \approx N_2(\rho_2^*)$, we have
\begin{align} 
    \frac{\sigma_{\mathcal{F}(\rho)}}{\mathcal{F}(\rho)} &\approx \sqrt{ \frac{1}
    {2 N_1(\rho_1^*)} }. 
\end{align} 
Second, when we are probing much further into the ``tail'' of the distribution of
one detector, e.g., $\Lambda_1 / N_1(\rho_1^*) \gg \Lambda_2 /
N_2(\rho_2^*)$, we have
\begin{align}\label{eq:sigerror} 
    \frac{\sigma_{\mathcal{F}(\rho)}}{\mathcal{F}(\rho)} &\approx \sqrt{ \frac{1}
{N_1(\rho_1^*)}.  } \end{align}
In both cases the fractional error is related to the inverse of the 
single-detector counts, not the combined counts $N(\rho)$ as one might
naively expect.  A similar contribution to the uncertainty in false alarm 
rate estimation, due to Poisson counting errors for single-detector events,
was found in \cite{Was:2009vh}. 
We note that a number of approximations were made to derive our ``rule of 
thumb'', though we show the level of agreement between this estimate and the 
results of the MDC analysis in Section~\ref{sec:results}. 

%
%
\section{Background estimation algorithms\label{sec:estimation}}
%
%

%
%
\subsection{Standard offline analysis: false alarm probability (FAP) via inverse false alarm rate (IFAR)\label{sec:ihopeapproach}}

We now describe the time slide method implemented in the all-sky 
LIGO-Virgo search pipeline for 
\ac{CBC}~\cite{Abbott:2009qj,Keppel:thesis,Colaboration:2011np,Babak:2012zx,Capano2011}
and indicate how the method has been adapted for the simplified 
high-statistics study presented in this paper. 

Each coincident event obtained in the search is characterized by its 
estimated coalescence time and binary component masses, and in addition
by the values of \ac{SNR} $\rho$ and the signal-based chi-squared test 
$\chi^2$ \cite{Allen:2004gu} in each detector, which together are 
intended to achieve separation of signals from non-Gaussian noise 
transients.  The event ranking statistic used, $\rho_c$, is the 
quadrature sum of re-weighted \acp{SNR} $\hat{\rho}_i(\rho_i,\chi^2_i)$~\cite{Babak:2012zx,Colaboration:2011np} 
over participating detectors $i$.\footnote{In real data the search may 
be divided into event bins determined by the component masses and 
participating interferometers~~\cite{Abbott:2009qj,Keppel:thesis}; 
however the present study does not attempt to simulate these 
complications.}
Exactly the same coincidence test is performed in the time-shifted 
analyses as in the search, resulting in a set of values $\{\rho_{c,b}\}$ 
from time-shifted events, considered as background samples.\footnote{In
real data an additional time clustering step is performed on the search
and on each time-shifted analysis in order to reduce the number of 
strongly-correlated coincident events separated by short intervals 
($\lesssim 1$\,s) resulting from the multiplicity of filter templates.  
In this study, however, single-detector events are already uncorrelated
by construction thus such clustering is not performed.}

With the search performed over a duration $T$ of two- or more-detector 
coincident data, and time-shifted analyses covering a total duration 
$T_b \equiv sT$, defining a background multiplier $s$, the \emph{estimated 
\ac{FAR}} of a candidate event having ranking statistic $\rho_c^\ast$ 
is calculated as the observed rate of louder background events over 
the time-shifted analyses:
\begin{equation}\label{eq:ihope_FAR}
 \textrm{FAR}(\rho_c^\ast) \equiv \frac{ \sum_{\{\rho_{c,b}\}} 
   \Theta(\rho_{c,b} - \rho_c^\ast) }{T_b} \equiv 
   \frac{n_b(\rho_c^\ast)}{T_b},
\end{equation}
where $\Theta(x) = 1$ if $x>0$ and $0$ otherwise.
$n_b$ is the number of events louder than $\rho_c^\ast$.
Note that the \ac{FAR} may equal zero for a high enough threshold 
$\rho_c^\ast$. 

The test statistic used to determine \ac{FAP} is inverse
FAR (IFAR), i.e.\ $1/\textrm{FAR}$; thus a false alarm corresponds
to obtaining a given value of $n_b/T_b$ or lower under the null 
hypothesis. 

Consider ranking the $\rho_c$ value of a single search event 
relative to a total number $N_b$ of time-shifted background event 
values.  Under the null hypothesis every ranking position is equally
probable, thus the probability of obtaining a count $n_b$ or smaller
of background events is 
$P_0(n_b\text{ or less}|1) = (1+n_b)/N_b$.  Since $n_b$ decreases
monotonically with increasing $\rho_c$, if a search event has a 
$\rho_c$ value equal to or greater than a given threshold 
$\rho_c^\ast$, the number of louder background events $n_b(\rho_c)$
must be equal to or less than $n_b(\rho_c^\ast)$.  Thus we may also
write\footnote{Note that a statistic value slightly \emph{below}
$\rho_c^\ast$ may also map to the same number of louder background 
events $n_b(\rho_c^\ast)$, thus the condition $\rho_c\geq\rho_c^\ast$
is more restrictive than $n_b \leq n_b(\rho_c^\ast)$.} 
\begin{equation}
 P_0(\rho_c\geq\rho_c^\ast|1) \leq \frac{1+n_b(\rho_c^\ast)}{N_b}.
\end{equation} 
Then, if there are $k$ such search events due to noise, the 
probability of at least one being a false alarm above the threshold
$\rho_c$ (implying an estimated IFAR as large as $T_b/n_b$) is 
\begin{equation}\label{eq:ihope_multicoinc_nomarg} 
 P_0(\text{1 or more}\,\geq \rho_c^\ast|k) = 1 - \left(1 - 
P_0(\rho_c\geq \rho_c^\ast|1)\right)^k. 
\end{equation} 

The implementation is simplified by considering the 
largest possible number of time-shifted analyses, such that a pair of 
single-detector triggers coincident in one analysis cannot be coincident 
for any other time shift.  This implies that the relative time shifts 
are multiples of $2\delta t$, and the maximum number of time-shifted 
analyses is $s=T/(2\delta t)-1$.   
The resulting time-shifted coincidences are then simply 
all possible combinations of the single-detector triggers, minus those 
coincident in the search (``zero-lag''), since every trigger in detector 
1 will be coincident with every trigger in detector 2 either in
zero-lag or for some time shift.  Identifying $\rho_c^\ast$ with the 
loudest coincident search event value $\rho_{c,\max}$ we have
\begin{equation} \label{eq:ihope_nlouderplusone}
 1+n_b(\rho_{c,\max}) = 
 1 + \sum_i \sum_j \Theta(\rho_{1,i}^2+\rho_{2,j}^2 - \rho_{c,\max}^2), 
\end{equation}
where the sums run over all single-detector 
triggers $\{\rho_{1i}\}$, $\{\rho_{2j}\}$, $i=1\ldots \Lambda_1$, 
$j=1\ldots \Lambda_2$. 

So far we have worked with specific values for the number of search 
events due to noise $k$ and time-shifted background events $N_b$, 
however these are not known in advance and should be treated as 
stochastic (see also~\cite{DentDCC}). 
We assume that $N_b$ is large enough that we can neglect its 
fluctuations, but we model $k$ as a Poisson process with mean rate 
$\mu = \langle N_b\rangle/s \simeq N_b/s$.  (In fact we know that 
$k+N_b = \Lambda_1\Lambda_2$, the product of single-detector trigger
counts, thus we assume that $\Lambda_{1,2}$ are large 
Poisson-distributed numbers and $s\gg1$.) 
We then marginalize over $k$ using the Poisson prior:  
\[
 p(k|\mu) = \frac{\mu^{k}e^{-\mu}}{k!}.
\]
After marginalization the dependence on $\mu$ vanishes to obtain 
\begin{equation} \label{eq:ihope_multicoinc_FAN}
  \mathcal{F}(\rho_c^\ast) = p(\rho_{c,\max}) \simeq 
  1 - \exp\left( -\frac{1+n_b(\rho_{c,\max})}{s} \right).
\end{equation}
Thus, false alarms louder than $\rho_{c,\max}$ arising by random 
coincidence from our sets of single-detector triggers are 
approximated by a Poisson process with expected number 
$(2\delta t/T)(1+n_b(\rho_{c,\max}))$. 
For this \ac{MDC}, the values of the coincidence window and analysis
time chosen imply $s\simeq 10^{7}$, giving a limit $p \gtrsim 10^{-7}$ 
to any \ac{FAP} estimate.  We have 
verified that the $p$-value of Eq.~\eqref{eq:ihope_multicoinc_FAN}
is distributed uniformly on $(0,1]$ for MDC data sets containing 
uncorrelated noise triggers.  

So far we have considered the case where all single-detector 
triggers are kept in constructing the background values.  To 
implement the case of removing zero-lag coincident triggers, we 
simply exclude these from the sums over pairs of triggers 
on the RHS of 
Eq.~\eqref{eq:ihope_nlouderplusone}.

%
%
\subsection{All possible coincidences (APC) approach\label{sec:newapproach}}

\newcommand{\combinedSNR}{\rho}

The \ac{APC} approach is described in detail in \cite{Capano:2015}. Here we 
provide a brief synopsis.

To estimate the \ac{FAP} of zero-lag triggers, we first find
the probability of getting a trigger from the background distribution
with combined SNR $\geq \combinedSNR$ in a single draw. When not removing
zero-lag triggers from the background estimate, this is:
\begin{equation}
\label{eqn:CollinSingleEstimate}
\overline{\mathcal{F}}(\combinedSNR) = P_0(\combinedSNR|1) = \frac{n_b(\rho)}{\Lambda_1 \Lambda_2-k}.
\end{equation}

Both background and zero-lag triggers are constructed by finding every
possible combination of triggers in detector 1 and detector 2.  Background
triggers are then any coincidence such that $\Delta t = |t_1 - t_2| > \delta t$,  while zero-lag triggers are those with $\Delta
t \leq \delta t$.  These can be found by adding the matrices $Z = X + Y$, where
$X_{ij} = \rho_{1,i}^2\, \forall j$ and $Y_{ij} = \rho_{2,j}^2\, \forall i$.
The elements of $Z$ are thus the $\combinedSNR^2$ of all possible combination
of triggers.

When removing zero-lag triggers from the background, the single detector
triggers that form the zero-lags are removed from the $X$ and $Y$ matrices
prior to finding $Z$.  This changes the denominator in Eq.~\eqref{eqn:CollinSingleEstimate} 
to $(\Lambda_1 -k)(\Lambda_2 - k)$. However, if $\Lambda_1, \Lambda_2
\gg k$, then the denominator is approximately $\Lambda_1\Lambda_2$ in either case; we use
this approximation in the following.

Since Eq.~\eqref{eqn:CollinSingleEstimate} is a measured quantity, it
has some uncertainty $\delta \overline{\mathcal{F}}$. This is given by:
\begin{equation}
\label{eqn:CollinErrEstimate}
\left(\frac{\delta\overline{\mathcal{F}}}{\overline{\mathcal{F}}}\right)^2 \bigg|_{\combinedSNR = \sqrt{\rho_1^2 + \rho_2^2}} = 
     \sum_{i=1,2}\left(\frac{\delta F_i(\rho_i)}{F_i(\rho_i)}\right)^2,
\end{equation}
%
where $F_i(\rho_i)$ is the estimated survival function in the $i$th detector,
given by:
\begin{equation}
F_i(\rho_i) = \frac{n_i(\rho_i)}{\Lambda_i}.
\end{equation}
Here, $n_i(\rho_i)$ is the number of triggers in the $i$th detector with SNR
$\geq \rho_i$.  We estimate $\delta F_i$ by finding the range of $F_i$ for which
$n_i$ varies by no more than one standard deviation.  Using the Binomial
distribution this is (similar to equation \ref{eq:ihope_multicoinc_nomarg}):
\begin{equation}
\substack{\max \\ \min} F_i = \frac{\Lambda_{i} (2n_{i} + 1) \pm \sqrt{4\Lambda_in_{i}(\Lambda_i-n_{i}) + \Lambda^2_{i}}}{2\Lambda_i(\Lambda_i+1)}.
\end{equation}
The error is thus:
\begin{equation}
\pm \delta F_i = \mp F_i \pm \substack{\max \\ \min} F_i.
\end{equation}
This error estimate can be asymmetric about $F_i$; to propagate to $\delta
\mathcal{F}$, we use $+(-)\delta F_1$ and $+(-)\delta F_2$ to find $+(-)
\delta \mathcal{F}$.

Equation~\eqref{eqn:CollinSingleEstimate} estimates the probability of getting
a trigger with combined SNR $\combinedSNR$ in a \emph{single} draw from the
background distribution. If we do $k$ draws, the probability of getting one or
more events from the background with combined SNR $\geq \combinedSNR$ is:
\begin{equation}
\label{eqn:CollinEstimate}
\mathcal{F}(\combinedSNR) = 1 - (1 - \overline{\mathcal{F}}(\combinedSNR))^k,
\end{equation}
with error:
\begin{equation}
\label{eqn:CollinErrEstimate2}
\pm \delta \mathcal{F}(\combinedSNR) = k (1-\overline{\mathcal{F}})^{k-1}(\pm \delta \overline{\mathcal{F}}).
\end{equation}
Thus, if we have two detectors with $\Lambda_1$ and $\Lambda_2$ triggers, $k$ of which form
zero-lag, or correlated, coincidences, then we can estimate the probability
(and the uncertainty in our estimate of the probability) that each trigger was
drawn from the same distribution as background, or uncorrelated, coincidences
using Eqs.~\eqref{eqn:CollinSingleEstimate}--\eqref{eqn:CollinErrEstimate2}.
The smaller this probability is for a zero-lag coincidence, the less likely
it is that that coincidence was caused from uncorrelated sources.  Since
gravitational waves are expected to be the only correlated source across
detectors, we use this probability as an estimate for the \ac{FAP}.

As this study is concerned with just the loudest zero-lag events, it is
useful to evaluate the smallest \ac{FAP} that can be estimated
using this method, and its uncertainty. From Eq.~\eqref{eqn:CollinSingleEstimate}, 
the smallest single-draw $\overline{\mathcal{F}}$ that
can be estimated is $(\Lambda_1 \Lambda_2)^{-1}$. By definition, this occurs at the largest
combined background SNR, $\combinedSNR^\dagger$. If the combined SNR of the
loudest zero-lag event is not $> \combinedSNR^\dagger$, then
$\combinedSNR^\dagger$ must be formed from the largest SNRs in each detector,
so that $n_i = 1$. Assuming $\Lambda_1, \Lambda_2 \gg 1$, then from Eqs.
\eqref{eqn:CollinEstimate} and \eqref{eqn:CollinErrEstimate} we find:
\begin{equation}
\label{eqn:CollinMinEstimate}
\min \overline{\mathcal{F}} \pm \delta\overline{\mathcal{F}} \underset{N_{1,2} \gg 1}{\approx}
    \frac{k}{\Lambda_1\Lambda_2}\left[1 \pm 
    \left\{\begin{array}{c} 2.3 \\ 0.87\end{array}\right\}\right].
\end{equation}
If the combined SNR of the loudest zero-lag is $> \combinedSNR^\dagger$, then
we cannot measure its \ac{FAP}. In this case, we use Eq.~\eqref{eqn:CollinMinEstimate}
to place an upper limit on $\overline{\mathcal{F}}$.

Determining the $\overline{\mathcal{F}}$ for every zero-lag trigger can
require storing and counting a large number of background triggers. To save
computational time and storage requirements, we reduce the number of background
triggers that have $\mathcal{F} >$ some fiducial $\mathcal{F}_0$ by a factor of
$\mathcal{F}/\mathcal{F}_0$ for each order of magnitude increase in
$\mathcal{F}$. We then apply a weight of $\mathcal{F}/\mathcal{F}_0$ to the
remaining background triggers when finding $\mathcal{F}$ for the zero-lag.
For this study, $\mathcal{F}_0$ was chosen to be $10^{-5}$. Thus, between
$\mathcal{F} = 10^{-4}$ and $10^{-5}$, 1 out of every $10$ background triggers
was kept, with a weight of $10$ applied to the remaining. Likewise, between
$\mathcal{F} = 10^{-3}$ and $10^{-2}$, 1 out of every $100$ background triggers
was kept, with a weight of $100$ applied to the remaining; etc. This
substantially reduces the number of background triggers that need to be counted
and stored; e.g., for $\lambda_1\lambda_2 = 10^8$, only $\sim 5000$ background 
triggers are needed, a saving of about 5 orders of magnitude.  The trade-off is our
accuracy in measuring the \ac{FAP} is degraded for triggers with
$\mathcal{F} > \mathcal{F}_0$.  This is assumed to be acceptable in a real
analysis, since triggers with larger $\mathcal{F}$ are, by definition, less
significant.\footnote{In retrospect, this background degradation was not really
necessary for this study, since we were only interested in the \ac{FAP}
 of the loudest zero-lag event in each realisation. However, we
wished to keep the analysis method as similar as possible to what would be done
in a real analysis.} 


%
%
\subsection{The gstlal approach\label{sec:gstapproach}}

The method to estimate the \ac{FAP} of coincident events based on the
likelihood ratio ranking statistic described in~\cite{Cannon2013FAR} was
modified for this test to use a single parameter, $\rho_c$.  The \ac{FAP}
for a single coincident event can be found as
\begin{align}
P_0(\rho_c) &= \int_{\Sigma_{\rho_c}} \prod_i p(\rho_{i}) \, d \rho_i,
\end{align}
where $p(\rho_{i}) \, d \rho_i$ are the probability densities of getting an event
in detector $i$ with SNR $\rho_{i}$, and $\Sigma_{\rho_c}$ is a surface of
constant coincident SNR.  The distributions $p(\rho_{i}) \, d \rho_i$ are measured
by histogramming the single detector SNR values either with or without
the coincident events included. 
To get the cumulative distribution for
a single event we have
\begin{align}
P_0(\rho_c > \rho_c^*|1) &= 1 - \int_0^{\rho_c^*} P_0(\rho_c) \, d \rho_c.
\end{align}
The multiple event \ac{FAP} is found in the same way as
\eqref{eq:ihope_multicoinc_nomarg}.

Notice that for this \ac{MDC}, an artificial lower boundary of $10^{-6}$ is set, as 
the participant decided any estimation below it is subject to excessive uncertainty 
and thus not reliable. 

\section{Results\label{sec:results}}
%
%
To achieve our aims of comparing the estimation from the `coincidence 
removal' and `all samples' modes, we will examine multiple properties of the submitted
challenge results.  We first examine the self-consistency of each set of
results for each simulation in Sec.~\ref{sec:ppplots}.  For experiments in the
absence of signals, the fraction of realisations with an estimated \ac{FAP}
smaller than a certain threshold should be identical to the value of that
threshold; we denote this property as self-consistency. 
In Sec.~\ref{sec:direct} we then investigate the accuracy of the
\ac{FAP} estimates by direct comparison with the exact calculated values for
each realisation in each simulation.   
In Sec.~\ref{sec:box} we select certain range of data and compare the median and mean of estimate with the exact value for both modes.
In Sec.~\ref{sec:ROC} we then construct \ac{ROC} plots for
each experiment as a way to compare estimates via their detection efficiency at fixed false positive rates.
Finally in Sec.~\ref{sec:error} we address the
general issue of \ac{FAP} estimate precision and attempt to extrapolate our
findings to predict the likely uncertainties rephrased on significance estimates for \ac{GW}
detection. 
%
%
The challenge was attempted by 3 different teams using a similar but 
independently implemented algorithms (see Sec.~\ref{sec:estimation}).  
Each algorithm was operated in 2 modes, one in which zero-lag triggers were
included in the background estimate and the other in which they were removed.
For each realisation of each experiment this gives us 6 \ac{FAP} estimates to
compare.  In the main text we include only plots from selected simulations that
highlight the main features of the comparisons; all other plots can be found in
Appendix~\ref{sec:additional}.

%
%
\subsection{Self consistency tests: $p$-$p$ plots\label{sec:ppplots}}
\begin{figure*}
  \centering
  \vspace*{-0.5cm}
  \begin{subfigure}[b]{\columnwidth}
    \centering
    \includegraphics[width=\columnwidth]{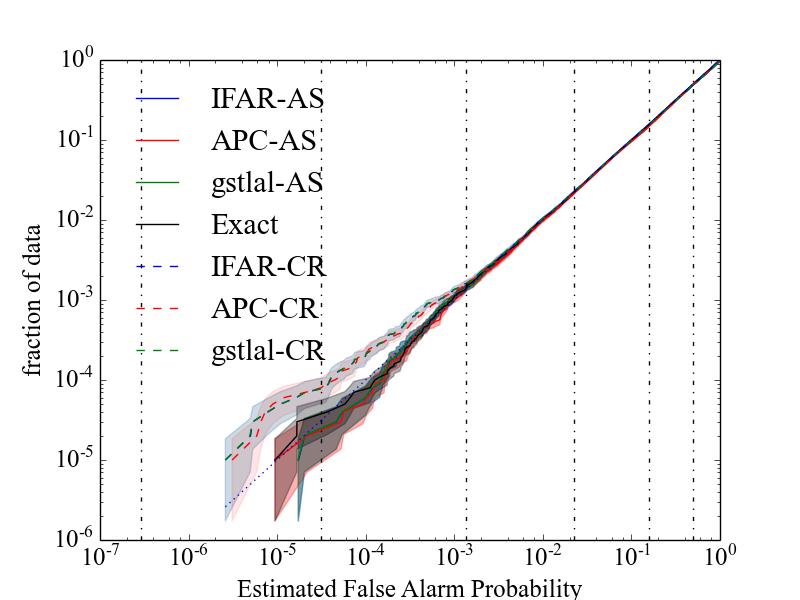}
    \caption{Experiment 1}
    \label{fig:1PP}
  \end{subfigure}
  \begin{subfigure}[b]{\columnwidth}
    \centering
    \includegraphics[width=\columnwidth]{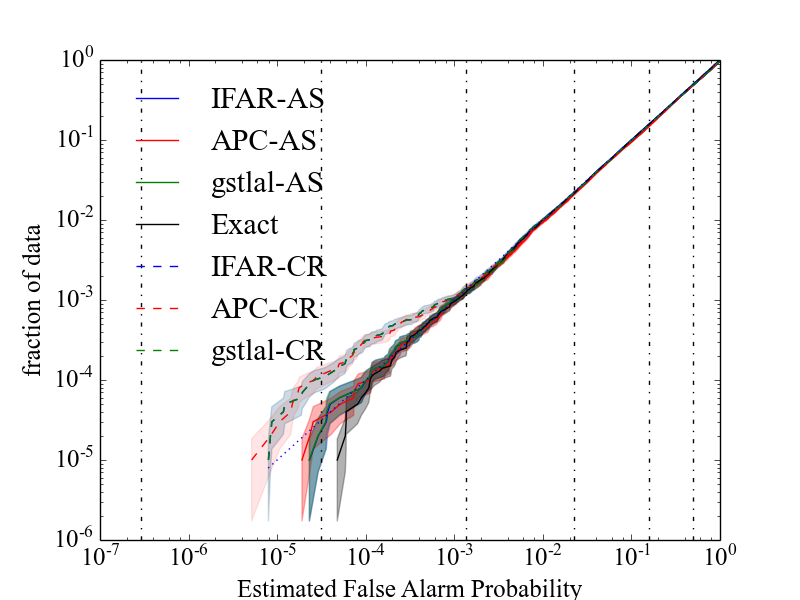}
    \caption{Experiment 3}
    \label{fig:3PP}
  \end{subfigure}\\
  \begin{subfigure}[b]{\columnwidth}
    \centering
    \includegraphics[width=\columnwidth]{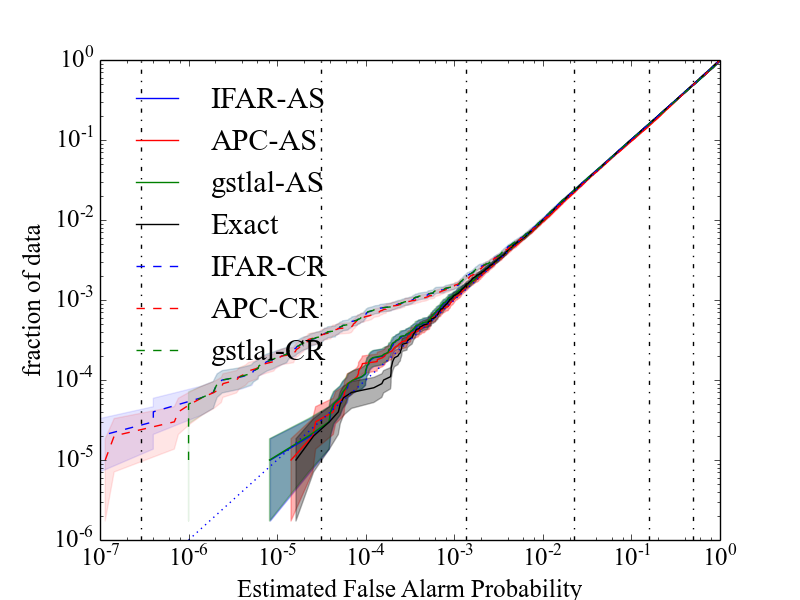}
    \caption{Experiment 12}
    \label{fig:12PP}
  \end{subfigure}
  \begin{subfigure}[b]{\columnwidth}
    \centering
    \includegraphics[width=\columnwidth]{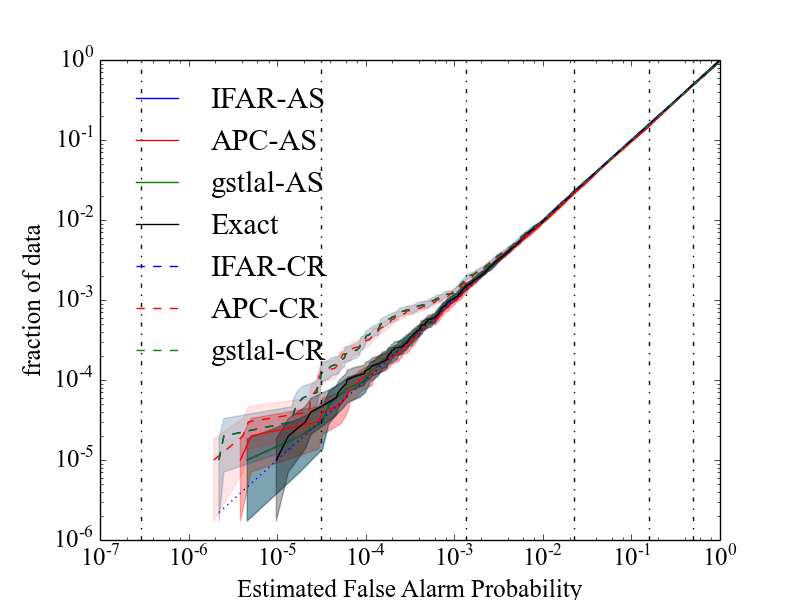}
    \caption{Experiment 14}
    \label{fig:14PP}
  \end{subfigure}
  \caption{Plots of estimated \ac{FAP} value versus the fraction
of events with that value or below (known as a $p$-$p$ plot).  If the estimate is self-consistent we would expect the value to be
representative of its frequency of occurrence; the diagonal line 
indicates a perfectly self-consistent \ac{FAP} estimate.  
We show results for the four experiments where the triggers were generated from 
background only. The solid
lines are the results obtained for our three algorithms in `all samples' mode
while the dashed lines are for the `coincidence removal' mode of operation.
Shaded regions include the uncertainty due to Poisson counting noise.  Vertical
dashed lines indicate the \ac{FAP} associated with integer
multiples of Gaussian standard deviations, i.e.\ the equivalent of
$n\,\sigma$ confidence. 
}
\label{fig:PPplots}
\end{figure*}

%
%
In Fig.~\ref{fig:PPplots} we show the relationship between the estimated 
\ac{FAP} values and their cumulative frequency of occurrence.  When
the zero-lag coincidences are drawn from the background distribution from which
the \ac{FAP} values are derived then we expect the curves to
trace the diagonal.  The figure shows results for the 4 experiments (1, 3, 12
and 14) for which there were only background triggers. As we probe lower 
\ac{FAP} values (i.e., rarer events) we begin to see counting noise
due to the finite number of rare events.  However, we see a marked difference
between the `coincidence removal' and `all samples' modes and no discernible
differences between algorithms.  In all cases the `all samples' mode stays
consistent with the diagonal within the expected fluctuations due to the finite
number of samples. The `coincidence removal' results, however, always
systematically overproduces very small numerical values of \ac{FAP}, 
with deviation from the expected behaviour for all values below ${\sim}10^{-3}$.

%
%
Experiments 1 and 3 were both designed to have simple background
distributions: the logarithms of their \ac{CDF} tails are linear in \ac{SNR}
with each detector having the same distribution, each experiment having a
different slope. Experiment 14 was designed to have an extreme background
distribution with multiple \ac{CDF} features. The behaviour of the $p$--$p$
plots in these 3 cases is very similar with the `coincidence removal' mode
deviating (by $\sim$1--2 standard deviations from the diagonal) for 
\acp{FAP} $<10^{-3}$. At exact \ac{FAP} values of $10^{-4}$ 
the `coincidence removal' mode tends to assign ${\sim}3$ times as many
realisations with an estimated \ac{FAP} at or below this value.
By contrast the `all samples' mode remain consistent throughout within the
1--$\sigma$ counting uncertainties.  For experiment 12, intended to have a
`realistic' background distribution, deviation from the diagonal occurs at
approximately the same point (\ac{FAP}$\,\sim 10^{-3}$) for `coincidence
removal'; here, for an estimated \ac{FAP} of $10^{-4}$, there are
${\sim}7$ times the number of estimated values at or below this level.  The
discrepancy in this case and the previous 3 experiments cannot be accounted
for by counting uncertainty over experiments.

%
%
The deviations seen for the `coincidence removal' case do not have direct
implications for point estimates of \ac{FAP} in specific
realisations; they also do not indicate a bias in those estimates in the sense
of a systematic \emph{mean} deviation of the estimated \ac{FAP}
away from the exact value.  The result does however indicate that for rare
events in a background-only dataset, using a `coincidence removal' mode gives a
greater than $\mathcal{F}$ chance of obtaining an event of estimated \ac{FAP}
 $\mathcal{F}$.  This result is also expected to hold for
experiments where the majority of realisations do not contain foreground
triggers, i.e.\ those with `low' signal rates. 

We may understand the onset of systematic discrepancies between the two 
modes as follows.  The change in estimated significance due to removal of 
coincident triggers will
only have a major effect -- comparable to the estimated value itself -- when 
much of the estimated background (louder than the loudest zero-lag event)
is generated by the triggers that are being removed.  This is 
likely only to be the case when the loudest event itself contains one of 
the loudest \emph{single-detector} triggers.  Thus, the probability of a
substantial shift in the estimate due to removal is approximately 
that of the loudest trigger in a single detector forming a random 
coincidence; for the parameters of this \ac{MDC} this probability is 
$~2\lambda_1\lambda_2\delta_t/T \simeq 10^{-3}$.


\subsection{Direct comparison with exact false alarm probability (FAP)\label{sec:direct}}
%
%

%

In this section, we show the direct comparison of estimated \ac{FAP} 
values with the exact \ac{FAP}.  In a parameter
estimation problem we may consider both the accuracy and precision of the
estimates as figures of merit: ideally the spread of estimated values compared
to the exact value should be small and the estimated values should concentrate
around the exact value. The estimated values could be influenced by a number of
factors including random fluctuations in the statistics of triggers, structures
like hidden tails could bias the estimates, and there may be contamination from
a population of foreground triggers.  Where possible we attempt to understand
the performance of the algorithms in terms of these factors, and to quantify
their influences.

Although the comparison shows obvious difference between the `all samples' and
`coincidence removal' modes, readers are reminded that the quantitative result
does not necessarily reflect the behaviour in an actual \ac{GW} search.  In
Figs.~\ref{fig:cmp14}--\ref{fig:cmp11} the estimation shows a large scatter in
estimated \ac{FAP} values below $10^{-3}$; this value is, though subject to 
the design of the \ac{MDC}, which does not attempt to model all aspects of
a \ac{CBC} search on real data.

\subsubsection{Low foreground rate}
%
%
To compare estimates of \ac{FAP} we display them for each
realisation, plotting $\log \mathcal{F}$ since the occurrence of very low
values is crucial for detection claims (either true or false).  In these
figures, a perfect estimation would lie on the line $y=x$; if an algorithm
provides an underestimate by assigning a smaller \ac{FAP}, it
will fall below the diagonal line; an overestimate 
would lie above the diagonal line.

\begin{figure*}
  \centering
  \vspace*{-0.5cm}
  \begin{subfigure}[b]{\columnwidth}
    \centering
    \includegraphics[width=\columnwidth]{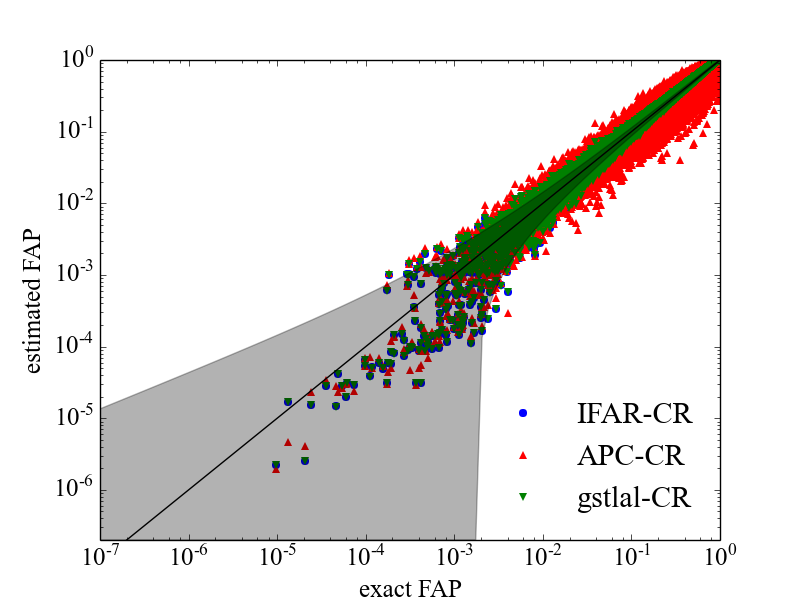}
    \caption{Direct comparison with `coincident removal'}
    \label{fig:cmp14rm}
  \end{subfigure} 
  \begin{subfigure}[b]{\columnwidth}
    \centering
    \includegraphics[width=\columnwidth]{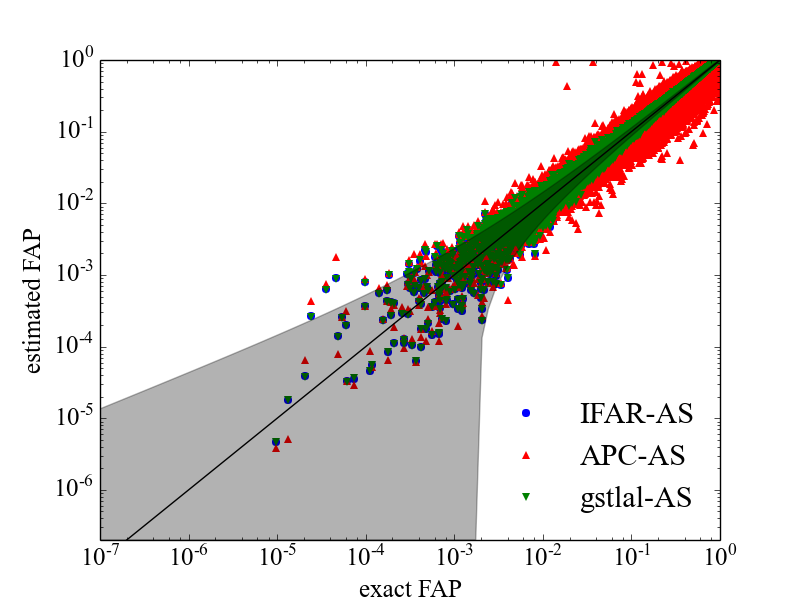}
    \caption{Direct comparison with `all samples'}
    \label{fig:cmp14nonrm}
  \end{subfigure}
  \caption{Direct comparisons of \ac{FAP} estimates with the
exact \ac{FAP} for experiment 14 (containing no signal events).  
In both plots the majority of blue points are masked by the green points 
since these methods provide closely matching results.  The estimates are
concentrated on the diagonal in both `coincidence removal' and `all samples'
cases.  However, in the `coincidence removal' mode, all algorithms place the
majority of points under the diagonal for exact \ac{FAP} values
$({<}10^{-3})$, indicating a non-conservative estimate.\label{fig:cmp14}}
\end{figure*}

For the experiments with no signal triggers or low foreground rate, the
triggers are at most slightly contaminated by foreground signals, so the
estimation of the \ac{FAP} should be correspondingly unaffected by their
presence.  Where there are no foreground triggers present, even the extreme
backgrounds, e.g.\ experiment 14, shown in Fig.~\ref{fig:cmp14}, don't appear
to adversely affect the estimation and the spread is relatively small around
the diagonal line.  However, in the `coincidence removal' mode, for all
algorithms there is a tendency to \emph{underestimate} $\mathcal{F}$ for small
values.  This is not conservative, i.e.\ it is over-optimistic, in the sense
that underestimating $\mathcal{F}$ claims that the experiment outcome is
rarer than they are in reality (in the absence of signal). 

%
%
In Fig.~\ref{fig:cmp14} we also see that gstlal estimation is very close to
that of the IFAR approach. For other experiments (see
App.~\ref{sec:AppDirComp}), their results do show small discrepancies most
notably in their different lower limits for estimated \ac{FAP}
values. 

For all \ac{APC} results, the estimation method used was designed such that
only for rare events (those with low \ac{FAP} values) were the
results computed to the highest accuracy (hence the large spread in \ac{FAP}
estimation seen in Fig.~\ref{fig:cmp14}).  This is motivated by the fact that
the astrophysical events that we are ultimately interested in will necessarily
have small \ac{FAP} values, and by the computational load of
the challenge itself.  

\subsubsection{Medium foreground rate}
%
%
The experiments with ``medium'' foreground rate have an average of $\sim$half
the realisations containing a foreground coincidence.  Foreground triggers 
are drawn from a long-tailed astrophysical distribution in \ac{SNR} and are 
likely to be loud if present. In such realisations,
any bias in the estimation of \ac{FAP} 
$\mathcal{F}$ due to this ``contamination'' would be in the direction of
overestimation.  
This kind of bias is considered to be conservative since it would 
underestimate the rarity of possible astrophysical events under the null 
hypothesis.

\begin{figure*}
  \centering
  \vspace*{-0.5cm}
  \begin{subfigure}[b]{\columnwidth}
    \centering
    \includegraphics[width=\columnwidth]{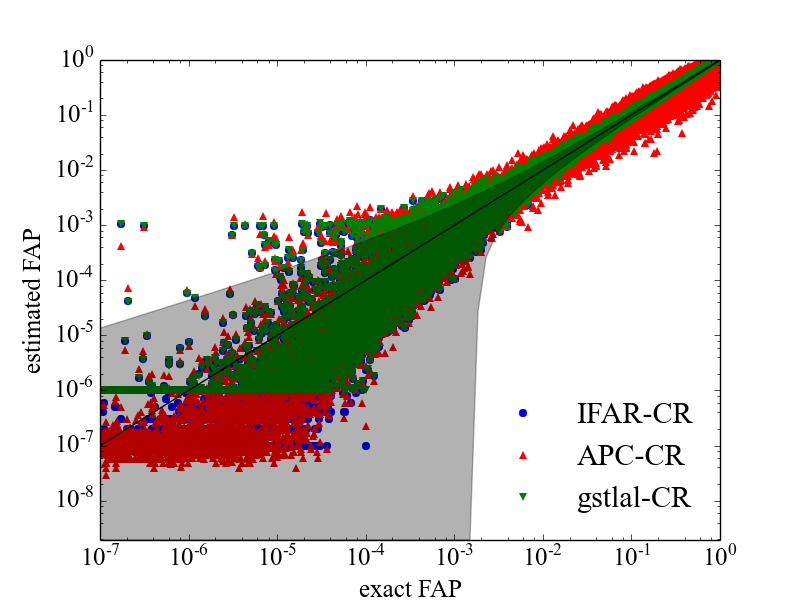}
    \caption{Direct comparison with `coincidence removal'}
    \label{fig:cmp9rm}
  \end{subfigure}%
  \begin{subfigure}[b]{\columnwidth}
    \includegraphics[width=\columnwidth]{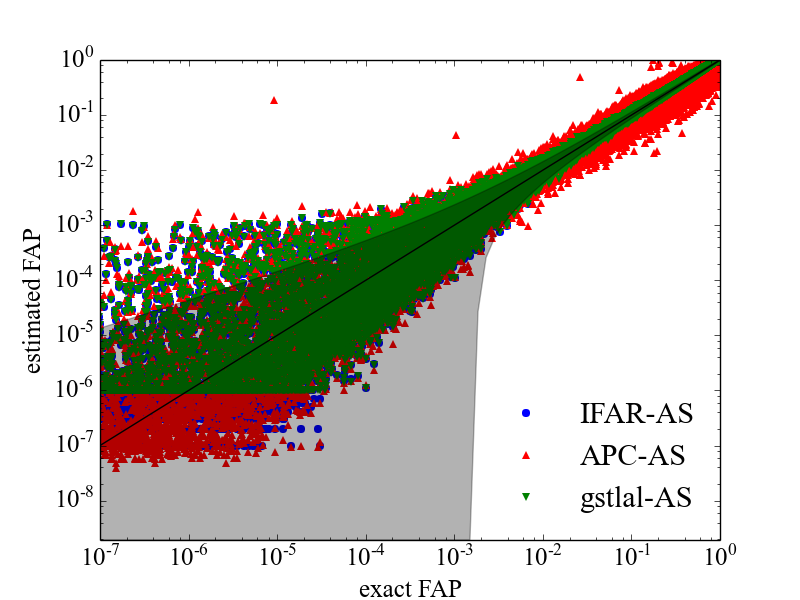}
    \caption{Direct comparison with `all samples'}
    \label{fig:cmp9nonrm}
  \end{subfigure}
  \caption{Direct comparisons of \ac{FAP} estimates with the
exact \ac{FAP} for experiment 9. The medium level foreground
rate in this case leads to a number of realisations containing
signals, resulting in a  
larger vertical spread.  Different algorithms fix different lower boundary values
for the estimated \ac{FAP}, visible in the sharp lower
edge in estimated $\mathcal{F}$ of gstlal results. The shaded region represents the expected uncertainty from Eq.~\ref{eq:sigerror}.\label{fig:cmp9}}
\end{figure*}

\begin{figure*}
  \centering
  \vspace*{-0.5cm}
  \begin{subfigure}[b]{\columnwidth}
    \centering
    \includegraphics[width=\columnwidth]{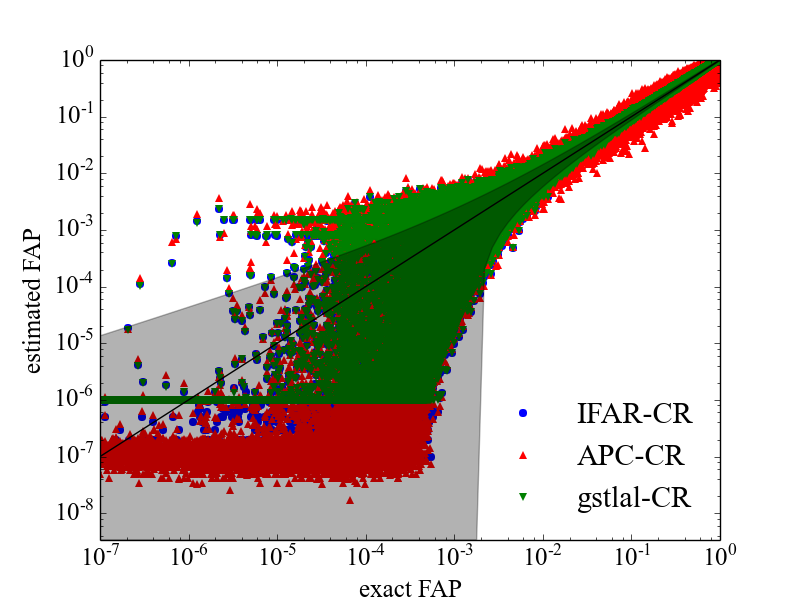}
    \caption{Direct comparison with `coincidence removal'}
    \label{fig:cmp6rm}
  \end{subfigure}%
  \begin{subfigure}[b]{\columnwidth}
    \includegraphics[width=\columnwidth]{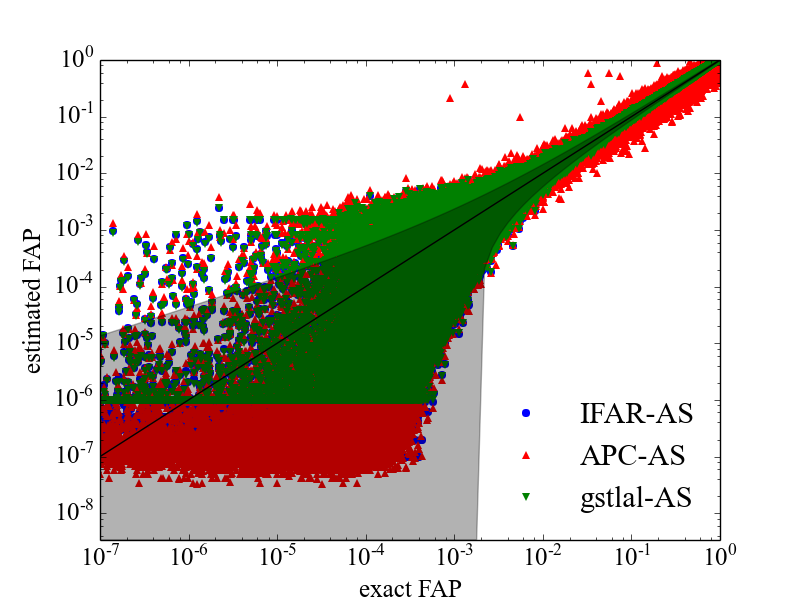}
    \caption{Direct comparison with `all samples'}
    \label{fig:cmp6nonrm}
  \end{subfigure}
  \caption{Direct comparisons of \ac{FAP} for experiment 6. The
    presence of a platform feature in the tail of the distribution
    causes the spread in estimates values to be wider than for experiment
    9. The shaded region represents the expected uncertainty from Eq.~\ref{eq:sigerror}.\label{fig:cmp6}}
\end{figure*}

%
%
We use results from experiments 9 and 6, shown in Figs.~\ref{fig:cmp9}
and~\ref{fig:cmp6} resp., as examples of a medium foreground rate.  We see
greater variance in the \ac{FAP} estimates in the low \ac{FAP} region, in comparison to experiments with zero or low foreground
rates.  This increased variance is clearly caused by the presence of foreground
events since nearly all points below $10^{-5}$ on the x-axis are due to
signals.  We again see general agreement between algorithms (with the exception
of \ac{APC} at high $\mathcal{F}$ values) however there is now evidence of
discrepancies at the lowest values of estimated \ac{FAP}.  This
is mostly due to the different choices of lowest estimatable value between
algorithms, and is independent of the \ac{MDC} dataset.  There is now also
stronger indication that `coincidence removal' shifts estimates to lower values
compared to the `all samples' mode.\footnote{ 
Note that since these plots have logarithmic scales, a very small difference in 
$\mathcal{F}$ may have a large apparent effect at low values.}
Further investigation of this systematic difference between modes will be 
made in Sec.~\ref{sec:box}.

\begin{figure}
\centering
\includegraphics[width=0.8\columnwidth]{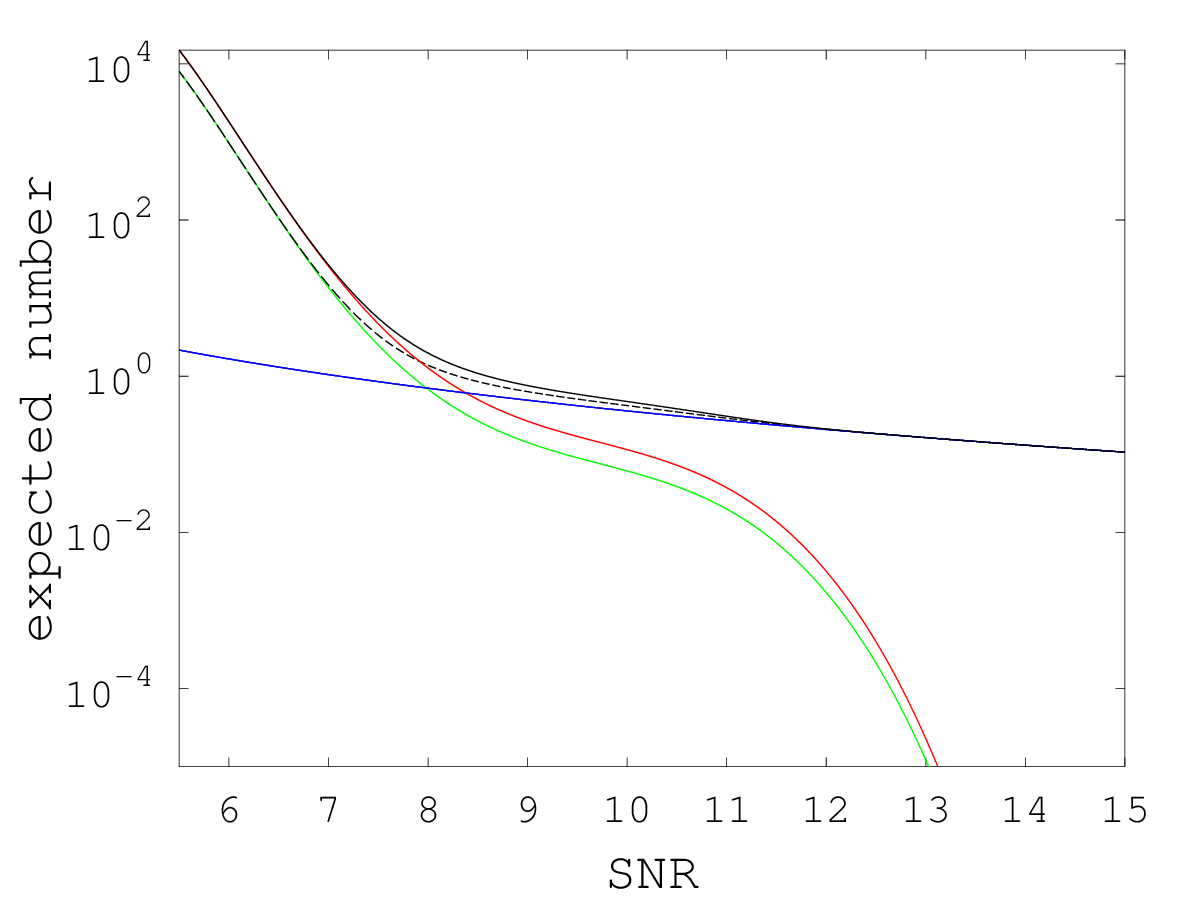}
    \caption{Reverse \ac{CDF} of trigger \acp{SNR} for experiment 6.  The red
and green curves represent the two individual detectors, while the blue curve
represents the astronomical signals.  The black lines represent the combined
distribution of both background and foreground triggers.\label{fig:back6}}
\end{figure}

%
%
The experiment shown in Fig.~\ref{fig:cmp6} has a background distribution with 
a shallow `platform' (seen in Fig.~\ref{fig:back6}), classed as an 
`extreme' background.  
Here we see similar behaviour to experiment 9, but with even greater 
variation in estimated \ac{FAP} values, spanning ${\sim}4$ orders of magnitude 
for all exact \ac{FAP} values below ${\sim}10^{-3}$. 

The platform feature ranging in \ac{SNR} between approximately 
$8 - 11$ contains on average less than $0.1$ triggers per realisation.  
Therefore in many cases the background at high \ac{SNR} is not well 
represented and could fool our algorithms towards underestimation of
$\mathcal{F}$, while in other cases the contamination due to foreground
triggers could mimic the background and lead to an overestimation of
$\mathcal{F}$.

\subsubsection{High foreground rate}
\begin{figure*}
  \centering
  \vspace*{-0.5cm}
  \begin{subfigure}[b]{\columnwidth}
    \centering
    \includegraphics[width=\columnwidth]{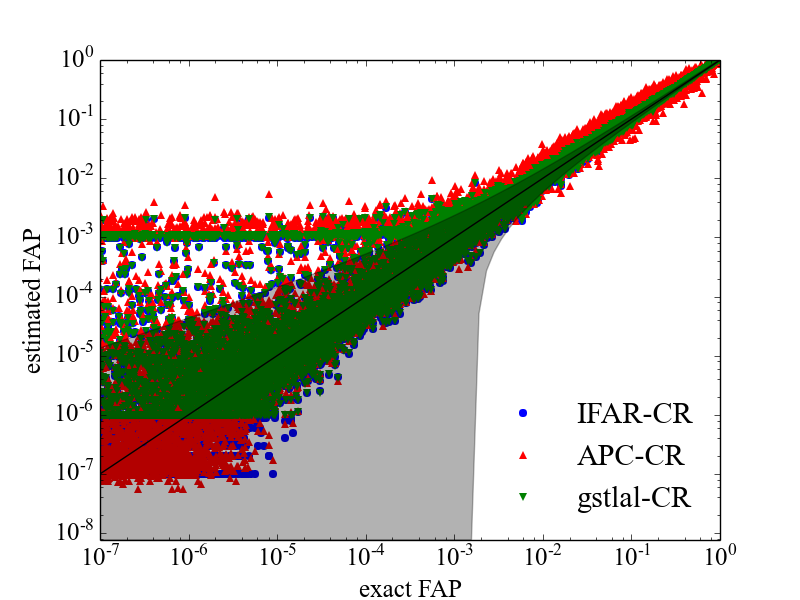}
    \caption{Direct comparison with `coincidence removal'}
    \label{fig:cmp11rm}
  \end{subfigure}%
  \begin{subfigure}[b]{\columnwidth}
    \centering
    \includegraphics[width=\columnwidth]{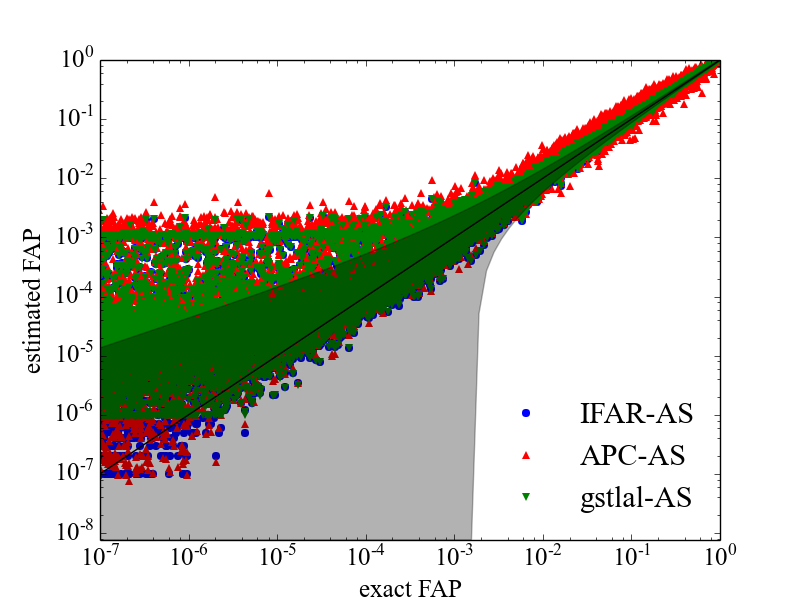}
    \caption{Direct comparison with `all samples'}
    \label{fig:cmp11nonrm}
  \end{subfigure}
  \caption{Direct comparisons of \ac{FAP} for experiment 11. The
    high foreground rate in this case causes general shifts towards
    larger estimates for the \ac{FAP}.  The horizontal bar
    feature visible in both plots is also most prominent in this high
    rate case. The shaded region represents the expected uncertainty from Eq.~\ref{eq:sigerror}.\label{fig:cmp11}}
\end{figure*}

%
%
Experiments with ``high'' foreground rate have, on average, $>1$ foreground 
event per realisation.  Results from experiment 11 are shown in
Fig.~\ref{fig:cmp11} where, as is generally the case, there is good agreement
between algorithms but clear variation between `coincidence removal' and `all
samples' modes.  Compared with experiments with lower foreground rates, the
presence of many contaminating foreground signals shifts estimates to higher
values. For `coincidence removal' mode, this shift reaches the point where, in
this experiment, the bulk of the distribution now appears consistent with the
exact \ac{FAP} with relatively narrow spread.\footnote{We again remind the 
reader again that the comparison is presented on logarithmic axes.}
For the `all samples' mode the contamination due to signals is greater and so
is the corresponding shift to higher values of \ac{FAP} (i.e.\
to the conservative side); in fact, underestimates of $\mathcal{F}$ are
extremely rare.  For all algorithms and for both `coincidence removal' and `all
samples', as the foreground rate increases, a horizontal feature appears in 
these comparison plots, which we discuss in the following section.

\subsubsection{Horizontal bar feature}
%
%
In all high foreground rate scenarios, horizontal features
appear at ${\sim}10^{-3}$ in estimated \ac{FAP}, which are also marginally visible in medium rate experiments. The
process of \ac{FAP} estimation for the loudest coincident event
is based on collecting the fraction of all possible unphysical coincidences
which are louder. The estimation will be strongly biased when there exists a
foreground trigger in one detector that is extremely loud 
and either not found in
coincidence in zero-lag, or coincident with a trigger with very low \ac{SNR}. 
In such cases it 
is highly likely that when performing background estimation it would result
in background coincidences which are louder than the loudest zero-lag event (the details of
this process are specific to each algorithm).  Assuming a method that
makes use of all possible unphysical trigger combinations between detectors,
this corresponds to ${\sim}10^{4}$ louder background coincidences out of a
possible ${\sim}10^{8}$ in total. Considering an expected ${\sim}10$ zero-lag
coincidences this gives an estimated \ac{FAP} of these
asymmetric events as ${\sim}10^{-3}$.

%
%
In experiment 11 (Fig.~\ref{fig:cmp11}), there are $\sim 650$ such
realisations.  For $\sim 500$ of them, the cause is a single astrophysical
signal appearing as an extremely loud trigger in one detector, while for the
other detector the combination of antenna pattern and non-central $\chi^2$
random fluctuations results in a sub-threshold \ac{SNR} and is hence not
recorded as a trigger. The remaining $\sim 150$ events also have very loud
\acp{SNR} in one detector, but in these cases the counterpart in the second
detector appears only as a relatively weak trigger. 
When foreground events appear with
asymmetrical \acp{SNR} between the two detectors, removing coincident triggers
from the background estimate could mitigate overestimation occurring in such
cases; while for the ${\sim}500$ realisations that contain a single loud
foreground trigger which does not form a coincidence, overestimation will occur
regardless of the method used.

\subsubsection{Uncertainty estimate}
Throughout figure \ref{fig:cmp14} - \ref{fig:cmp11}, a shaded region was plotted 
which represents the uncertainty predicted from Eq. \ref{eq:sigerror}.  In the 
derivation of Eq. \ref{eq:sigerror} several simplifying assumptions were used, 
thus some discrepancy between the theoretical expectation and the actual spread
is not surprising. 
However, this expression does capture the order of magnitude of the uncertainty, 
so as a ``rule of thumb'' it serves as a guide in the estimation of uncertainty for 
the \ac{FAP}.

%
%
\subsection{Box plots \label{sec:box}}
%
%
In this section we characterise the estimated \ac{FAP} 
values in more detail by conditioning them on the value of exact \ac{FAP}. 
We take decade-width slices in exact \ac{FAP}
and for each summarise the distributions of estimated
values. For a given experiment, algorithm, and mode, we isolate those results
corresponding to the chosen \ac{FAP} decade and take the ratio
of estimated to exact \ac{FAP} value.  We then calculate and
plot a central box, median point, mean value, whiskers and individual outliers.
 
The central box contains the central half of all estimated values, covering
from $25\%$ to $75\%$ of the total sorted values, i.e.\ between the first and
the third quartile, also known as the \ac{IQR}. The
box is divided by a vertical line identifying the $50\%$ (median) quartile.
When the distribution is relatively concentrated, and the most extreme samples
lie within $1.5\times$ the \ac{IQR}, then the whisker ends at the most extreme
samples. Otherwise the whisker is drawn to be $1.5\times$ the \ac{IQR}, and
outliers beyond the whisker are drawn as individual points. We also indicate
on these plots the mean value of ratio for each distribution.

%
%
Since we are more interested in the region of low \ac{FAP}
values, where detection of foreground signals will likely occur, we take
bins corresponding to values of exact \ac{FAP} between
$(10^{-5}$--$10^{-4})$, $(10^{-4}$--$10^{-3})$, and $(10^{-3}$--$10^{-2})$.  For
each bin, we draw box plots on the ratio between the estimated and exact
\ac{FAP} value, using a logarithmic x-axis scale.  
The vertical purple line
corresponds to where the log of the ratio is zero, meaning that the estimation
and exact \ac{FAP} are identical; left hand side means the
estimated \ac{FAP} value is smaller than the actual value, which
translates to an underestimation of \ac{FAP}. In all plots the
vertical axis gives the experiment index ranging from the lowest foreground
rates to the highest. For each index there are 3 coloured boxes associated with
each algorithm.  Figures are divided into 2 plots, one for the `coincidence
removal' mode and the other for `all samples'.

\subsubsection{False alarm probability range $10^{-3}$--$10^{-2}$}
\begin{figure}
  \centering
  \vspace*{-0.5cm}
  \begin{subfigure}[b]{\columnwidth}
    \centering
    \includegraphics[width=\columnwidth]{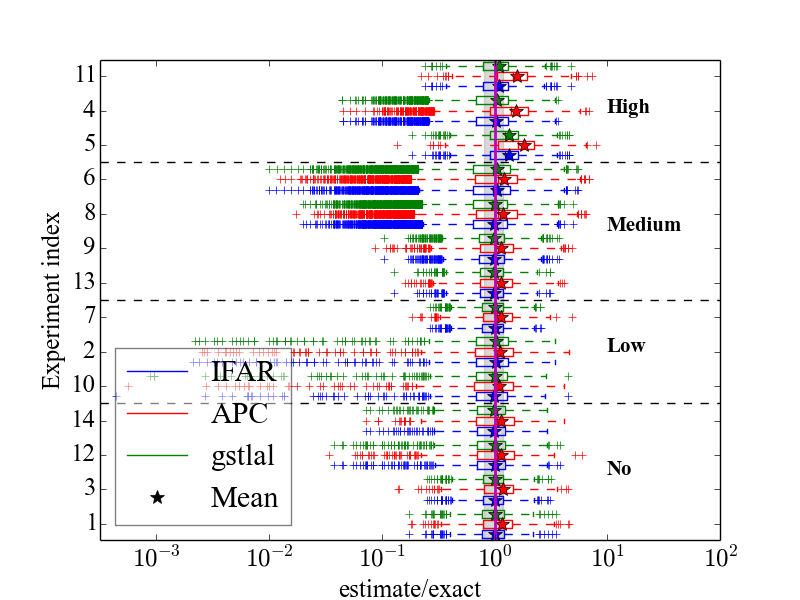}
    \caption{Box plots based on `coincidence removal'. }
    \label{fig:box-3rm}
  \end{subfigure}\\
  \begin{subfigure}[b]{\columnwidth}
    \centering
    \includegraphics[width=\columnwidth]{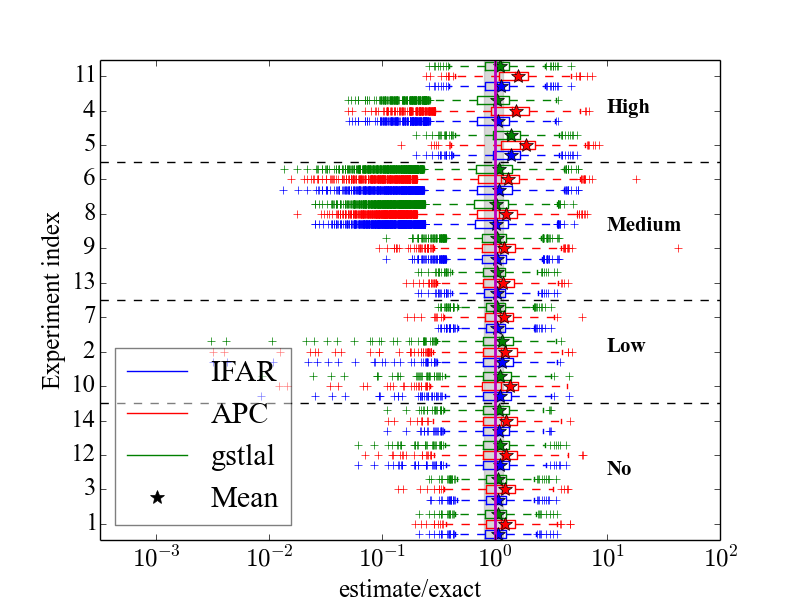}
    \caption{Box plots based on `all samples'}
    \label{fig:box-3nonrm}
  \end{subfigure}
  \caption{Box plots of the ratio of estimated to exact
    \ac{FAP} value, for exact \acp{FAP} between $10^{-3}$ 
    and $10^{-2}$. The shaded region represents the expected uncertainty from Eq.~\ref{eq:sigerror}.}\label{fig:box-3}
\end{figure}

%
%
In Fig.~\ref{fig:box-3} we see relatively tight and symmetric distributions for
all algorithms when considering the \ac{IQR} with strong agreement specifically
between the gstlal and IFAR algorithms.  We remind the reader that the \ac{APC}
algorithm was
not optimised at high \ac{FAP} values and hence shows very
slightly broader distributions. We note that the extrema of the \ac{FAP}
 ratios in most experiments range symmetrically over 
$\lesssim\!\pm 1$
order of magnitude.  However, for some experiments, most notably 12, 10, 2, 8, 6 and 4 
there are large deviations in the extrema towards 
underestimates of \ac{FAP}.  Such deviations would be classed as 
non-conservative, i.e.\ events are estimated as more rare than indicated by 
the exact calculation.  This effect is somewhat reduced for the `all 
samples' mode, most evidently for experiments 2 and 10. 

%
%
The points identified with star symbols in Fig.~\ref{fig:box-3} show the means
of the distribution of ratios. 
In general, the means for the `coincidence removal' mode are slightly more 
consistent with the expected vertical line than for the `all samples' mode.  
This trend will be amplified in subsequent sections as we investigate lower 
values of exact \ac{FAP}.  For this $(10^{-3}$--$10^{-2})$ region
we note that, for reasons discussed earlier, the means computed from
\ac{APC} tend to overestimate the expected value.

\subsubsection{False alarm probability range $10^{-4}$--$10^{-3}$}
\begin{figure}
  \centering
  \vspace*{-0.5cm}
  \begin{subfigure}[b]{\columnwidth}
    \centering
    \includegraphics[width=\columnwidth]{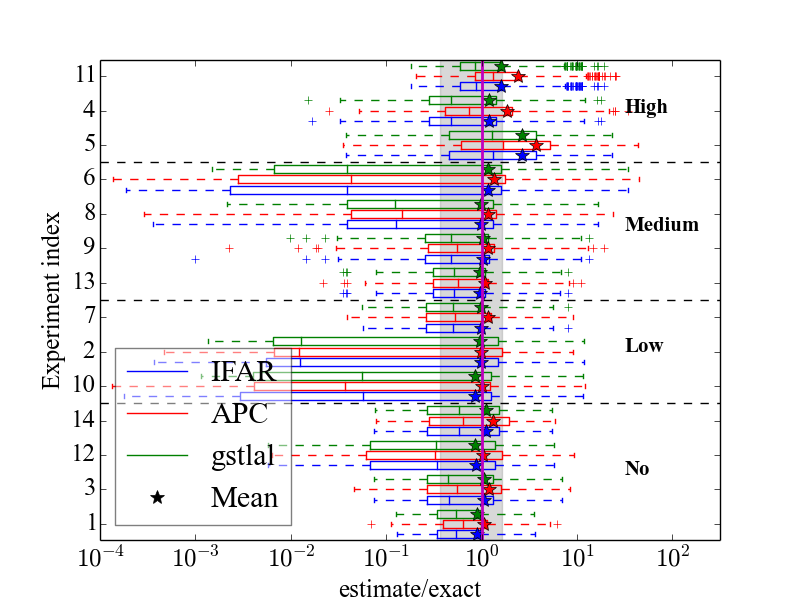}
    \caption{Box plots based on `coincidence removal'}
    \label{fig:box-4rm}
  \end{subfigure}\\
  \begin{subfigure}[b]{\columnwidth}
    \centering
    \includegraphics[width=\columnwidth]{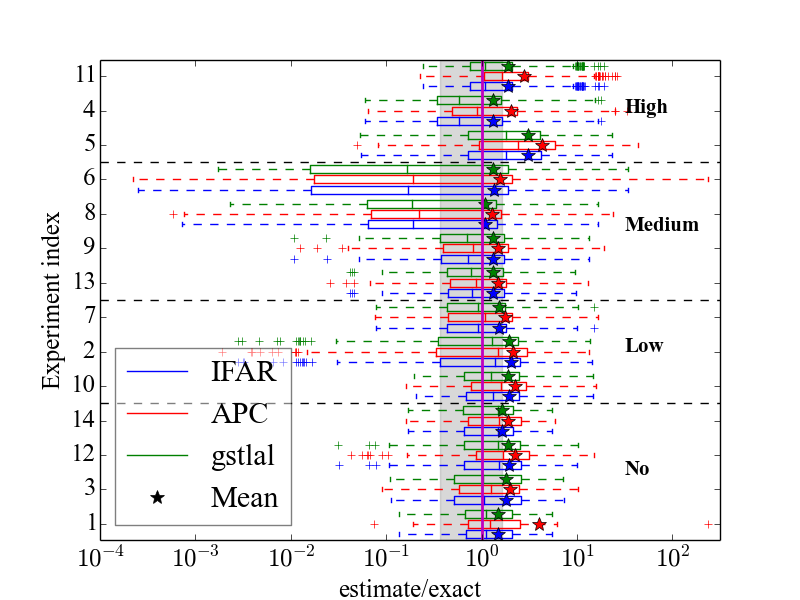}
    \caption{Box plots based on `all samples'}
    \label{fig:box-4nonrm}
  \end{subfigure}
  \caption{Box plots of the ratio of estimated to exact
    \ac{FAP} value, for exact \acp{FAP} between $10^{-4}$
    and $10^{-3}$. The shaded region represents the expected uncertainty from Eq.~\ref{eq:sigerror}.\label{fig:box-4}}
\end{figure}

%
%
As we move to lower \ac{FAP} ranges, shown in
Fig.~\ref{fig:box-4}, we start to see the effects of having lower numbers of
results.  By definition, for experiments with no foreground we expect 
to see a factor of ${\approx}10$ fewer
results in the decade $(10^{-4}$--$10^{-3})$ than in the decade 
 $(10^{-3}$--$10^{-2})$, implying larger statistical fluctuations due to the
reduced number of samples.  We also see intrinsically broader 
distributions, as the estimation methods themselves are 
constrained by the infrequency of loud, low-\ac{FAP} events. As
seen in previous figures of merit, results differ only slightly between
algorithms with the largest differences coming from the issue of inclusion or
removal of coincident triggers.

%
%
Overall, we see ranges in the extrema spanning $\pm 1$ order of magnitude for 
both `coincidence removal' and `all samples' modes.  However, for
experiments 10, 2, 6, and 8 the lower extrema extend to $\sim 4$
order of magnitude below the expected ratio for the `coincidence removal' mode.
This behaviour is mitigated for the `all samples' mode: note that
for experiment 10 the extrema are reduced to be
consistent with the majority of other experiments. In general it is
clear that the \acp{IQR} for the `coincidence removal' mode are broader in
logarithmic space than for `all samples'.  This increase in
width is always to lower values of the ratio, implying underestimation of
the \ac{FAP}. This trend is also exemplified by the locations of 
the median values: for the `coincidence removal' mode, low foreground rates yield 
medians skewed to lower values by factors of ${\sim}2$--$200$.  For the 3 high
foreground rate experiments the \acp{IQR} and corresponding medians
appear consistent with the exact \ac{FAP}.  For the `all samples'
mode the \acp{IQR} and medians are relatively symmetric about the
exact \ac{FAP} and the \acp{IQR} are in all cases narrower than for
the `coincidence removal' mode. 

%
%
In this \ac{FAP} range the difference in distribution means 
between the `coincidence removal' and `all samples' modes becomes more obvious. 
The removal mode results consistently return
mean estimates well within factors of $2$ for all no, low and medium 
foreground rates.  For high foreground rates they consistently overestimate the
means by up to a factor of $\sim 3$.  For the `all samples' mode there is a
clear overestimate of the ratio (implying a conservative overestimate of the
\ac{FAP}) for all experiments irrespective of foreground rate or
background complexity.  This overestimate is in general a factor of ${\sim}2$.
Note though that the estimates from both modes for the three
high foreground rate experiments are very similar in their distributions and
means.

\subsubsection{False alarm probability range $10^{-5}$--$10^{-4}$}
\begin{figure}
  \centering
  \vspace*{-0.5cm}
  \begin{subfigure}[b]{\columnwidth}
    \centering
    \includegraphics[width=\columnwidth]{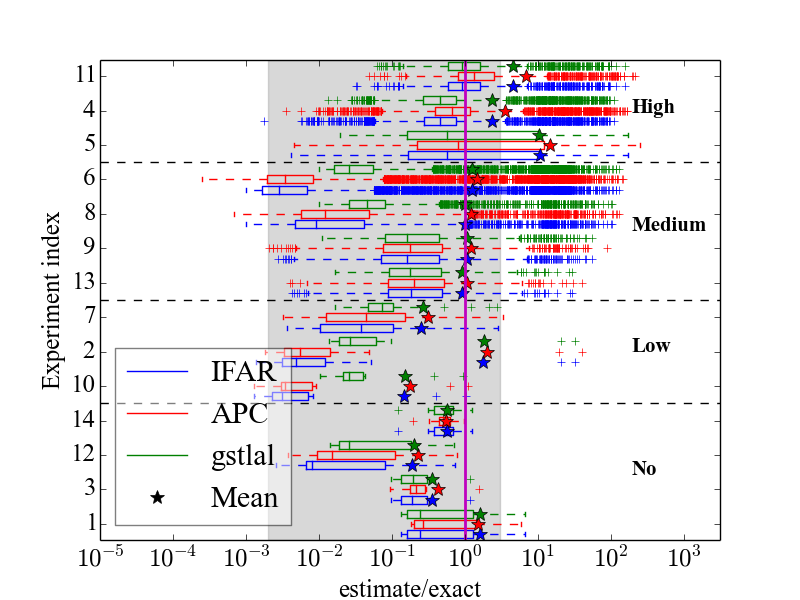}
    \caption{Box plots based on `coincidence removal'}
    \label{fig:box-5rm}
  \end{subfigure}\\
  \begin{subfigure}[b]{\columnwidth}
    \centering
    \includegraphics[width=\columnwidth]{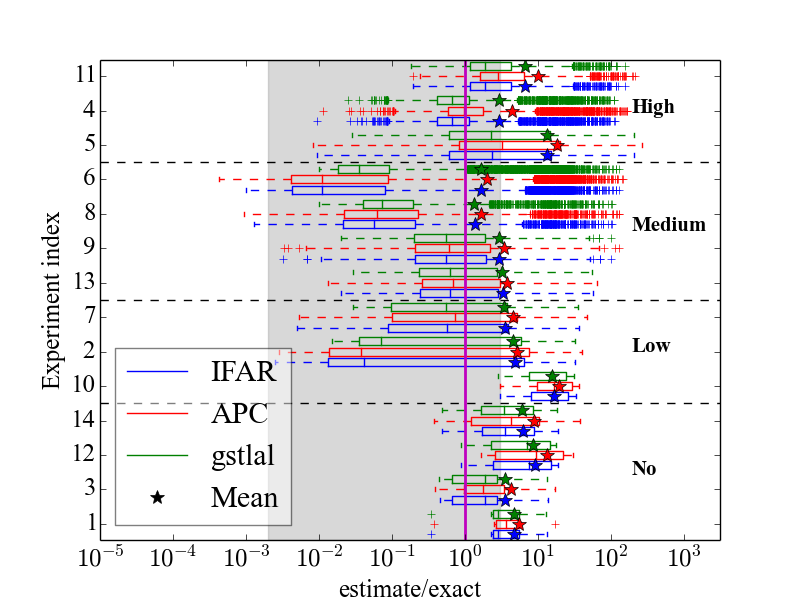}
    \caption{Box plots based on `all samples'}
    \label{fig:box-5nonrm}
  \end{subfigure}
  \caption{Box plots for the ratio between estimated and exact
\ac{FAP} values for \acp{FAP} between $10^{-5}$ and
    $10^{-4}$. The shaded region represents the expected uncertainty from Eq.~\ref{eq:sigerror}.\label{fig:box-5}}
\end{figure}

%
%
In this \ac{FAP} range the uncertainties and variation in the
results are strongly influenced by the low number of events present at such low
\ac{FAP} values. Among all the experiments with no foreground rate suffer the most.
Nonetheless, in Fig.~\ref{fig:box-5} we see
similarities between algorithms and striking differences between
`coincidence removal' and `all samples' modes.  Firstly, in all cases the
variation in extrema is comparable, in this case spanning ${\sim\pm}3$ orders
of magnitude.  The \acp{IQR} are broadly scattered and in many cases do not
intersect with the exact \ac{FAP}.  This is not indicative of poor estimation
but indicative of highly non-symmetric distributions.

%
%
For no and low foreground rates there is a marked difference between results from
`coincidence removal' and `all samples' modes.  For `coincidence removal' all
distributions are skewed to low values which is also a characteristic for the
medium foreground rate experiments.  For example, in experiment 12 there are no
estimates in any of the realisations in this range that overestimate the 
\ac{FAP}.  Removal methods in general in this range of very low \ac{FAP} 
for low and medium foreground rates provide \acp{IQR} of
width ${\sim}1$ order of magnitude with medians shifted by between
${\sim}1$--$2$ orders of magnitude below the exact values. For `all samples'
mode, all no and low foreground experiments (with the exception of experiment 2)
provide conservative (over)estimates of the \ac{FAP} ratio with
\acp{IQR} and extrema ranges spanning ${<}1$ and ${\sim}1$ order of magnitude
respectively. 
We see that for experiment 10 there are no `all samples' estimates in any
realisation that underestimate the \ac{FAP}.  With `all samples'
there is then a marked change as we move to medium level foreground rates and
the distributions become relatively symmetric in log-space with all medians
lower than, but mostly within a factor of 2 of, the exact \ac{FAP}. Experiments
6 and 8 both have medium level foreground rates and give rise to results that
are very similar between `coincidence removal' and `all samples' results and
that exhibit highly skewed distributions to low values with long distribution
tails extending to high values.  This trend of similarity is then continued for
high foreground rates where there is little distinction between either
algorithm of `coincidence removal' mode. In these cases however, the
distributions appear relatively symmetric in log-space with reasonably well
constrained \acp{IQR}.

%
%
Considering the means of the distributions, we
see similar behaviour but with more variation than in the previous
\ac{FAP} ranges.  Starting with `all samples' mode there is
consistent conservative bias in the mean of the ratio of estimated to
exact \ac{FAP}.  For low/no and high foreground rates this bias is
${\sim}1$ order of magnitude which reduces to a factor of ${\sim}3$
overestimate for medium level foregrounds.  For the `coincidence removal' mode,
no and low foreground rates return distribution means that are scattered
symmetrically about the exact \ac{FAP} with a variation of ${\sim}1$
order of magnitude.  For all medium level foregrounds including
experiments with low, medium and high background complexity, the mean
estimates are very tightly localised around the exact \acp{FAP} with
variations of $10$'s of percent.  For high foreground rates the means
obtained from both `coincidence removal' and `all samples' modes are all
consistently overestimates of the exact \ac{FAP} by up to ${\sim}1$
order of magnitude. 

%
%
By looking at the bulk distribution in the box plots, it seems that `coincidence
removal' will generally underestimate the \ac{FAP}, while `all
samples' is systematically unbiased over all 14 experiments.  However, note
that if we only look at the mean values, then `all samples' modes almost always
overestimate the \ac{FAP}, while `coincidence removal' is
generally consistent with the exact \ac{FAP}.  This indirectly
proves that `coincidence removal' mode is a mean-unbiased estimator. 
For a significant event, the exact \ac{FAP} is very close to 
zero, thus any difference due to underestimation may be very small in
linear space though not necessarily small in log space, while
overestimation could bias the value with a large relative deviation.  
When the \ac{FAP} is very small, the estimation uncertainty 
(variance) is large relative to the exact value; then, since estimated values 
are bounded below by zero, in order to achieve a mean-unbiased estimate a large
majority of estimated values are necessarily below the exact value, i.e.\
underestimates. In other words, the distribution is highly skewed.

%
%
\subsection{\ac{ROC} analyses\label{sec:ROC}}
%
%

The \ac{FAP} value is a measure of the rarity of observed
events, but in this section we treat the estimated \ac{FAP}
as a test statistic.  This allows us to use \ac{ROC} plots to compare the
ability to distinguish signal from noise for each method. In practice this
involves taking one of our experiments containing $10^5$ realisations and, as a
function of a variable threshold on our test statistic (the \ac{FAP}), 
computing the following.  The \ac{FPR} is the fraction of loudest
events due to background that had estimated \ac{FAP} values below the 
threshold. 
The \ac{TPR} is computed as the fraction of loudest events due to the
foreground that had estimated \acp{FAP} below the threshold.  For each
choice of threshold a point can be placed in the \ac{FPR}-\ac{TPR} plane
creating an \ac{ROC} curve for a given test-statistic.  Better performing
statistics have a higher \ac{TPR}
at a given \ac{FPR}. 
A perfect method 
would recover $100\%$ of the signals while incurring no false alarms,
corresponding to a \ac{ROC} curve that passes through the upper left corner.  A
random classifier assigning uniformly distributed random numbers to the 
\ac{FAP} would lead to identical \ac{FPR} and \ac{TPR}, yielding a
diagonal \ac{ROC} curve.

%
%
\begin{figure}
  \centering
  \vspace*{-0.5cm}
  \includegraphics[width=\columnwidth]{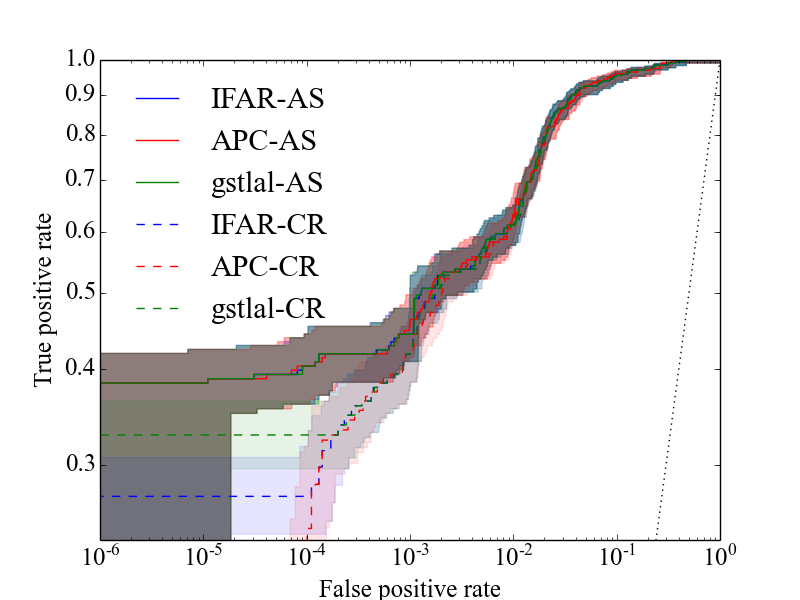}
  \caption{\ac{ROC} plot for experiment 2. The error bars in both horizontal
and vertical directions are calculated with a binomial likelihood under a
68\% credible interval. Solid lines correspond to `all samples', while dashed
lines corresponds to `coincidence removal'. The dotted line represents the expected performance of random guess, no rational analysis would perform worse than it.\label{fig:ROC_2low}}
\end{figure}
Error regions for the \ac{FPR} and \ac{TPR} are computed using a 
binomial likelihood function. 
In general, as can be seen in our \ac{ROC} plots, as the foreground
rate goes up, the more events are available to compute the \ac{TPR},
reducing the vertical uncertainties.  Conversely, the more background events 
are available, the smaller the horizontal uncertainties. 

In the following subsections we focus on the experiments where there are
clear discrepancies, leaving cases where there is agreement between
methods to Appendix~\ref{sec:AppROC}.  We stress that \ac{ROC} curves 
allow us to assess the ability of a test-statistic to distinguish between
realisations where the loudest event is foreground vs.\ background; 
however they make no direct statement on the accuracy or precision of 
\ac{FAP} estimation.

\subsubsection{Low foreground rate}
%
%
There are 3 experiments, 2, 7 and 10, that have low foreground rates.
The \ac{ROC} curve for experiment 2 in Fig.~\ref{fig:ROC_2low} exhibits a 
number of interesting features.  Firstly, there is
general agreement between algorithms; deviations are only visible between
`coincidence removal' and `all samples' modes of operation. At a \ac{FPR} of
${\sim}10^{-3}$ and below, the `all samples' mode appear to achieve higher
\acp{TPR}, when accounting for their respective uncertainties, by $\sim 10\%$.
This indicates that in this low-rate case, where $\approx 1$ in 1000 loudest
events were actual foreground signals, the `all samples' mode is more efficient
at detecting signals at low \acp{FPR}.  We can identify all experiments that
show such deviations, and all have tail features or obvious asymmetry between
the two detectors' background distributions, combined with a low to medium
foreground rate. 

\subsubsection{Medium foreground rate}
%
%
Experiments 6, 8, 9 and 13 contain medium foreground rates and
collectively show two types of behaviour.  Experiments 9 and 13 show
general agreement between algorithms and `coincidence removal' and `all samples'
modes (see 
Figs.~\ref{fig:ROC_9} and~\ref{fig:ROC_13}).  Experiments 6 and 8 show
similar deviations to those seen in the $p$--$p$ plots in
Section~\ref{sec:ppplots}. This similarity is not surprising since the
vertical axes of the $p$-$p$ plots are identical to horizontal axes of
the \ac{ROC} plots, with the distinction that they are computed on
background-only and background-foreground experiments respectively.
\begin{figure}
  \centering
  \vspace*{-0.5cm}
  \includegraphics[width=\columnwidth]{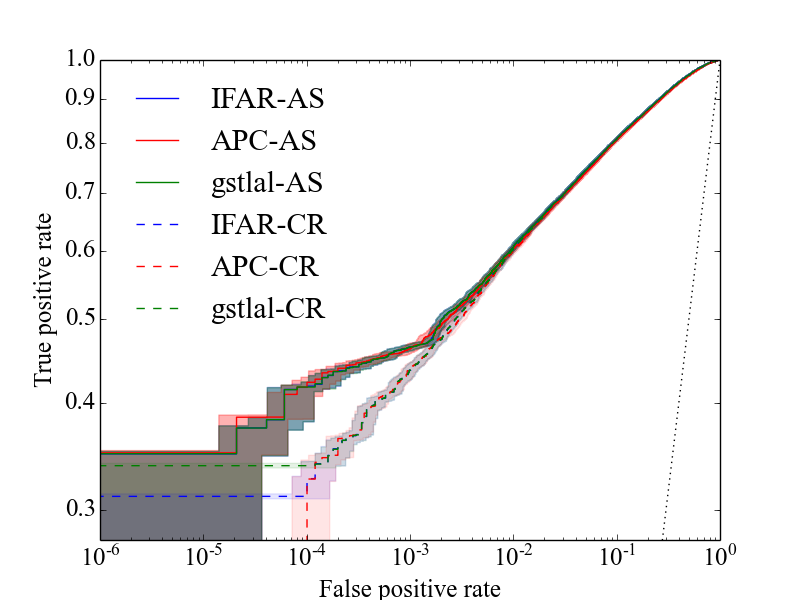}
  \caption{\ac{ROC} plot for experiment 8. The error bars in both horizontal
and vertical direction are calculated with a binomial likelihood under a 68\%
credible interval. Solid lines correspond to `all samples', while dashed lines
corresponds to `coincidence removal'. The dotted line represents the expected
performance of random guess, no rational analysis would perform worse than it.\label{fig:ROC_8med}}
\end{figure}

Here we focus on Experiment 8 shown in Fig.~\ref{fig:ROC_8med} which
contained realistic but slightly different background distributions in
each detector.  As seen in the low-foreground example there is good
agreement between algorithms but differences between `coincidence removal' and
`all samples' modes.  In this case, due to the increased number of
foreground events, this difference is more clearly visible and the
disscrepancies are inconsistent with the estimated uncertainties.  So for
medium foreground rates we conclude that as a detection statistic,
the `all samples' mode obtains higher \acp{TPR} at fixed 
\ac{FPR} for low values of \ac{FAP}.  We remind the reader that
detection claims will be made at low \ac{FAP} values,
although the absolute values appearing in our \ac{MDC} may not be 
representative of those obtained by algorithms in a realistic analysis.
%

\subsubsection{High foreground rate}
\begin{figure}
  \centering
  \vspace*{-0.5cm}
  \includegraphics[width=0.9\columnwidth]{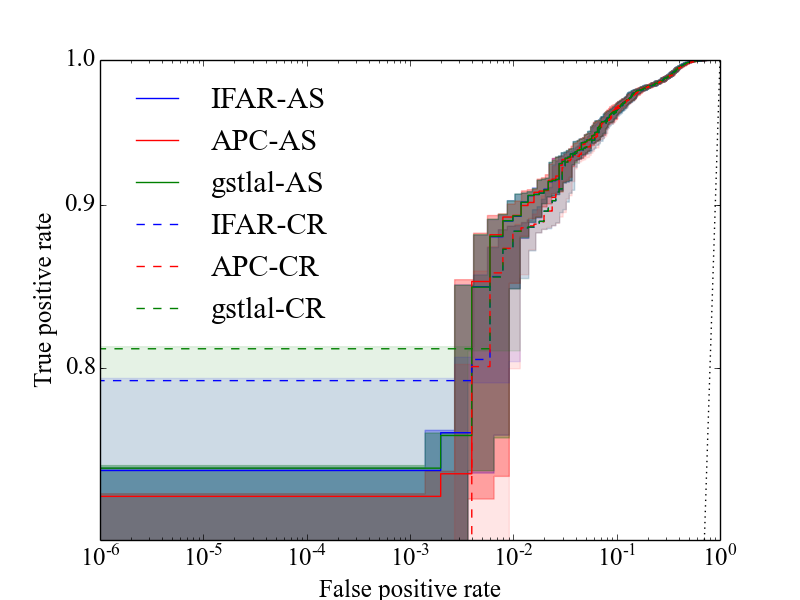}
  \caption{\ac{ROC} plot for experiment 4. The error bars in both horizontal
and vertical direction are calculated with a binomial likelihood under a 68\%
credible interval. Solid lines correspond to `all samples', while dashed lines
correspond to `coincidence removal'. The dotted line represents the expected performance of random guess, no rational analysis would perform worse than it.\label{fig:ROC_4high}}
\end{figure}

%
%
The high rate experiments 4, 5 and 11 all show similar behaviour. Here we show
Fig.~\ref{fig:ROC_4high} as an example where we see general agreement (within
our estimated uncertainties) between algorithms and between the `coincidence 
removal' and `all samples' modes.  The high rates used result in ${>}90\%$ of 
realisations containing a loudest coincident event from the foreground.  
The 3 experiments were examples of all 3 levels of background complexity 
respectively and the
results indicate that in terms of detection efficiency, all algorithms and
approaches perform equally well at any fixed \ac{FPR}. 
This high rate scenario is most likely to be relevant
in the epochs following the first direct detection. 

%
%
\subsection{Uncertainty in estimation\label{sec:error}}
%
%
From the results presented in the previous sections we can conclude that the
relative uncertainty in the estimation of \ac{FAP} increases as
\ac{FAP} decreases.  As shown in
Figs.~\ref{fig:cmp14},~\ref{fig:cmp9},~\ref{fig:cmp6}, and~\ref{fig:cmp11},
with the exception of \ac{APC} results, which are designed to be accurate only
at low \ac{FAP} values, both other estimation algorithms show
larger spread as the \ac{FAP} value goes lower.  Specific
features in the background distributions would vary the actual spread, but the
trend is consistent.  When the value of the exact \ac{FAP} is as
small as $10^{-4}$, the relative uncertainty can exceed $100\%$; since the
estimated \ac{FAP} can't be negative, the errors in 
estimated \ac{FAP} are not symmetrically distributed. 
%
%

Any claims of \ac{GW} detection will necessarily be made at low values of 
\ac{FAP} and most likely at low or medium level foreground rate.  Using
Fig.~\ref{fig:box-5} and focusing on the low and medium foreground rate experiments
10, 2, 7, 13, and 9 it is clear from both `coincidence removal' and `all
samples' modes that a single loudest event measurement of \ac{FAP} 
will be drawn from a very broad distribution. For `all samples'
mode, in all but experiment 10 for medium or low foregrounds, 
the \ac{IQR} looks  
symmetric mostly consistent with the true value within ${\pm}1$ order of
magnitude.  For the equivalent `coincidence removal' mode results, the 
\ac{IQR} is much smaller than its counterpart in the `all sample' mode in
log-space, and hence more precise.  The extrema between approaches are
comparable but the bulk of the distribution is more concentrated in the
`coincidence removal' case.  Note however that the median 
and IQR for the `coincidence removal' mode are both uniformly below the exact 
value, in some cases by orders of magnitude.

%
%
Our results show that the uncertainty in the \ac{FAP} value can be predicted 
by the ``rule of thumb'' Eq.~\ref{eq:sigerror}, derived
in Sec.~\ref{sec:sigerror}.  The scope of the \ac{MDC} described in this work
was designed to probe multiple background types and foreground levels with
large numbers of realisations. The number of triggers simulated for each
realisation does not match the number that would be present in a realistic
\ac{GW} search; nor does the coincidence test correspond to a real search
in which many thousands of distinct, but nontrivially correlated templates are 
used.  
Hence, the \ac{FAP} value at which the uncertainty in our
estimates approaches $100\%$ can be very different in reality than the $10^{-3}$
value seen in our simulations. 
We expect that Eq.~\ref{eq:sigerror} will serve as a guide in estimating
uncertainties due to different realisations of noise in the \ac{FAP} values 
produced within realistic \ac{GW} searches. 

\section{Discussion\label{sec:discussion}}
%
%
We have designed and generated an \ac{MDC} to evaluate the performance of
methods to determine the significance of transient \ac{GW} candidate events,
simulating wide ranges of qualitatively different background distributions 
and of possible signal rates.
We compared the results obtained
via three different algorithms, each operating in two different modes: 
estimating the distribution and rate of background events by considering 
either `coincidence removal': excluding single-detector triggers found to be 
coincident in the search from the background estimate; or `all samples': 
including all single-detector triggers in the background estimate.
These background estimates were then used to assign a false alarm probability (FAP)
 to the loudest coincident event in each realisation of each 
experiment in the \ac{MDC}. Our methods for comparison of these results 
include self-consistency checks via the use of $p$-$p$ plots, for those
experiments not containing any foreground signals; direct comparison of
estimates with the exact value of \ac{FAP} resulting
from the generating background distributions; a box-plot comparison
of the result distributions from each experiment; and finally an \ac{ROC}
analysis to identify detection efficiency at fixed \ac{FPR}. 

Based on these comparison analyses we find the following key conclusions:
\begin{enumerate}[a.]
%
%
\item The results of all experiments indicate a good consistency among all
  three algorithms; disagreements only occur between the modes of `coincidence
  removal' and `all samples' for low values of false alarm probability (FAP).
%
%
\item For all except high foreground rates, the `coincidence removal' mode 
  is more likely to underestimate the \ac{FAP} than 
  overestimate, though producing an unbiased mean value; 
  while the `all samples' mode tends to overestimate, especially for smaller
  \acp{FAP}.
%
%
\item For high foreground rates, both the `coincidence removal' and `all samples'
  modes overestimate \ac{FAP} as indicated by the mean over 
  realisations, while the 
  `coincidence removal' mode has a larger fraction of underestimated \ac{FAP}.
%
%
\item We only observe extreme underestimation of \ac{FAP} from
  either complex or realistic background distributions. When the foreground rate
  is not high, or the background distributions have no tail or asymmetry between
  detectors there is evidence that the `coincidence removal' mode can
  underestimate the \ac{FAP}.
%
%
\item Due to different detector responses and random noise fluctuations, an 
  astrophysical event may induce a very loud trigger in one detector and a 
  sub-threshold trigger in the other. This would lead to a systematic 
  overestimation of \ac{FAP} for all algorithms and modes, as 
  shown as Fig.~\ref{fig:cmp11}.
%
%
\item The evaluation of \ac{FAP} is found to be 
  entirely self-consistent only 
  for the `all samples' mode. In this \ac{MDC}, the `coincidence removal' mode 
  would claim a fraction of $10^{-4}$ realisations containing only noise 
  events to have \ac{FAP} $10^{-5}$, hence the estimated 
  \ac{FAP} for this mode does not have a frequentist 
  interpretation at low values. 
  Such a deviation, however, is expected to occur at far lower
  \ac{FAP} values for a real analysis of \ac{GW} data.
%
%
\item In general, \ac{FAP} estimates computed using `all samples'
  were found to be more effective at distinguishing foreground events from
  background at fixed \ac{FPR}. This was most notable in experiment 8
  which contained a medium level foreground and a realistic background.  
%
%
\item For all but high foreground rates, coincidence removal methods 
  have the merit of appearing to be unbiased estimators concerning the 
  mean of \ac{FAP} estimates.  However, the 
  distributions of these estimates are highly asymmetric, especially for 
  low \ac{FAP} values. Single realisations from `coincidence 
  removal' mode are consequently highly likely to have underestimated values 
  of \ac{FAP}.  By contrast, estimates from the `all 
  samples' mode show a conservative bias (i.e.\ towards overestimation) 
  concerning the mean over realisations; but for low \ac{FAP} 
  events, these estimates more likely to lie around the exact value or 
  above it.
%
%
\item The relative uncertainty in the estimation is larger when the \ac{FAP} is smaller. The relative uncertainty reaches $100\%$ when
  the \ac{FAP} is about $10^{-4}$, for the experiment 
  parameters chosen in this \ac{MDC}. This value depends on the expected number of
  coincident events and the number 
  of single detector triggers.
\end{enumerate}
%

At the time of writing this we eagerly await results from the Advanced 
detector era. 
While we are aiming to make several detections over the lifespan of
the upgraded Advanced facilities, we should bear in mind that the first
detection(s) may be relatively quiet and belong to a distribution with a low
astrophysical event rate. In this case, we recommend a sacrifice in 
possible accuracy of \ac{FAP} estimation in favour of 
conservatism. Considering also the self-consistency and relative efficacy
of the methods in distinguishing signal realisations from noise, we recommend
the use of the `all samples' mode of any of our 3 algorithms, anticipating 
that false alarms will likely be (conservatively) overestimated rather 
than underestimated for the first \ac{GW} detections.



\begin{acknowledgments}
  The authors have benefited from discussions with many \ac{LSC} and Virgo 
collaboration members and ex-members including Drew Keppel and Matthew West.
We acknowledge the use of LIGO Data Grid computing facilities for the generation
and analysis of the \ac{MDC} data.
C.~M.\ is supported by a Glasgow University Lord
Kelvin Adam Smith Fellowship and the Science and Technology Research Council (STFC)
grant No. ST/ L000946/1.
J.~V.\ is supported by the STFC grant No. ST/K005014/1.
\end{acknowledgments}

\section*{References}
\bibliography{masterbib,cbc-group}

\clearpage

\appendix

\section{Parameters of simulated trigger distributions}\label{sec:triggerParams}
In this section, we list the parameters used to define the background
distributions. Recall that we adopt a form of \ac{SNR}
distribution for the background triggers given by Eq.~\ref{eq:CDF}
using input polynomial coefficients $a_{i}$ as listed in 
Table~\ref{tab:paramsBkgd} for all 14 experiments. 

\begin{table}
  \begin{tabular}{ c || 
 c  c  c  c  c  c  c }
    & \multicolumn{7}{c}{Background parameters} \\
	Expt & $a_0$ & $a_1$ & $a_2$ & $a_3$ & $a_4$
        & $a_5$ & $a_6$  \\
        \cline{1-2} \cline{3-8}
	\multirow{2}{*}{1} & -10.0000 & -5 & 0 & 0 & 0 & 0 & 0 \\  
	                   & -10.0000 & -5 & 0 & 0 & 0 & 0 & 0 \\
	\multirow{2}{*}{2} & -7.2240 & -0.32 & 0.53 & -0.73 & 0.12 & 0.067 & -0.018 \\  
	                   & -7.2240 & -0.32 & 0.53 & -0.73 & 0.12 & 0.067 & -0.018 \\
	\multirow{2}{*}{3} & -7.8000 & -3.9 & 0 & 0 & 0 & 0 & 0 \\  
	                   & -7.8000 & -3.9 & 0 & 0 & 0 & 0 & 0 \\  
	\multirow{2}{*}{4} & -6.2800 & -0.3 & 0.5 & -0.7 & 0.1 & 0.07 & -0.02 \\  
	                   & -4.8800 & -1 & 0 & -0.6 & 0 & -0.04 & -0.05 \\  
	\multirow{2}{*}{5} & -8.4000 & -4.2 & 0 & 0 & 0 & 0 & 0 \\  
                           & -8.0000 & -4 & 0 & 0 & 0 & 0 & 0 \\  
	\multirow{2}{*}{6} & -8.0704 & -3 & 0.8 & 0.01 & -0.05 & 0.007 & -0.0004 \\  
	                   & -8.0704 & -3 & 0.8 & 0.01 & -0.05 & 0.007 & -0.0004 \\  
	\multirow{2}{*}{7} & -9.1072 & -4 & 0.7 & 0.09 & -0.05 & 0.005 & -0.0002 \\  
	                   & -9.6200 & -5.55 & -0.37 & 0 & 0 & 0 & 0 \\  
	\multirow{2}{*}{8} & -3.5040 & -1.4 & 0 & -0.16 & 0 & -0.034 & -0.026 \\
	                   & -4.6400 & -2 & 0 & -0.2 & 0 & -0.03 & -0.03 \\  
	\multirow{2}{*}{9} & -7.0000 & -3.5 & 0 & 0 & 0 & 0 & 0 \\  
	                   & -6.6000 & -3.3 & 0 & 0 & 0 & 0 & 0 \\  
	\multirow{2}{*}{10} & -2.4800 & -1 & 0 & -0.1 & 0 & -0.03 & -0.02 \\  
	                    & -5.8400 & -3 & 0 & -0.1 & 0 & -0.03 & -0.03 \\  
	\multirow{2}{*}{11} & -4.0800 & -1 & 0 & -0.3 & 0 & -0.05 & -0.03 \\  
	                    & -8.3200 & -4 & 0 & -0.1 & 0 & -0.035 & -0.025 \\  
	\multirow{2}{*}{12} & -3.5040 & -1.4 & 0 & -0.16 & 0 & -0.034 & -0.026\\
	                    & -4.6400 & -2 & 0 & -0.2 & 0 & -0.03 & -0.03 \\  
	\multirow{2}{*}{13} & -7.8000 & -3.9 & 0 & 0 & 0 & 0 & 0 \\  
	                    & -7.8000 & -3.9 & 0 & 0 & 0 & 0 & 0 \\  
	\multirow{2}{*}{14} & -6.2800 & -0.3 & 0.5 & -0.7 & 0.1 & 0.07 & -0.02 \\  
	                    & -4.8800 & -1 & 0 & -0.6 & 0 & -0.04 & -0.05 \\  
  \end{tabular}
  \caption{Parameters of our background model distributions.\label{tab:paramsBkgd}}
\end{table}

For the tails of the \acp{CDF}, the form is changed to a simpler
representation as defined in Eq.~\ref{eq:CDF} in order to make sure
that the background distribution is well behaved as the \ac{SNR}
rises. The corresponding parameters $b$, $C_{\rm SP}$, and
$\rho_{\rm SP}$ are listed in Table~\ref{tab:paramsTail}. Notice
that here the actual control parameter is $C_{\rm SP}$, while $b$
and $\rho_{\rm SP}$ are derived values which could be subject to
round-off error.
\begin{table}
  \begin{tabular}{ c | c || c | c | c   }
	Expt & IFO & $C_{\rm SP}$ & $b$ & $\rho_{\rm SP}$ \\
	\hline
	1 & 1 & 1e-10 & -5.0000 & 10.1052 \\
	1 & 2 & 1e-10 & -5.0000 & 10.1052 \\
	\hline
	2 & 1 & 1e-4 & -1.2690 & 9.9447 \\
	2 & 2 & 1e-4 & -1.2690 & 9.9447 \\
	\hline
	3 & 1 & 1e-10 & -3.9000 & 11.4041 \\
	3 & 2 & 1e-10 & -3.9000 & 11.4041 \\
	\hline
	4 & 1 & 1e-4 & -3.0350 & 10.0791 \\
	4 & 2 & 5e-4 & -5.1906 & 8.7474 \\
	\hline
	5 & 1 & 1e-10 & -4.2000 & 10.9823 \\
	5 & 2 & 1e-10 & -4.0000 & 11.2565 \\
	\hline
	6 & 1 & 1e-9 & -6.8668 & 13.0625 \\
	6 & 2 & 1e-9 & -6.8668 & 13.0625 \\
	\hline
	7 & 1 & 1e-9 & -4.3611 & 12.9193 \\
	7 & 2 & 1e-9 & -6.8728 & 9.2876 \\
	\hline
	8 & 1 & 2e-2 & -1.4415 & 7.7886 \\
	8 & 2 & 5e-3 & -2.0660 & 7.8256 \\
	\hline
	9 & 1 & 1e-9 & -3.5000 & 11.4209 \\
	9 & 2 & 1e-9 & -3.3000 & 11.7798 \\
	\hline
	10 & 1 & 5e-2 & -1.0889 & 8.0018 \\
	10 & 2 & 1e-3 & -3.0410 & 7.8544 \\
	\hline
	11 & 1 & 3e-4 & -7.1603 & 9.1251 \\
	11 & 2 & 1e-5 & -4.2931 & 8.2823 \\
	\hline
	12 & 1 & 2e-2 & -1.4415 & 7.7886 \\
	12 & 2 & 5e-3 & -2.0660 & 7.8256 \\
	\hline
	13 & 1 & 1e-9 & -3.9000 & 10.8137 \\
	13 & 2 & 1e-9 & -3.9000 & 10.8137 \\
	\hline
	14 & 1 & 1e-4 & -3.0350 & 10.0791 \\
	14 & 2 & 5e-4 & -5.1906 & 8.7474 \\
  \end{tabular}
  \caption{Parameters of background distributions (tail). 
\label{tab:paramsTail}}
\end{table}

The rate of both background triggers and foreground triggers are
controlled by parameters listed in Table~\ref{tab:paramsRate}.
Here $n$ is the predicted average coincidence number in one realisation,
while the measured $n$ is the actual value concluded from the data,
their consistency reflect our confidence in the generation of the mock data.
The \emph{AstroRate (all)} is the expected rate for astrophysical foreground events
in each realisation, but as only a fraction of them have large enough SNR to be
detectable, thus the \emph{AstroRate (loud)} represents such detectable event rate.
\begin{table}
  \begin{tabular}{ c || c | c | c | c | c | c }
	Expt  & $\lambda_1$ & $\lambda_2$ & $n$ & measured $n$ & AstroRate (loud) & AstroRate (all)\\
	\hline
	1 &  10500 & 10500 & 11.025 & 11.0098 & 0 & 0\\
	\hline
	2 &  11500 & 11500 & 13.225 & 13.2453 & 0.001 & 0.0022\\
	\hline
	3 &  9900 & 9900 & 9.801 & 9.7874 & 0 & 0\\
	\hline
	4 &  12000 & 9000 & 10.8 & 10.8023 & 2.96 & 6.41\\
	\hline
	5 &  9800 & 10100 & 9.898 & 9.8857 & 2.74 & 5.94\\
	\hline
	6 &  8000 & 15000 & 12 & 12.0195 & 0.548 & 1.19\\
	\hline
	7 &  10300 & 9900 & 10.197 & 10.2206 & 0.0011 & 0.0024 \\
	\hline
	8 &  10100 & 11100 & 11.211 & 11.2202 & 0.438 & 0.95 \\
	\hline
	9 &  9700 & 10600 & 10.282 & 10.2785 & 0.11 & 0.24\\
	\hline
	10 &  12000 & 10800 & 12.96 & 12.9552 & 0.0001 & 0.0003\\
	\hline
	11 &  9800 & 10700 & 10.486 & 10.4938 & 3.07 & 6.65\\
	\hline
	12 &  10100 & 11100 & 11.211 & 11.2047 & 0 & 0\\
	\hline
	13 &  9900 & 9900 & 9.801 & 9.7814 & 0.022 & 0.048\\
	\hline
	14 &  12000 & 9000 & 10.8 & 10.7857 & 0 & 0\\
    \end{tabular}
    \caption{Parameters for rates of both background and foreground triggers?\label{tab:paramsRate}}
\end{table}
\clearpage

%
%
\section{Additional results\label{sec:additional}}
%
%
\subsection{\ac{CDF} distribution of \ac{SNR}}\label{sec:CDFSNR}
\begin{figure}[hptb]
  \centering
  \includegraphics[width=0.75\columnwidth]{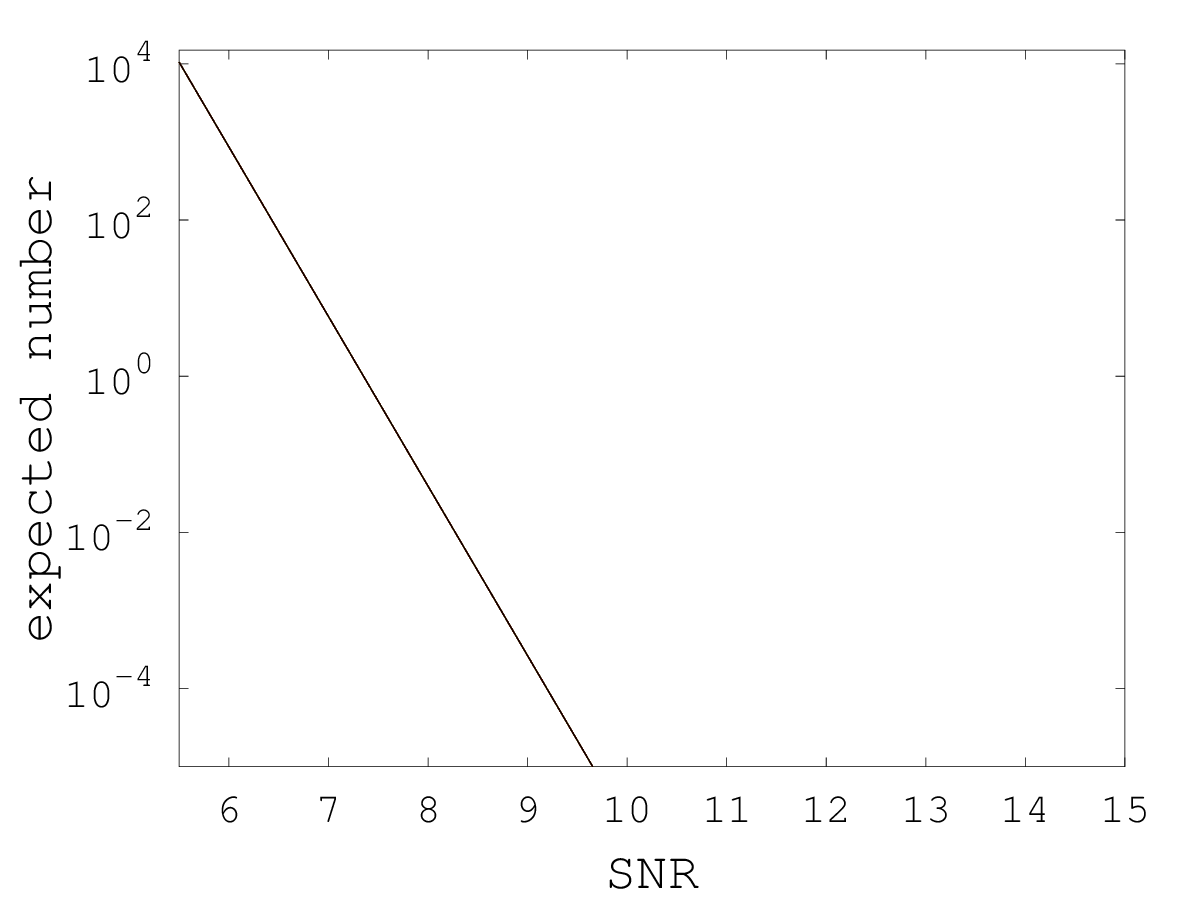}
  \caption{Reverse \ac{CDF} distribution of the triggers' SNR for experiment 1.
The red and green curves represent the two individual detectors,
while the blue curve represents the astronomical signals.
The black lines represent the combined distribution of both background
and foreground triggers.}
  \label{fig:back1}
\end{figure}

\begin{figure}[hptb]
  \centering
  \includegraphics[width=0.75\columnwidth]{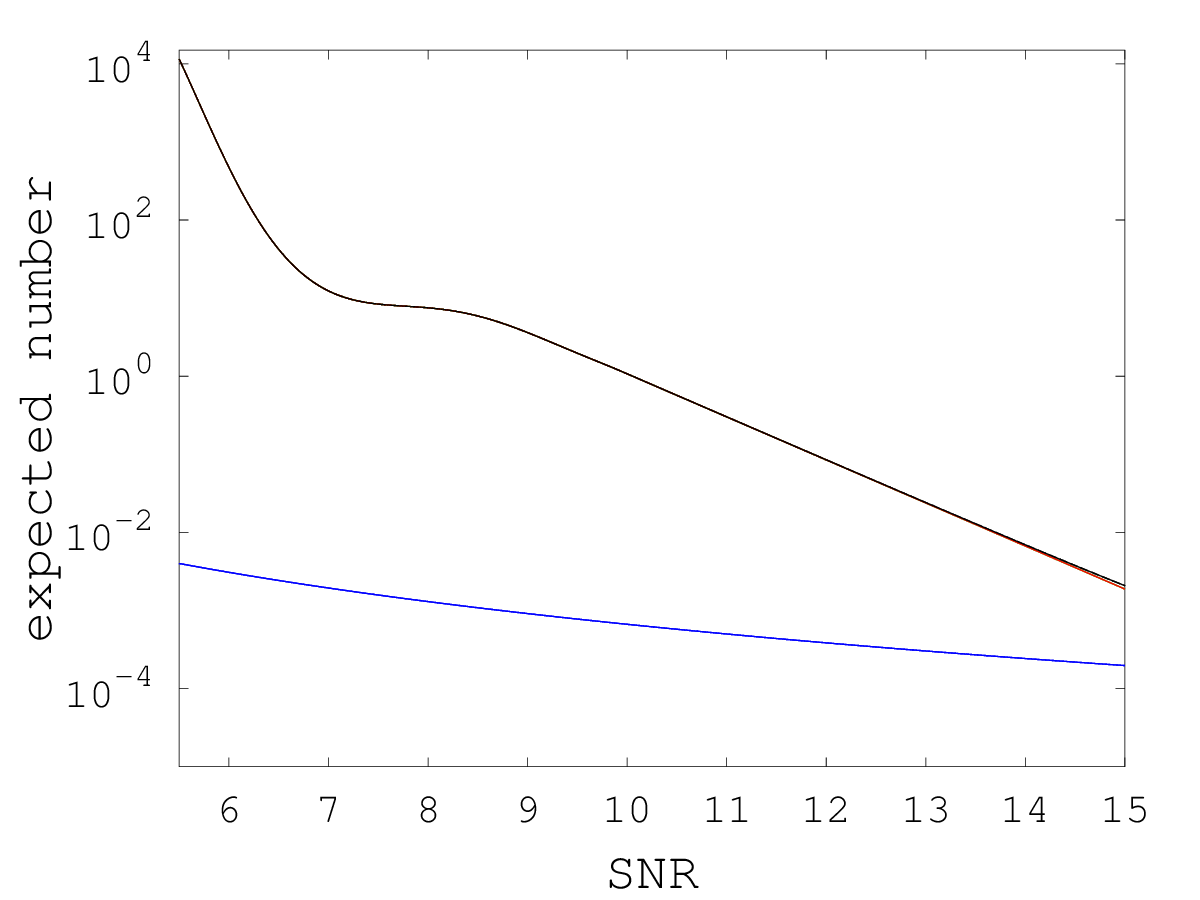}
  \caption{Reverse \ac{CDF} of the triggers' SNR for experiment 2:
colours assigned as in Fig.~\ref{fig:back1}.}
  \label{fig:back2}
\end{figure}

\begin{figure}[hptb]
  \centering
  \includegraphics[width=0.75\columnwidth]{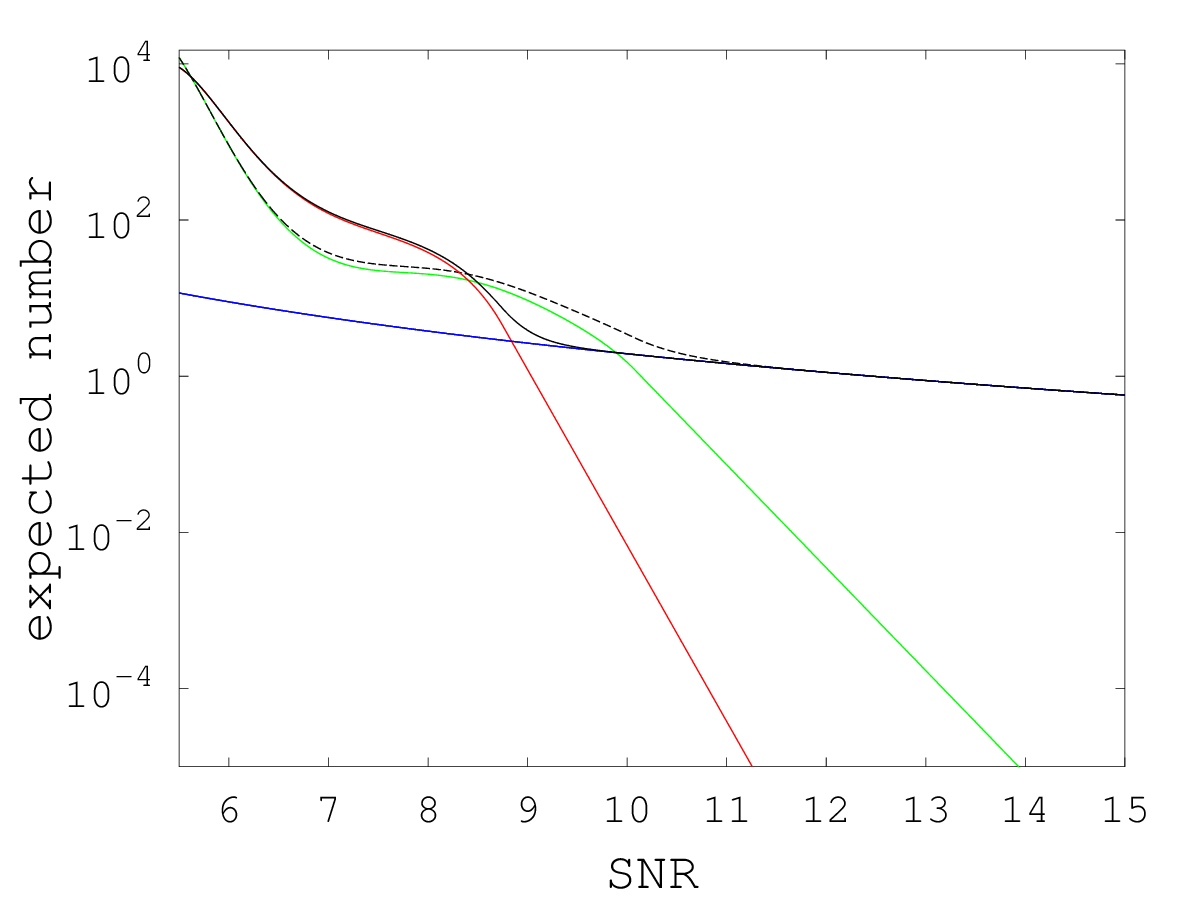}
  \caption{Reverse \ac{CDF} of the triggers' SNR for experiment 4:
colours assigned as in Fig.~\ref{fig:back1}.}
  \label{fig:back4}
\end{figure}

\begin{figure}[hptb]
  \centering
  \includegraphics[width=0.75\columnwidth]{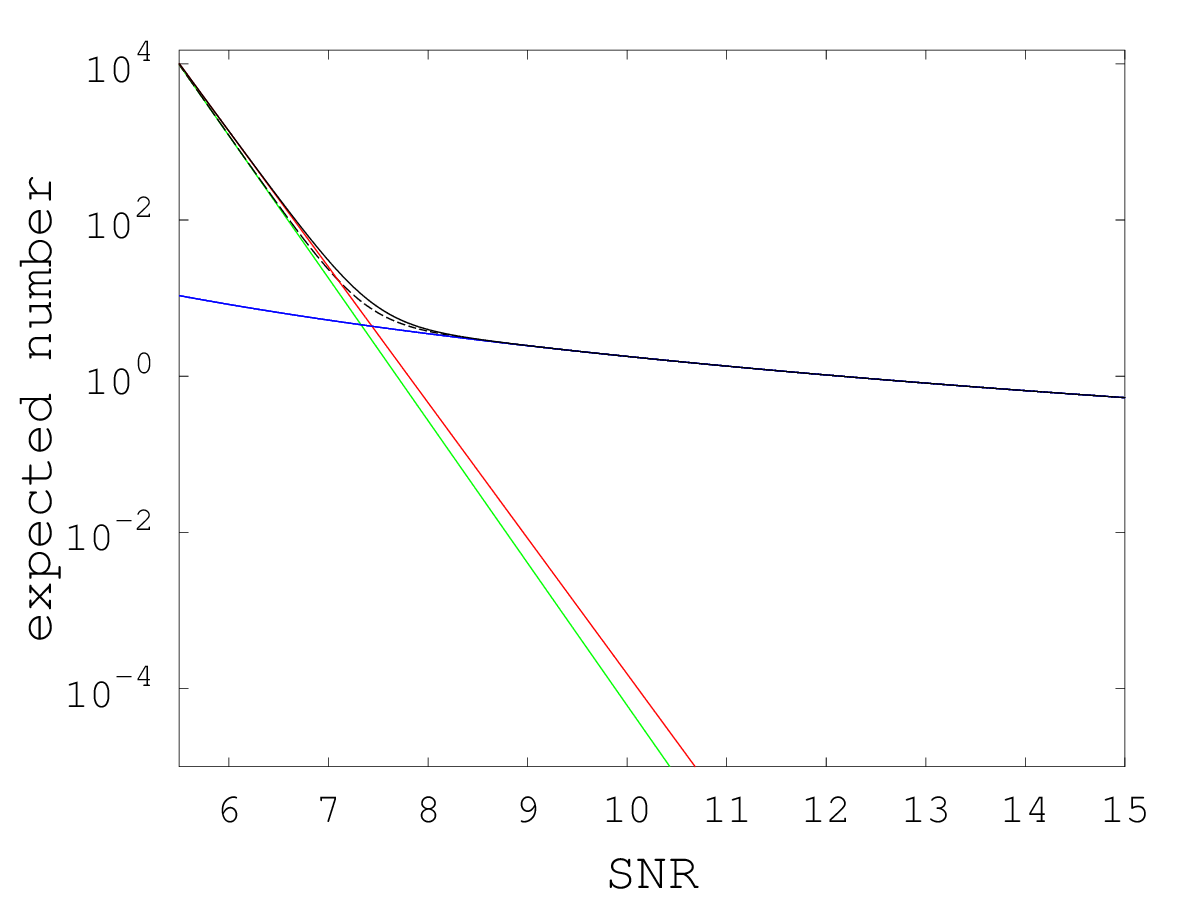}
  \caption{Reverse \ac{CDF} of the triggers' SNR for experiment 5:
colours assigned as in Fig.~\ref{fig:back1}.}
  \label{fig:back5}
\end{figure}

\begin{figure}[hptb]
  \centering
  \includegraphics[width=0.75\columnwidth]{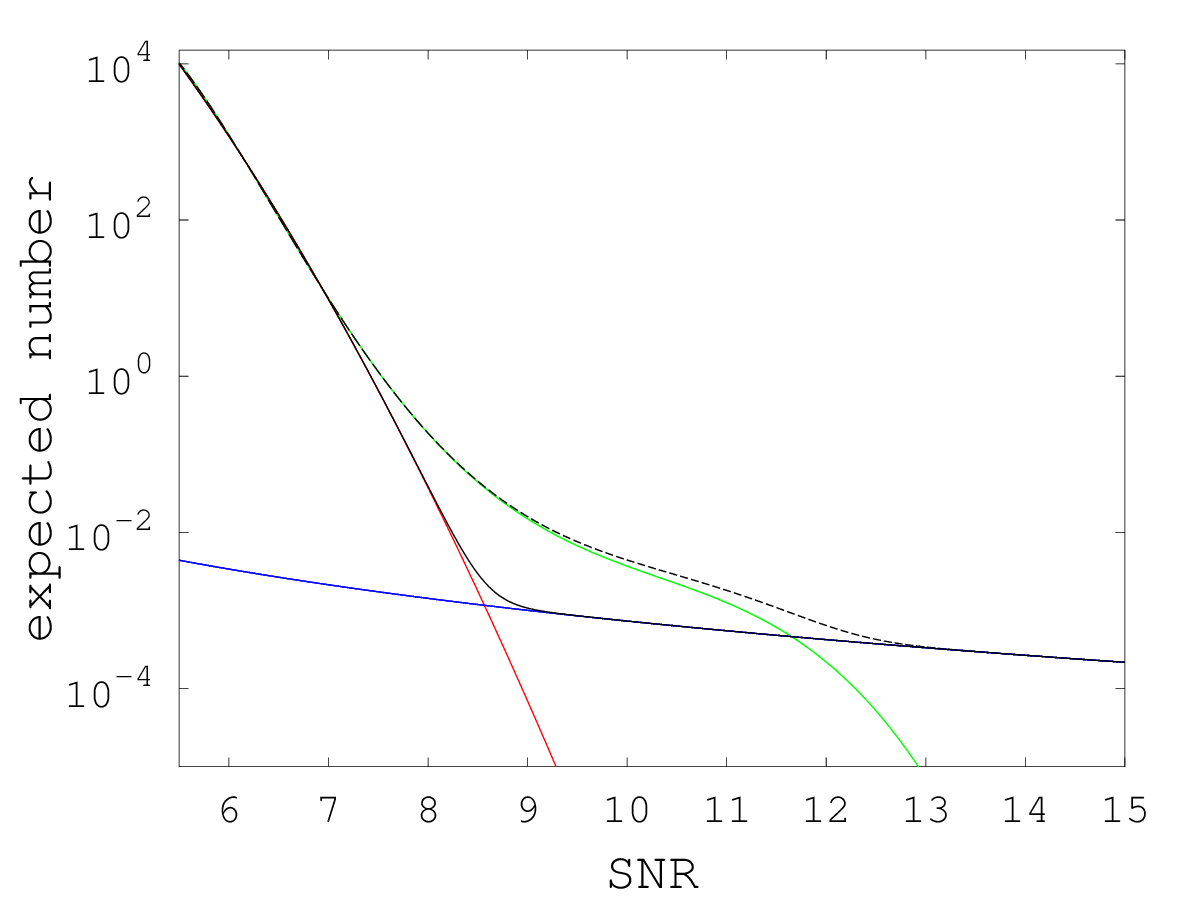}
  \caption{Reverse \ac{CDF} of the triggers' SNR for experiment 7:
colours assigned as in Fig.~\ref{fig:back1}.}
  \label{fig:back7}
\end{figure}

\begin{figure}[hptb]
  \centering
  \includegraphics[width=0.75\columnwidth]{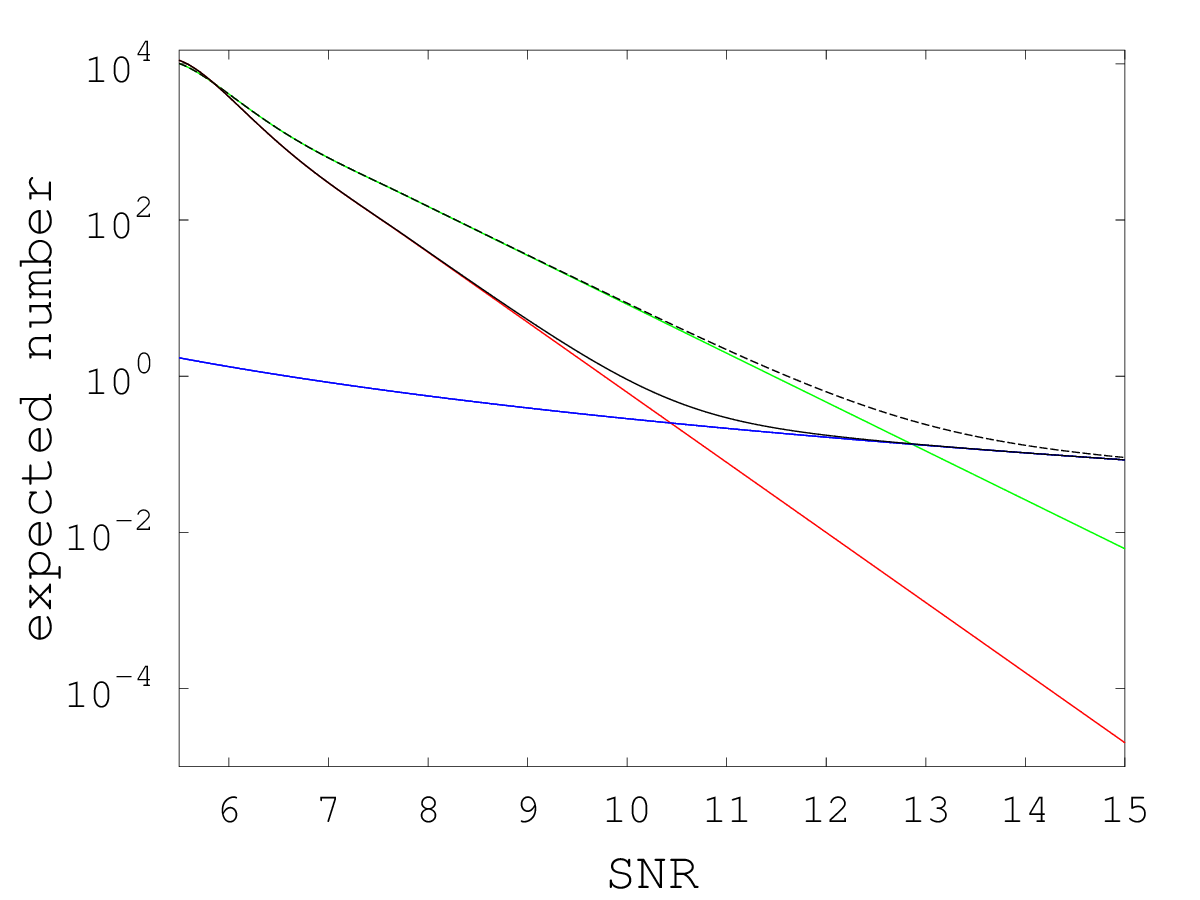}
  \caption{Reverse \ac{CDF} of the triggers' SNR for experiment 8:
colours assigned as in Fig.~\ref{fig:back1}.}
  \label{fig:back8}
\end{figure}

\begin{figure}[hptb]
  \centering
  \includegraphics[width=0.75\columnwidth]{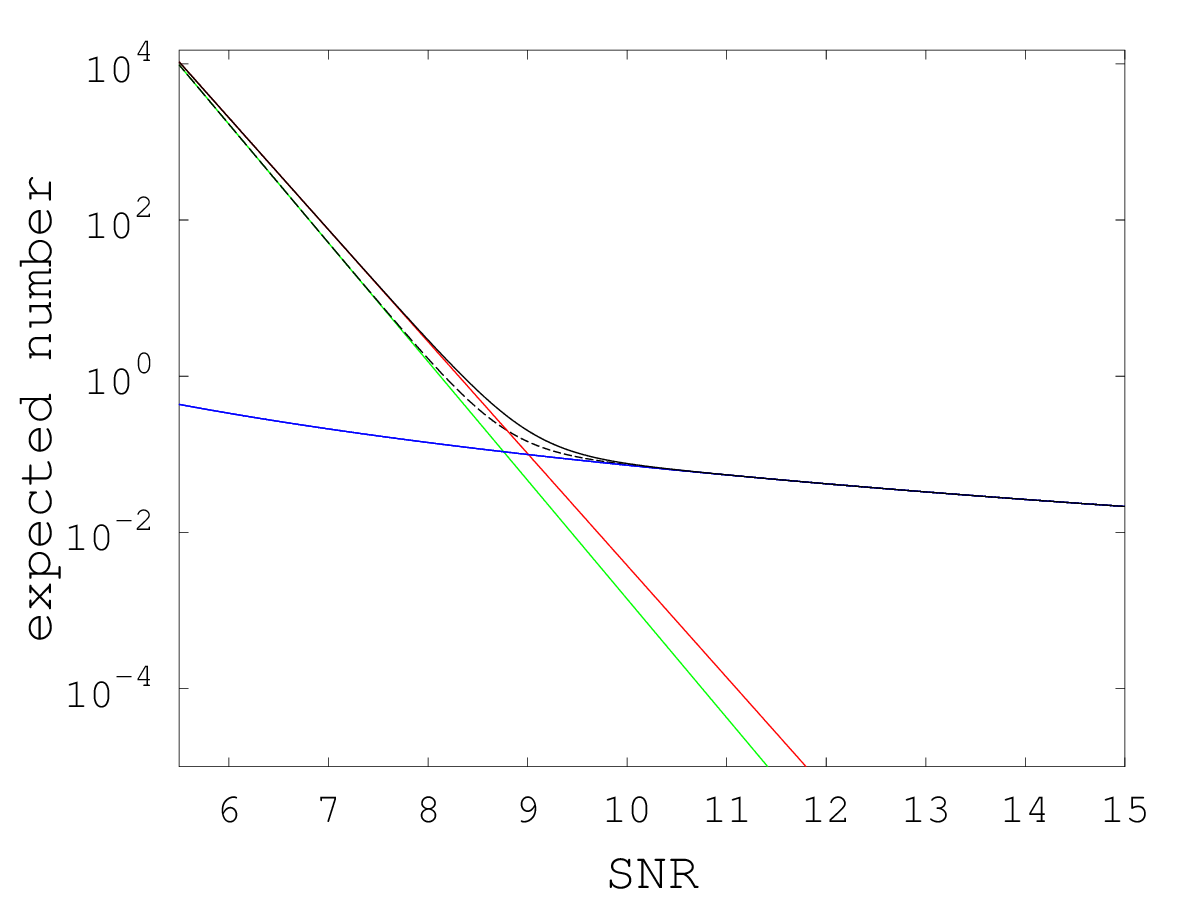}
  \caption{Reverse \ac{CDF} of the triggers' SNR for experiment 9:
colours assigned as in Fig.~\ref{fig:back1}.}
  \label{fig:back9}
\end{figure}

\begin{figure}[hptb]
  \centering
  \includegraphics[width=0.75\columnwidth]{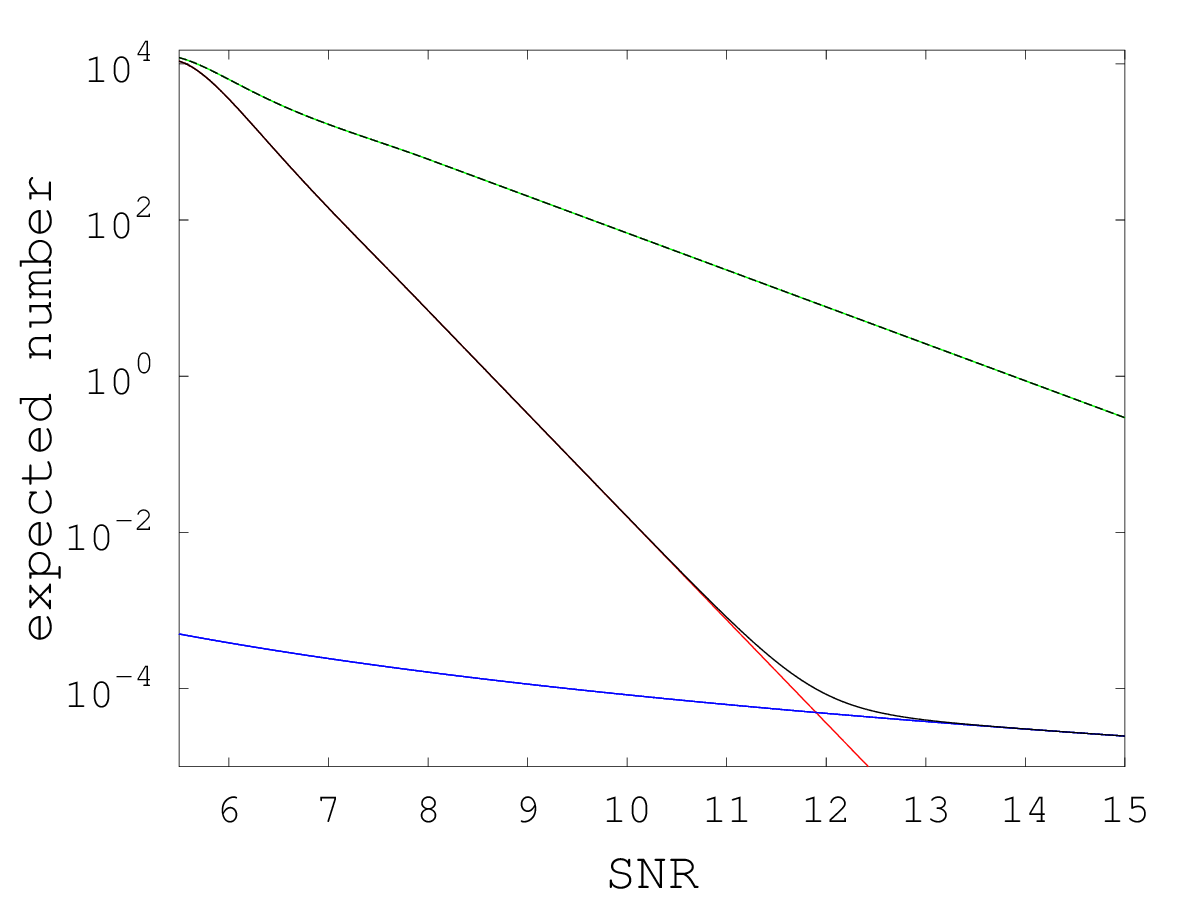}
  \caption{Reverse \ac{CDF} of the triggers' SNR for experiment 10:
colours assigned as in Fig.~\ref{fig:back1}.}
  \label{fig:back10}
\end{figure}

\begin{figure}[hptb]
  \centering
  \includegraphics[width=0.75\columnwidth]{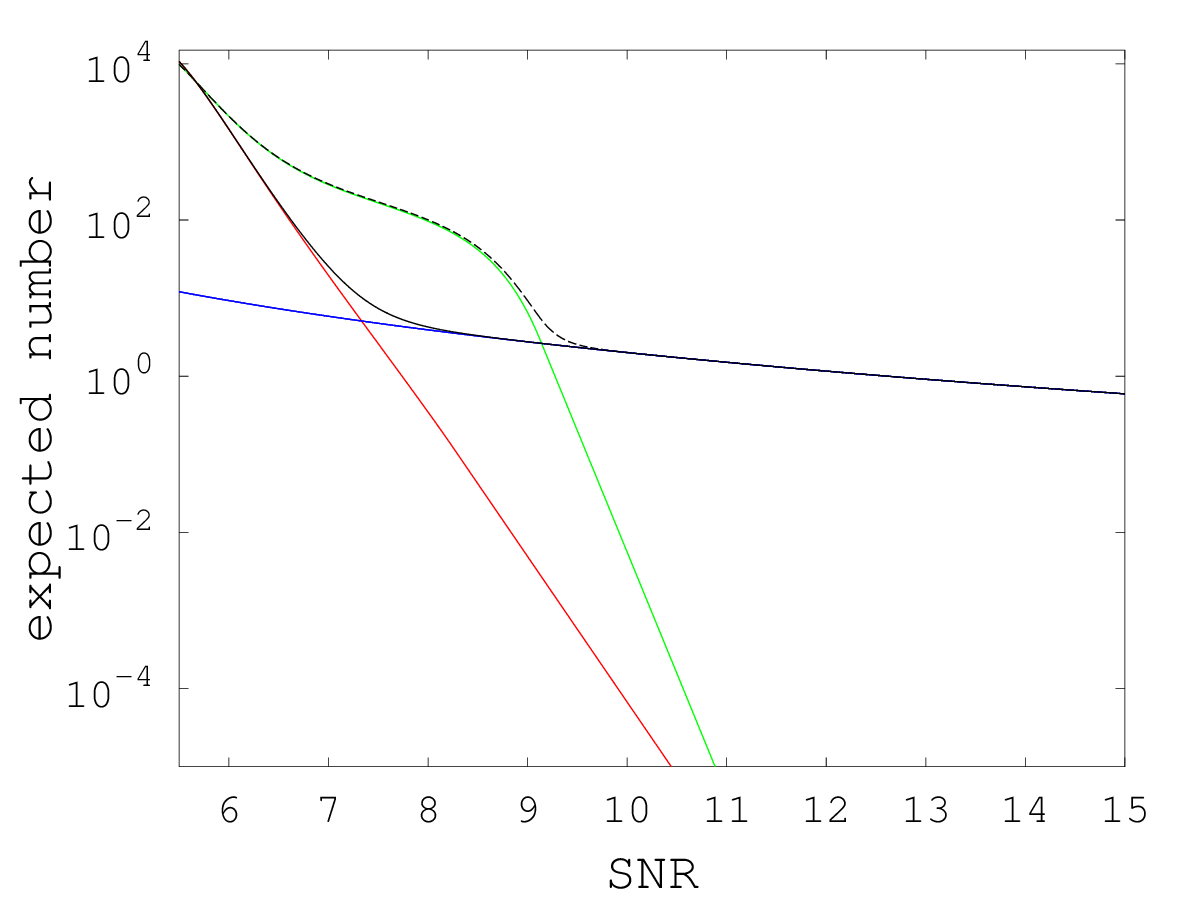}
  \caption{Reverse \ac{CDF} of the triggers' SNR for experiment 11:
colours assigned as in Fig.~\ref{fig:back1}.}
  \label{fig:back11}
\end{figure}

\begin{figure}[hptb]
  \centering
  \includegraphics[width=0.75\columnwidth]{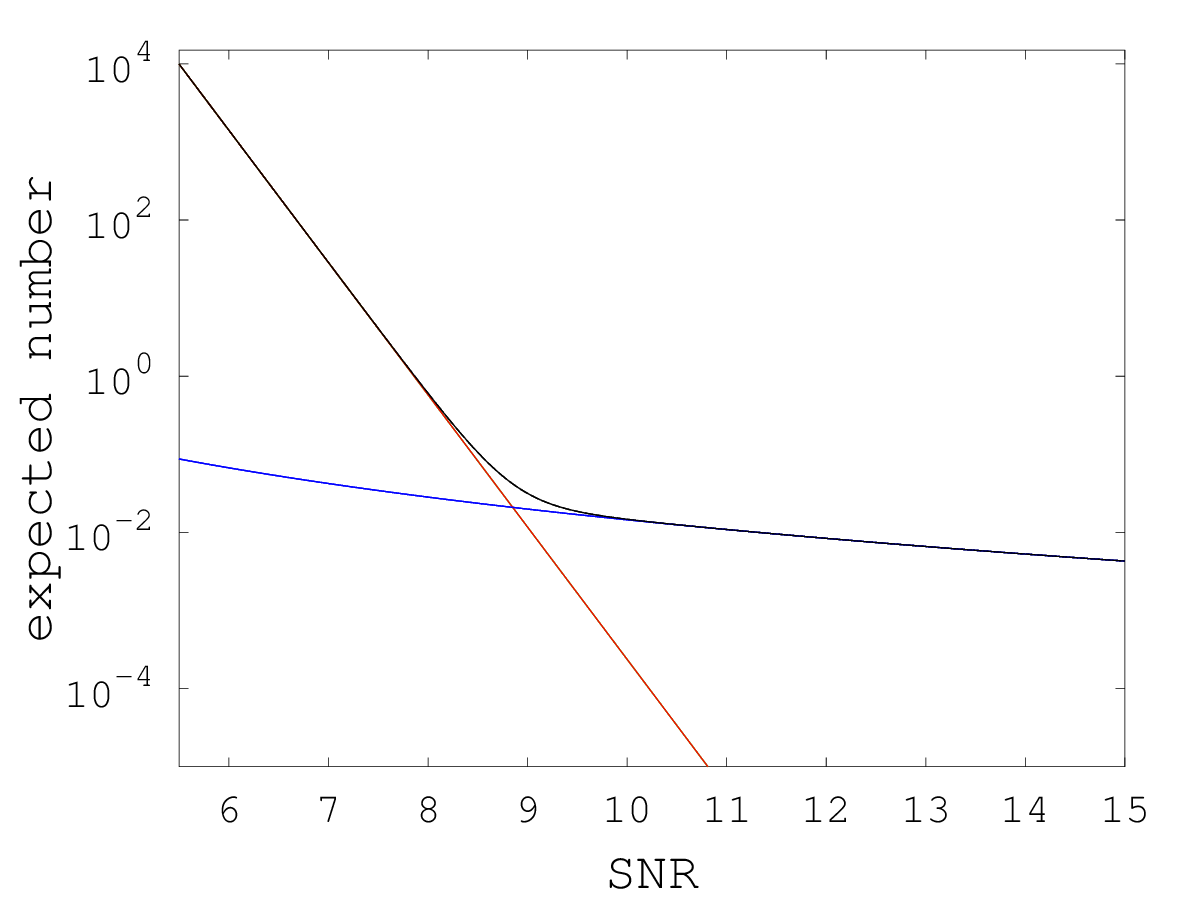}
  \caption{Reverse \ac{CDF} of the triggers' SNR for experiment 13:
colours assigned as in Fig.~\ref{fig:back1}.}
  \label{fig:back13}
\end{figure}

In this section, we show the reverse \ac{CDF} distribution of the 
triggers' \ac{SNR}.
For one experiment, two detectors could have different 
background distributions, but they share the same astronomical foreground 
distribution.
In figure \ref{fig:back1} to \ref{fig:back13}, two detectors' background SNR 
distribution is demonstrated. Background for two detectors are represented by red 
and green line, while the 
foreground distribution is shown as the blue line, and combined distribution is 
the black line.

\clearpage


\subsection{Direct comparison}\label{sec:AppDirComp}

\begin{figure*}
  \centering
  \begin{subfigure}[b]{\columnwidth}
    \centering
    \includegraphics[width=0.9\columnwidth]{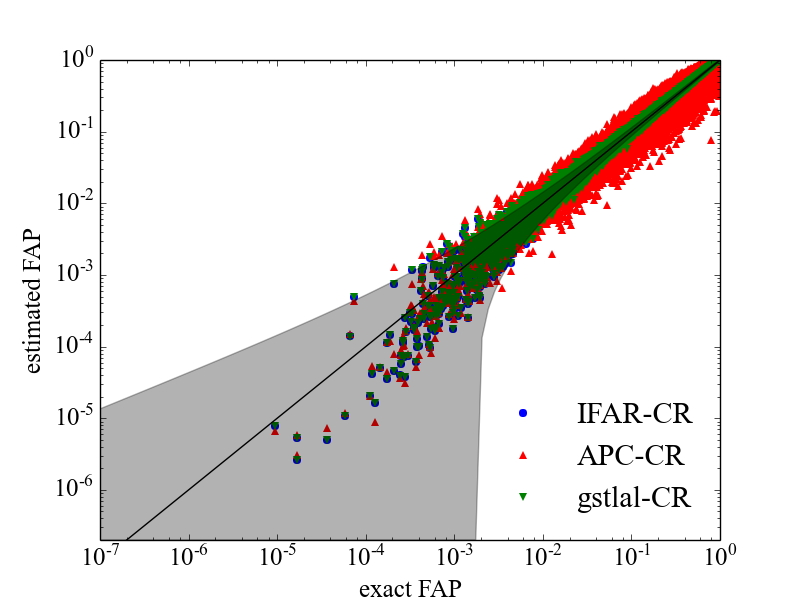}
    \caption{direct comparison with `coincidence removal'}
    \label{fig:cmp1rm}
  \end{subfigure}
  \begin{subfigure}[b]{\columnwidth}
    \centering
    \includegraphics[width=0.9\columnwidth]{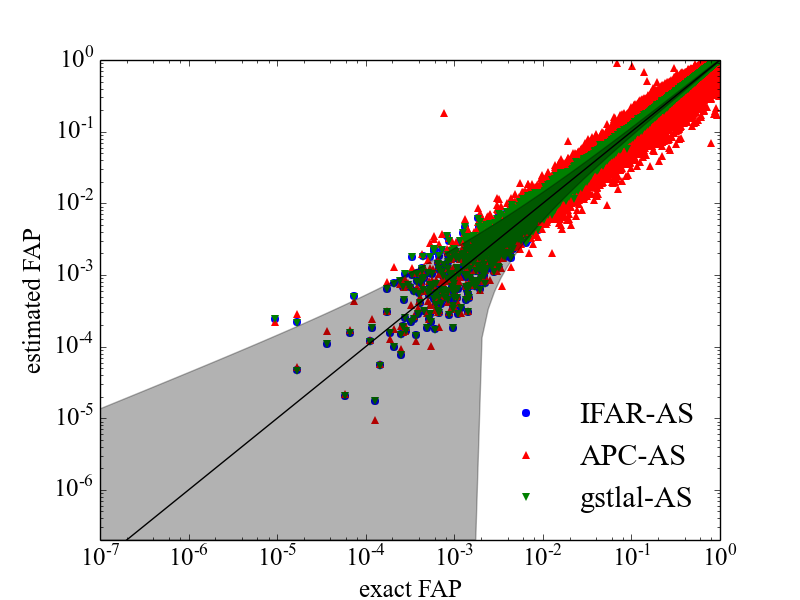}
    \caption{direct comparison with `all samples'}
    \label{fig:cmp1nonrm}
  \end{subfigure}
  \caption{Direct comparisons on experiment 1.\label{fig:cmp1}} 
\end{figure*}

\begin{figure*}
  \centering
  \vspace*{-0.5cm}
  \begin{subfigure}[b]{\columnwidth}
    \centering
    \includegraphics[width=0.9\columnwidth]{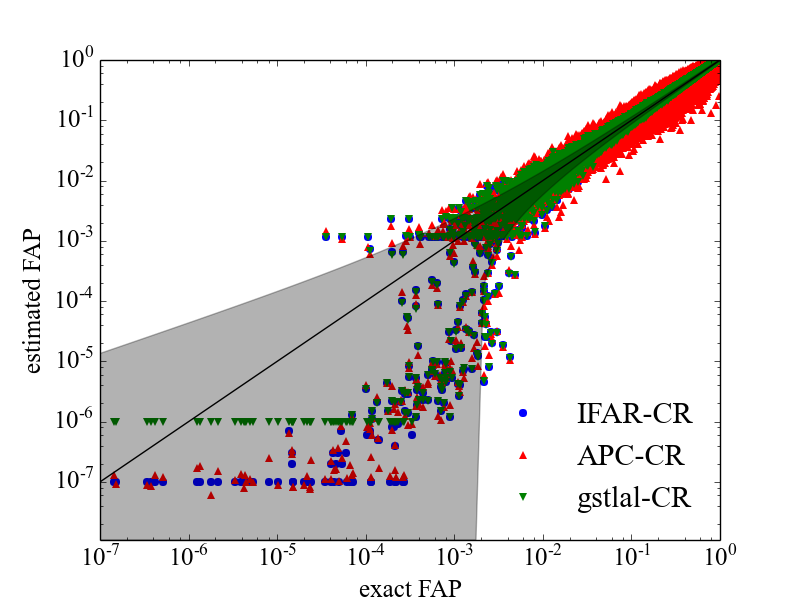}
    \caption{direct comparison with `coincidence removal'}
    \label{fig:cmp2rm}
  \end{subfigure}
  \begin{subfigure}[b]{\columnwidth}
    \centering
    \includegraphics[width=0.9\columnwidth]{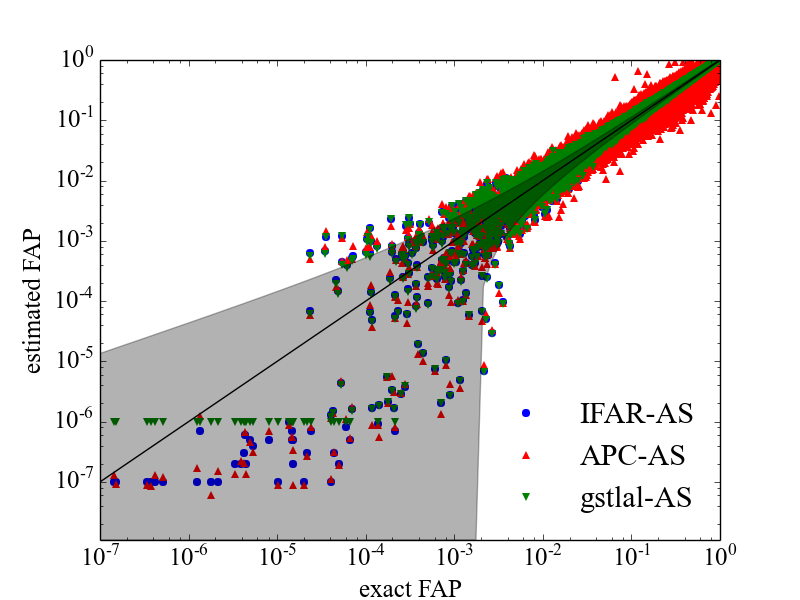}
    \caption{direct comparison with `all samples'}
    \label{fig:cmp2nonrm}
  \end{subfigure}
  \caption{Direct comparisons on experiment 2.
\label{fig:cmp2}}
\end{figure*}

\begin{figure*}
  \centering
  \begin{subfigure}[b]{\columnwidth}
    \centering
    \includegraphics[width=0.9\columnwidth]{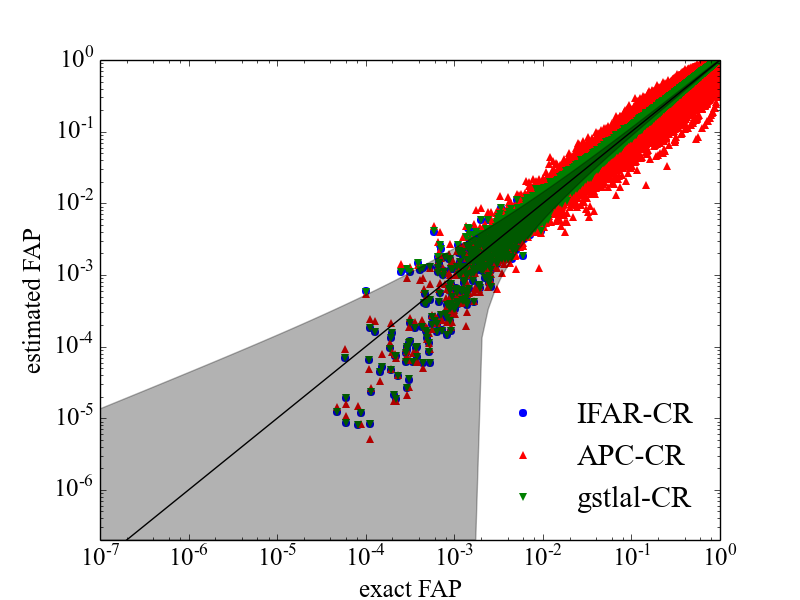}
    \caption{direct comparison with `coincidence removal'}
    \label{fig:cmp3rm}
  \end{subfigure}%
  \begin{subfigure}[b]{\columnwidth}
    \centering
    \includegraphics[width=0.9\columnwidth]{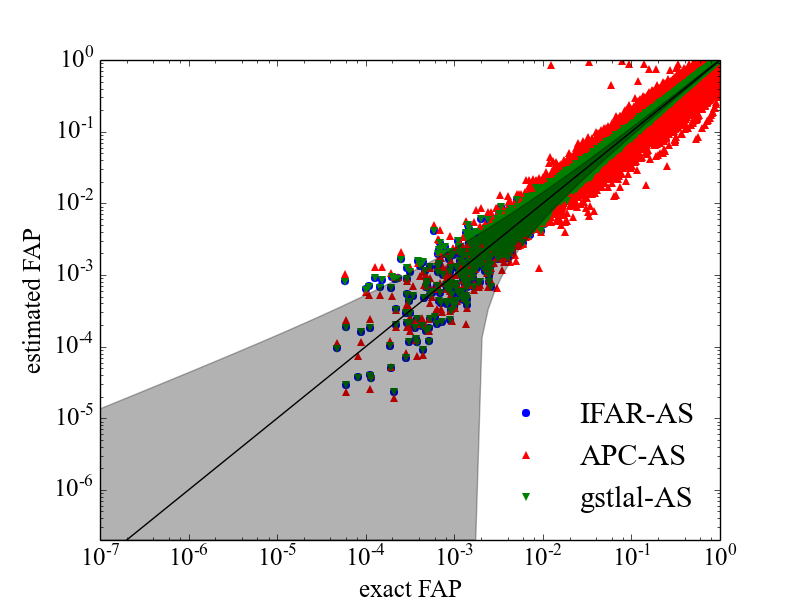}
    \caption{direct comparison with `all samples'}
    \label{fig:cmp3nonrm}
  \end{subfigure}
  \caption{Direct comparisons on experiment 3. 
\label{fig:cmp3}}
\end{figure*}

\begin{figure*}
  \centering
  \begin{subfigure}[b]{\columnwidth}
    \centering
    \includegraphics[width=0.9\columnwidth]{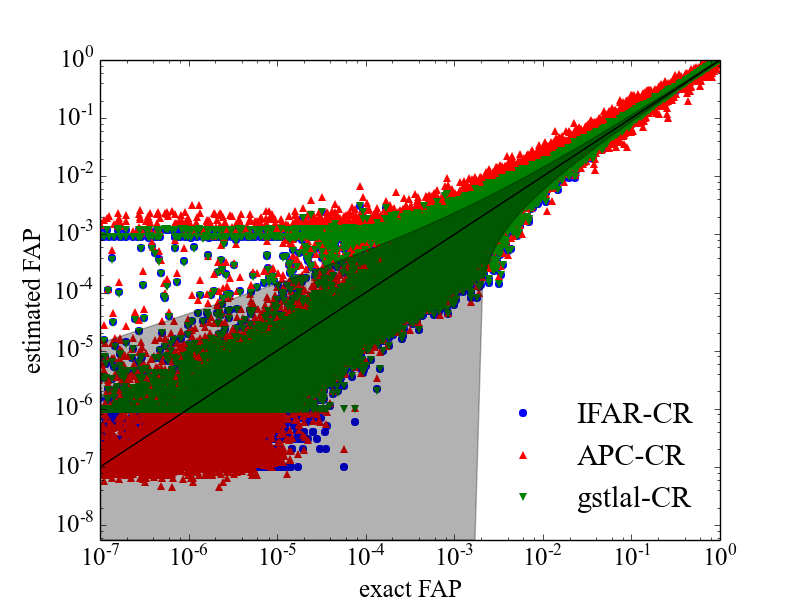}
    \caption{direct comparison with `coincidence removal'}
    \label{fig:cmp4rm}
  \end{subfigure}%
  \begin{subfigure}[b]{\columnwidth}
    \centering
    \includegraphics[width=0.9\columnwidth]{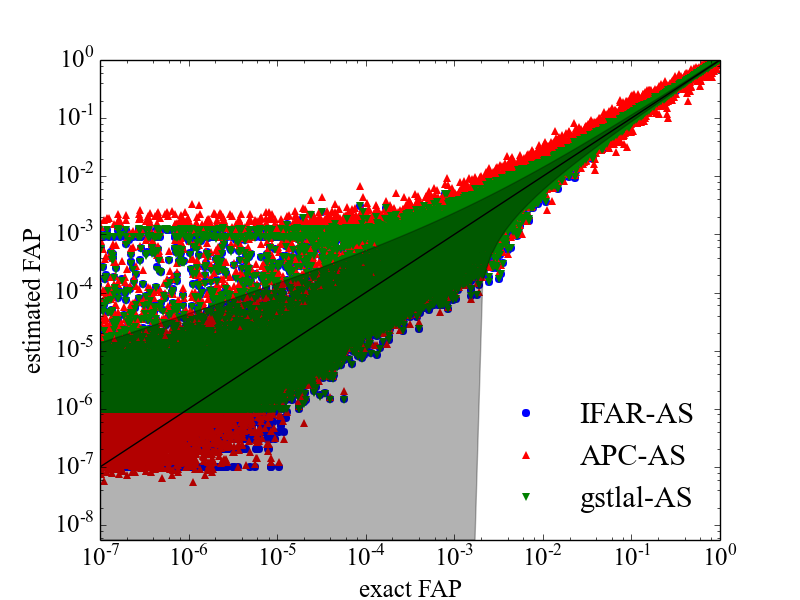}
    \caption{direct comparison with `all samples'}
    \label{fig:cmp4nonrm}
  \end{subfigure}
  \caption{Direct comparisons on experiment 4. 
\label{fig:cmp4}}
\end{figure*}

\begin{figure*}
  \centering
  \begin{subfigure}[b]{\columnwidth}
    \centering
    \includegraphics[width=0.9\columnwidth]{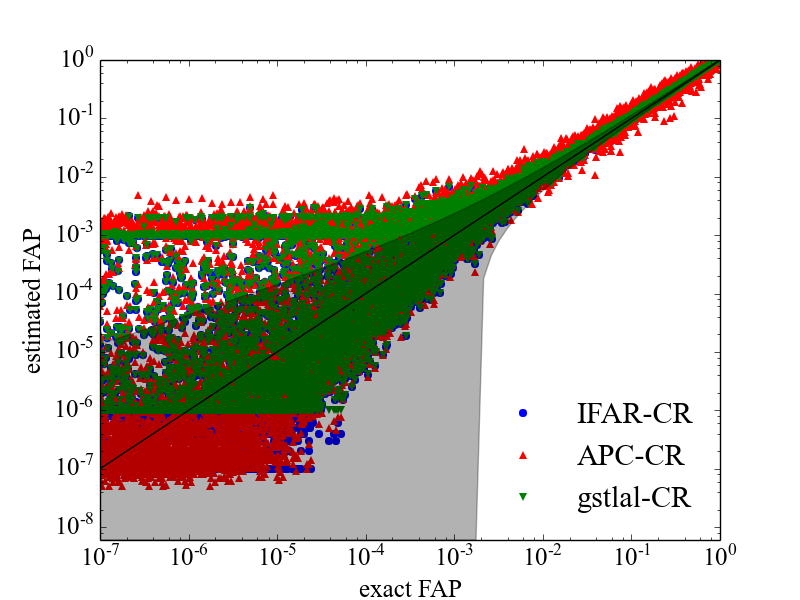}
    \caption{direct comparison with removal}
    \label{fig:cmp5rm}
  \end{subfigure}%
  \begin{subfigure}[b]{\columnwidth}
    \centering
    \includegraphics[width=0.9\columnwidth]{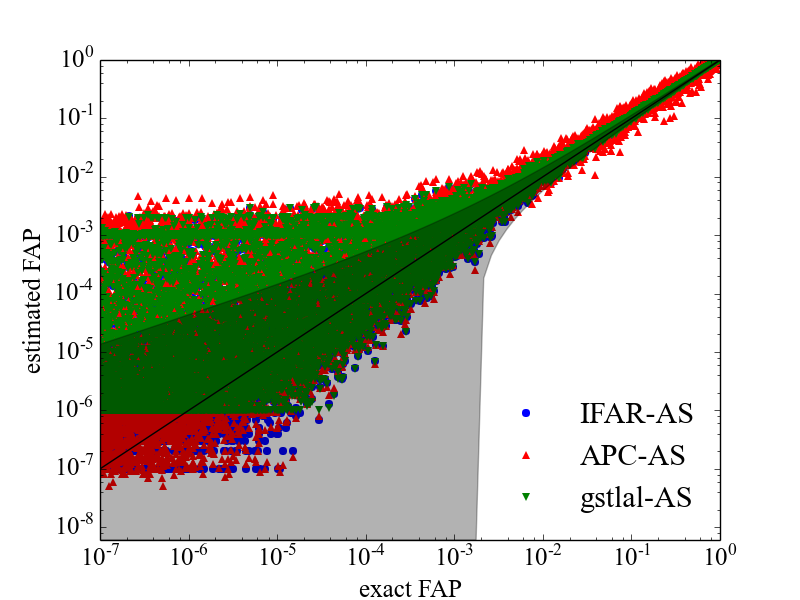}
    \caption{direct comparison with `all samples'}
    \label{fig:cmp5nonrm}
  \end{subfigure}
  \caption{Direct comparisons on experiment 5. 
\label{fig:cmp5}}
\end{figure*}

\begin{figure*}
  \centering
  \begin{subfigure}[b]{\columnwidth}
    \centering
    \includegraphics[width=0.9\columnwidth]{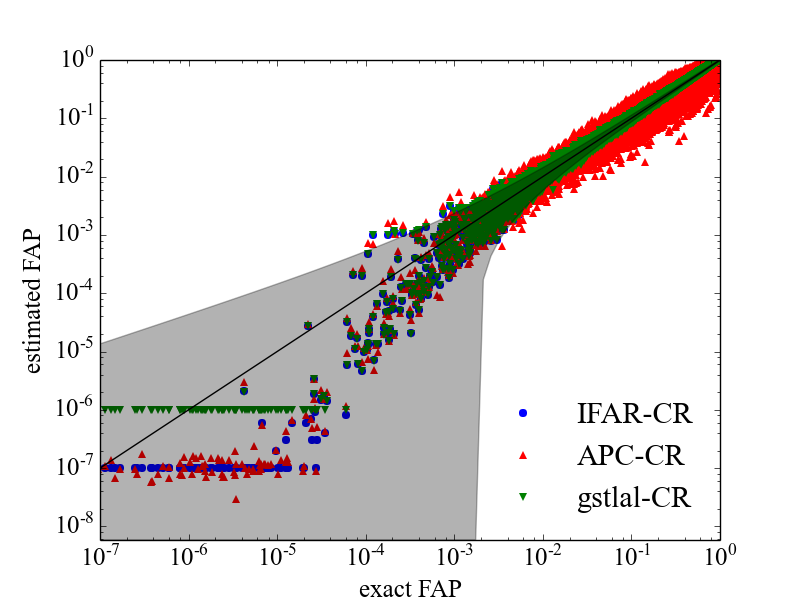}
    \caption{direct comparison with `coincidence removal'}
    \label{fig:cmp7rm}
  \end{subfigure}
  \begin{subfigure}[b]{\columnwidth}
    \centering
    \includegraphics[width=0.9\columnwidth]{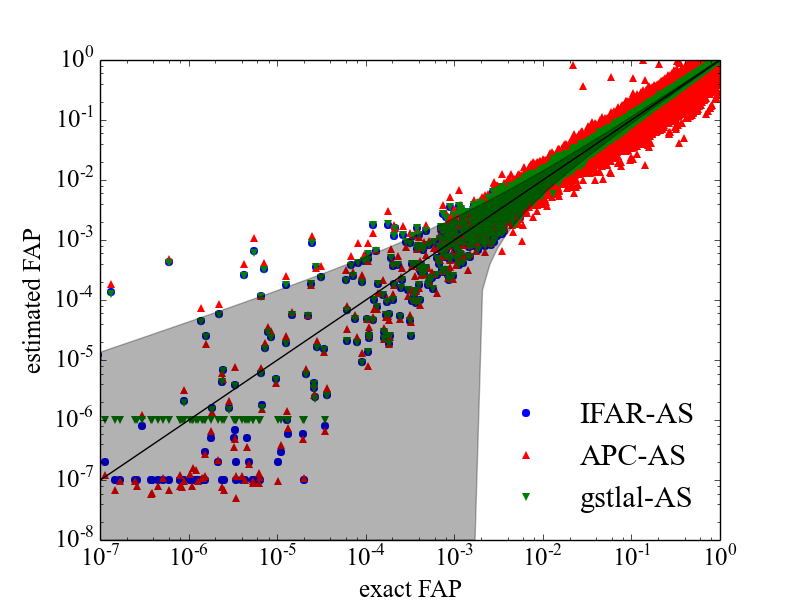}
    \caption{direct comparison with `all samples'}
    \label{fig:cmp7nonrm}
  \end{subfigure}
  \caption{Direct comparisons on experiment 7.
\label{fig:cmp7}}
\end{figure*}

\begin{figure*}
  \centering
  \begin{subfigure}[b]{\columnwidth}
    \centering
    \includegraphics[width=0.9\columnwidth]{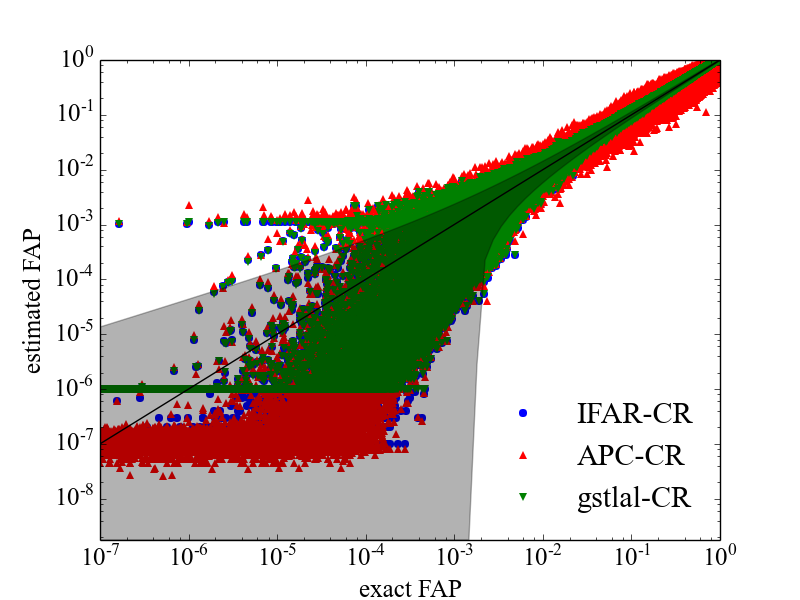}
    \caption{direct comparison with `coincidence removal'}
    \label{fig:cmp8rm}
  \end{subfigure}
  \begin{subfigure}[b]{\columnwidth}
    \centering
    \includegraphics[width=0.9\columnwidth]{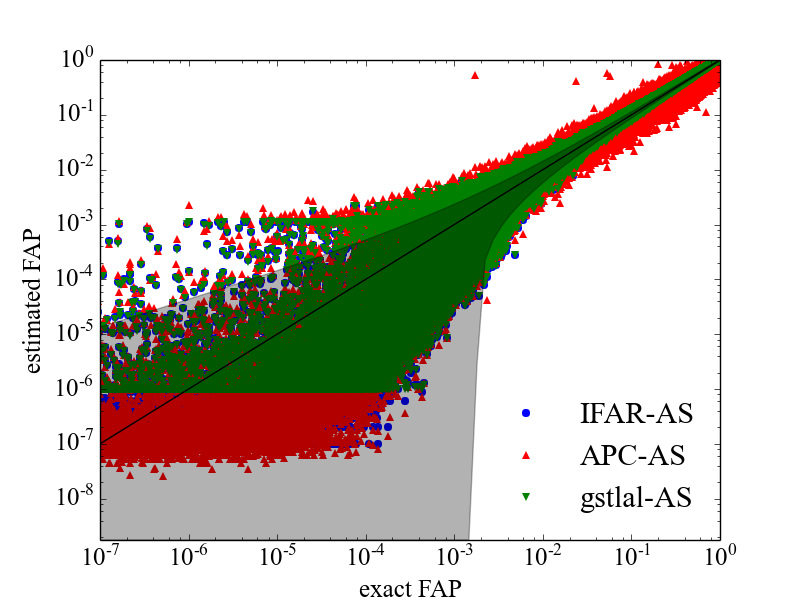}
    \caption{direct comparison with `all samples'}
    \label{fig:cmp8nonrm}
  \end{subfigure}
  \caption{Direct comparisons on experiment 8. 
\label{fig:cmp8}}
\end{figure*}

\begin{figure*}
  \centering
  \begin{subfigure}[b]{\columnwidth}
    \centering
    \includegraphics[width=0.9\columnwidth]{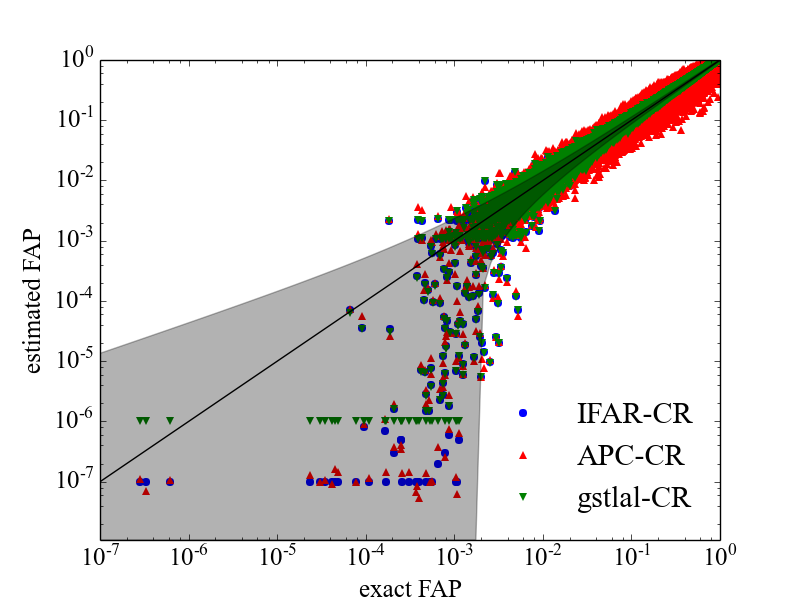}
    \caption{direct comparison with `coincidence removal'}
    \label{fig:cmp10rm}
  \end{subfigure}
  \begin{subfigure}[b]{\columnwidth}
    \centering
    \includegraphics[width=0.9\columnwidth]{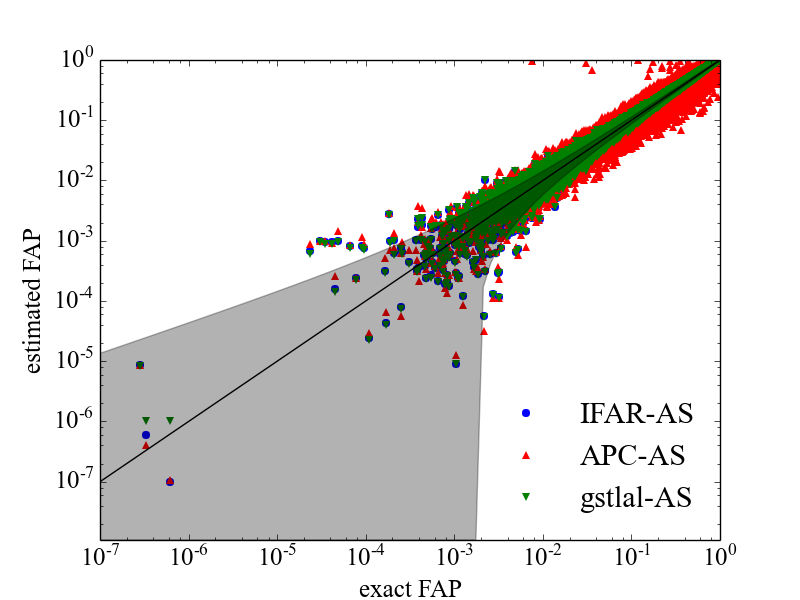}
    \caption{direct comparison with `all samples'}
    \label{fig:cmp10nonrm}
  \end{subfigure}
  \caption{Direct comparisons on experiment 10. 
\label{fig:cmp10}}
\end{figure*}

\begin{figure*}
  \centering
  \begin{subfigure}[b]{\columnwidth}
    \centering
    \includegraphics[width=0.9\columnwidth]{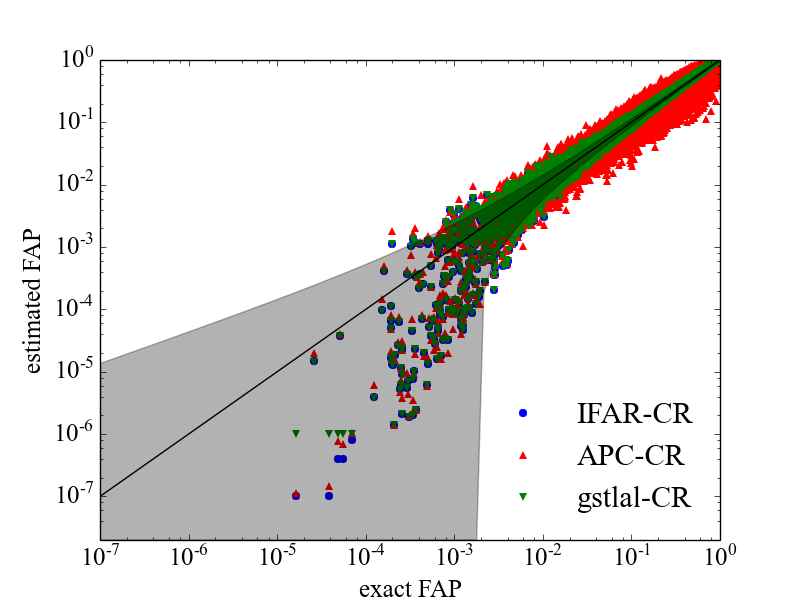}
    \caption{direct comparison with `coincidence removal'}
    \label{fig:cmp12rm}
  \end{subfigure}
  \begin{subfigure}[b]{\columnwidth}
    \centering
    \includegraphics[width=0.9\columnwidth]{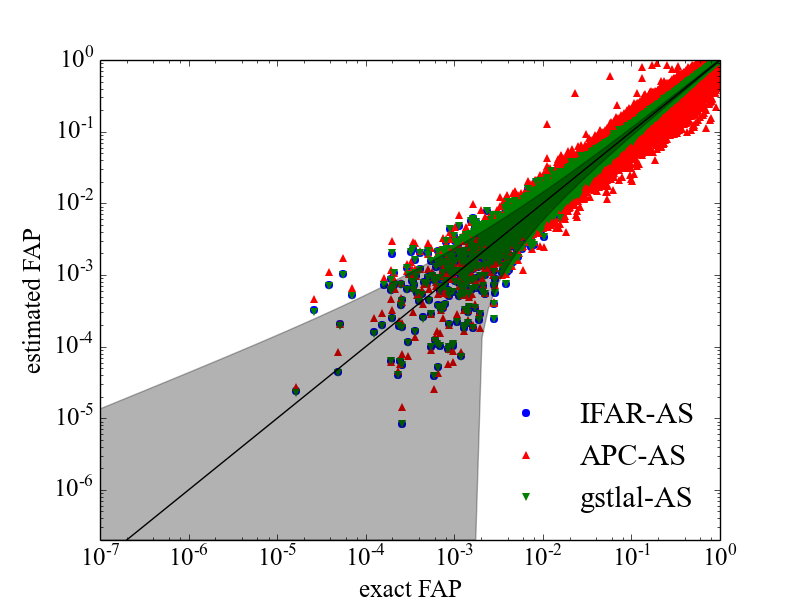}
    \caption{direct comparison with `all samples'}
    \label{fig:cmp12nonrm}
  \end{subfigure}
  \caption{Direct comparisons on experiment 12. 
\label{fig:cmp12}}
\end{figure*}

\begin{figure*}
  \centering
  \begin{subfigure}[b]{\columnwidth}
    \centering
    \includegraphics[width=0.9\columnwidth]{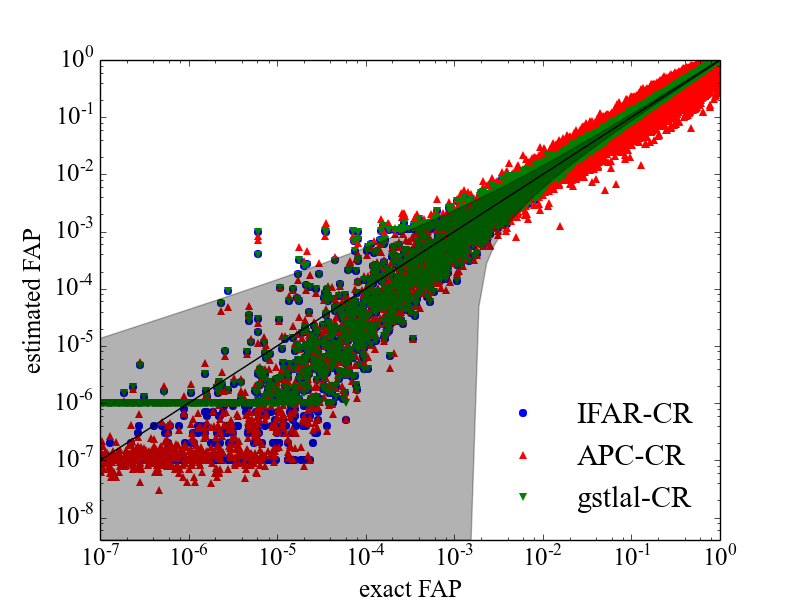}
    \caption{direct comparison with `coincidence removal'}
    \label{fig:cmp13rm}
  \end{subfigure}
  \begin{subfigure}[b]{\columnwidth}
    \centering
    \includegraphics[width=0.9\columnwidth]{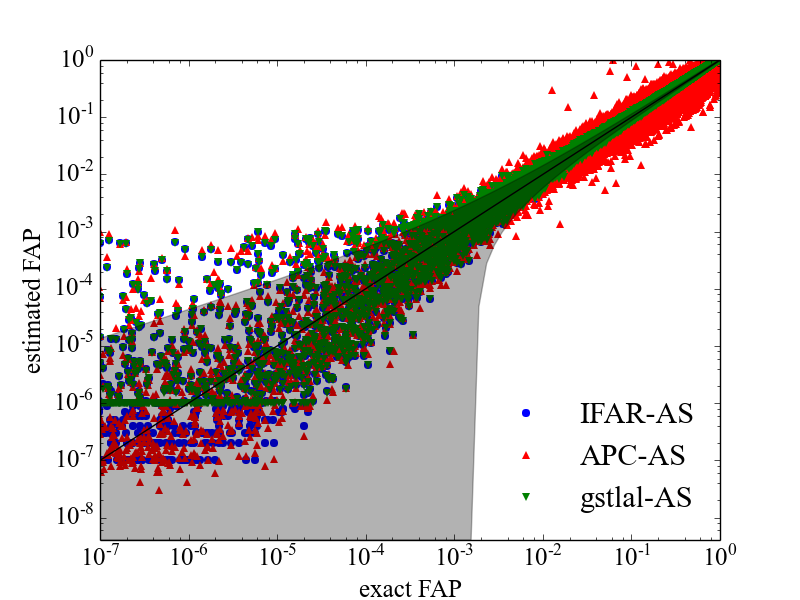}
    \caption{direct comparison with `all samples'}
    \label{fig:cmp13nonrm}
  \end{subfigure}
  \caption{Direct comparisons on experiment 13. 
\label{fig:cmp13}}
\end{figure*}

In
Figs.~\ref{fig:cmp1},~\ref{fig:cmp2},~\ref{fig:cmp3},~\ref{fig:cmp4},~\ref{fig:cmp5},~\ref{fig:cmp7},~\ref{fig:cmp8},~\ref{fig:cmp10},~\ref{fig:cmp12},
and~\ref{fig:cmp13} we present plots of the direct comparison between
actual \ac{FAP} and the \ac{FAP} estimations from all methods.

\clearpage

\subsection{\ac{ROC} plots}\label{sec:AppROC}
\begin{figure}[htbp]
  \centering
  \includegraphics[width=0.9\columnwidth]{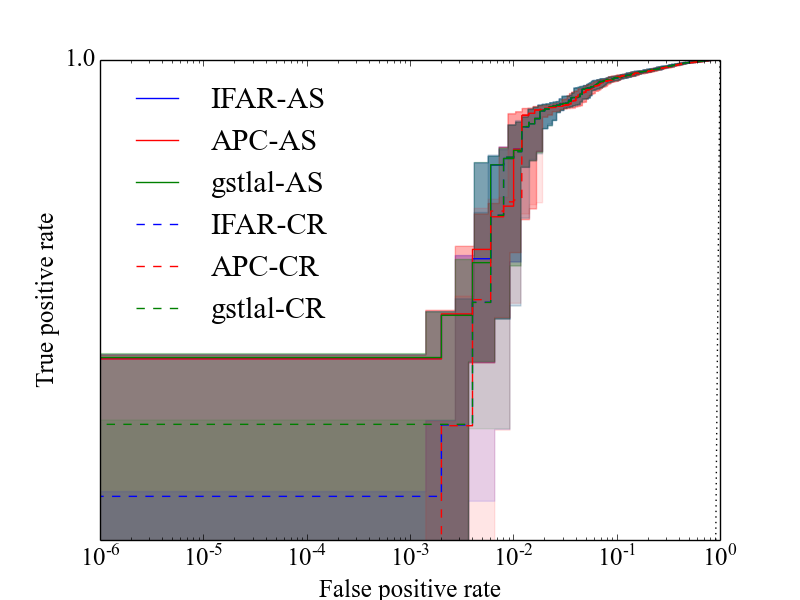}
  \caption{\ac{ROC} plot for experiment 5.}
  \label{fig:ROC_5}
\end{figure}

\begin{figure}[htbp]
  \centering
  \vspace*{-0.5cm}
  \includegraphics[width=0.9\columnwidth]{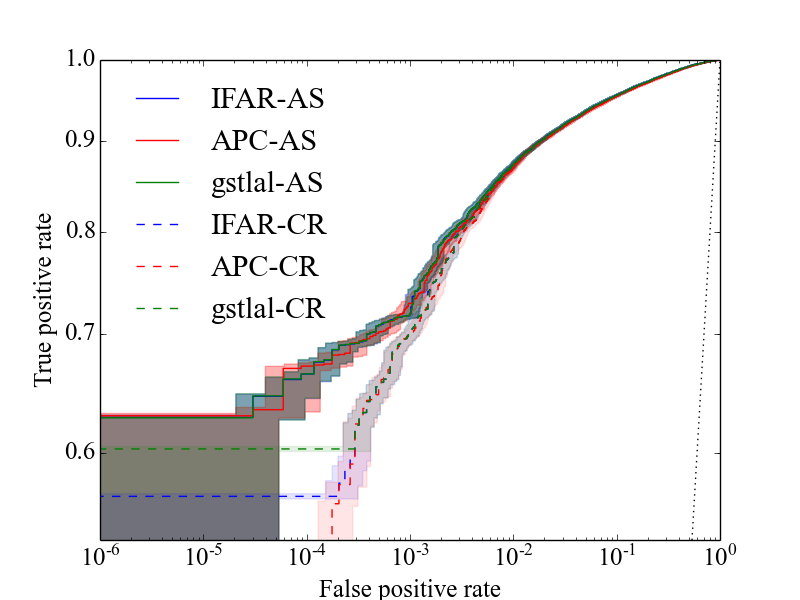}
  \caption{\ac{ROC} plot for experiment 6.}
  \label{fig:ROC_6}
\end{figure}

\begin{figure}[htbp]
  \centering
  \vspace*{-0.5cm}
  \includegraphics[width=0.9\columnwidth]{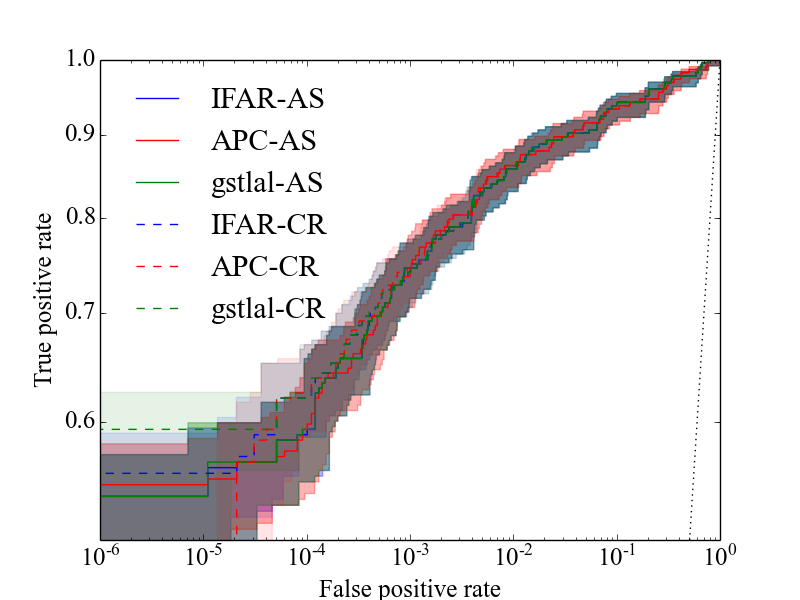}
  \caption{\ac{ROC} plot for experiment 7.}
  \label{fig:ROC_7}
\end{figure}

\begin{figure}[htbp]
  \centering
  \vspace*{-0.5cm}
  \includegraphics[width=0.9\columnwidth]{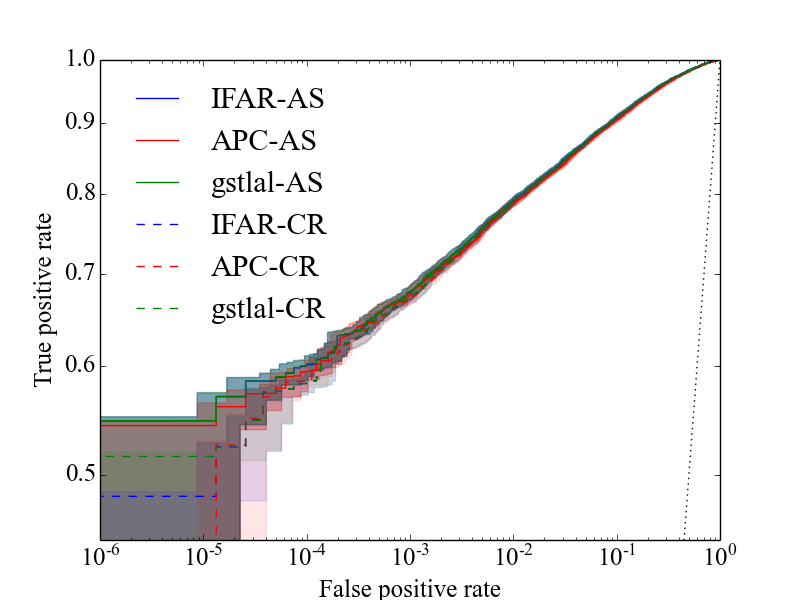}
  \caption{\ac{ROC} plot for experiment 9.}
  \label{fig:ROC_9}
\end{figure}

\begin{figure}[htbp]
  \centering
  \vspace*{-0.5cm}
  \includegraphics[width=0.9\columnwidth]{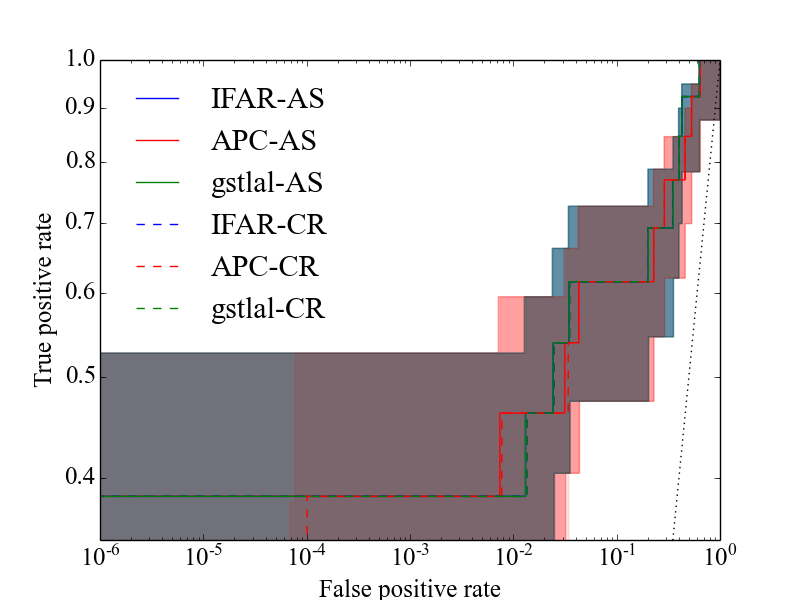}
  \caption{\ac{ROC} plot for experiment 10.}
  \label{fig:ROC_10}
\end{figure}

\begin{figure}[htbp]
  \centering
  \vspace*{-0.5cm}
  \includegraphics[width=0.9\columnwidth]{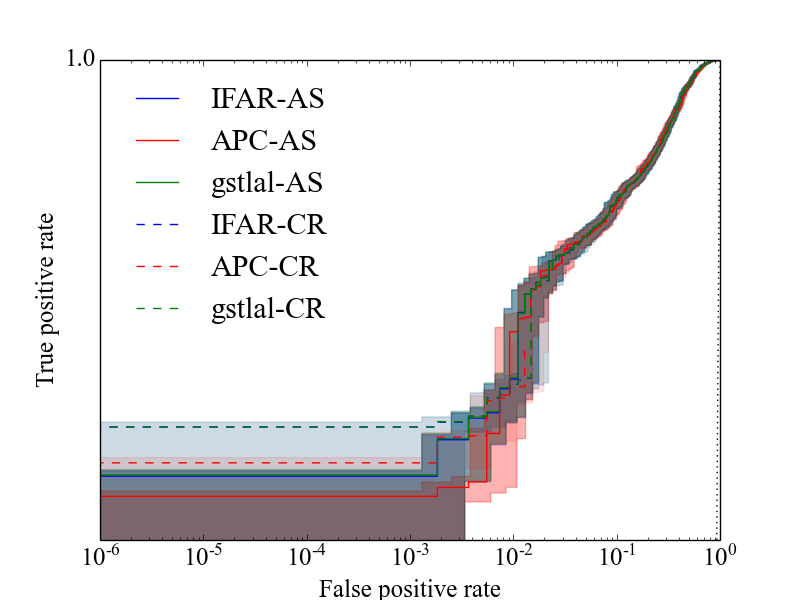}
  \caption{\ac{ROC} plot for experiment 11.}
  \label{fig:ROC_11}
\end{figure}

\begin{figure}[htbp]
  \centering
  \vspace*{-0.5cm}
  \includegraphics[width=0.9\columnwidth]{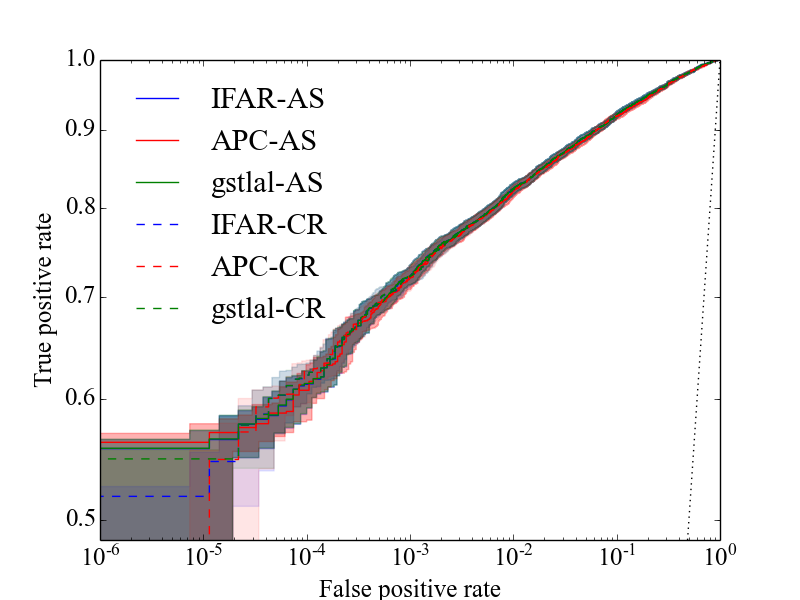}
  \caption{\ac{ROC} plot for experiment 13.}
  \label{fig:ROC_13}
\end{figure}

In this section we include all remaining \ac{ROC} plots. Note that it
is only possible to create a \ac{ROC} plot when there are foreground
events in the data, so we only show 7 \ac{ROC} plots,
Figs.~\ref{fig:ROC_5},~\ref{fig:ROC_6},~\ref{fig:ROC_7},~\ref{fig:ROC_9},~\ref{fig:ROC_10},~\ref{fig:ROC_11},
and~\ref{fig:ROC_13}, complementing those already shown in
Sec.~\ref{sec:ROC}.
 
\clearpage

\end{document}